\documentclass{aa}

\usepackage{graphicx}
\usepackage{siunitx}
\usepackage{lipsum}
\usepackage{bm}
\usepackage{xcolor}
\usepackage{upgreek}
\usepackage{enumitem}
\usepackage{makecell}
\usepackage{newtxtext}
\usepackage{anyfontsize}
\usepackage{placeins}
\usepackage[varvw]{newtxmath}
\usepackage{multirow}

\setlist[enumerate]{itemsep=1.5mm}

\newcommand{\rmn}[1]{\mathrm{#1}}
\newcommand{\bnabla}{\bm{\nabla}}

\newcommand{\bfit}[1]{\textbf{\textit{#1}}}

\DeclareSymbolFont{bmisymbols}{OML}{cmm}{b}{it}
\DeclareMathSymbol{\bupsilon}{0}{bmisymbols}{"1D}

\usepackage{xpatch}
\usepackage{hyperref}
\hypersetup{
	colorlinks=true,
	breaklinks=true,
	citecolor=blue,
	allcolors=blue,
	frenchlinks=true
}

\makeatletter
\xpatchcmd\NAT@citex
{%
	\@citea\NAT@hyper@{%
		\NAT@nmfmt{\NAT@nm}%
		\hyper@natlinkbreak{\NAT@aysep\NAT@spacechar}{\@citeb\@extra@b@citeb}%
		\NAT@date
	}%
}
{%
	\@citea
	\NAT@nmfmt{\NAT@nm}%
	\NAT@aysep\NAT@spacechar
	\NAT@hyper@{\NAT@date}%
}
{}{}
\xpatchcmd\NAT@citex
{%
	\@citea\NAT@hyper@{%
		\NAT@nmfmt{\NAT@nm}%
		\hyper@natlinkbreak{\NAT@spacechar\NAT@@open\if*#1*\else#1\NAT@spacechar\fi}%
		{\@citeb\@extra@b@citeb}%
		\NAT@date
	}%
}
{
	\@citea
	\NAT@nmfmt{\NAT@nm}%
	\NAT@spacechar\NAT@@open\if*#1*\else#1\NAT@spacechar\fi
	\NAT@hyper@{\NAT@date}%
}
{}{}
\makeatother

\makeatletter
\renewcommand*\aa@pageof{, page \thepage{} of \pageref*{LastPage}}
\makeatother

\begin{document} 
        
\title{Zooming in on cluster radio relics}
\subtitle{I. How density fluctuations explain the Mach number discrepancy, microgauss magnetic fields, and spectral index variations}

\author{Joseph Whittingham\inst{1,2}
    \and
    Christoph Pfrommer\inst{1}
    \and
    Maria Werhahn\inst{3}
    \and
    L\'{e}na Jlassi\inst{1,2}
    \and
    Philipp Girichidis\inst{4}
}

\institute{Leibniz-Institut f\"{u}r Astrophysik Potsdam (AIP), 
    An der Sternwarte 16, D-14482 Potsdam, Germany\\
    \email{jwhittingham@aip.de}
  \and
    Institut für Physik und Astronomie, Universität Potsdam, Karl-Liebknecht-Str. 24/25, 14476 Potsdam, Germany
  \and
    Max-Planck-Institut f\"{u}r Astrophysik , Karl-Schwarzschild-Str. 1, 85748 Garching, Germany
  \and
    Universit\"{a}t Heidelberg, Zentrum f\"{u}r Astronomie, Institut f\"{u}r Theoretische Astrophysik, Albert-Ueberle-Str. 2, 69120 Heidelberg, Germany \label{ITA}
}

\date{\today}

\abstract 
{
It is generally accepted that radio relics are the result of synchrotron emission from shock-accelerated electrons. However, current models are still unable to explain several aspects of their formation. In this paper, we focus on three outstanding problems: i) Mach number estimates derived from radio data do not agree with those derived from X-ray data, ii) cooling length arguments imply a magnetic field that is at least an order of magnitude larger than the surrounding intracluster medium (ICM), and iii) spectral index variations do not agree with standard cooling models. We used a hybrid approach to solve these problems; we first identified typical shock conditions in cosmological simulations and then used these to inform idealised shock-tube simulations, which can be run with a substantially higher resolution. We post-processed our simulations with the cosmic ray electron spectra code \textsc{Crest} and the emission code \textsc{Crayon+}, which allowed us to generate mock observables ab-initio. We observed that, upon running into an accretion shock, merger shocks generate a dense, shock-compressed sheet, which in turn runs into upstream density fluctuations. This mechanism directly gives rise to solutions to the three aforementioned problems: density fluctuations lead to a distribution of Mach numbers forming at the shock front. This flattens cosmic ray electron spectra, thereby biasing radio-derived Mach number estimates to higher values. We show that such estimates are particularly inaccurate in weaker shocks ($\mathcal{M} \lesssim 2$). Secondly, the density sheet becomes Rayleigh-Taylor unstable at the contact discontinuity, which causes turbulence and additional compression downstream. This amplifies the magnetic field from ICM-like conditions up to $\upmu$G levels. We show that synchrotron-based measurements are strongly biased by the tail of the distribution here too. Finally, the same instability also breaks the common assumption that matter is advected at the post-shock velocity downstream, thus invalidating laminar-flow-based cooling models.
}


\titlerunning{Zooming in on cluster radio relics -- I.}
\authorrunning{Whittingham et al.}
\maketitle

\section{Introduction}
\label{sec:intro}

Radio relics have presented a problem for theorists ever since \citet{mills1960} reported an unusually strong spectral flux density of 57 $\upmu$Jy in the vicinity of Abell S753\footnote{This source is known today as 1401-33.}. Follow-up observations \citep[see, in particular,][]{hoskins1970, mcadam1977, wall1979, subrahmanyan2003} were able to progressively resolve this emission into two distinct regions: one roughly circular and centred on the cluster and one thin and arc-like, sited at the periphery of the cluster. Today, these are recognised as a radio {halo} and a radio {relic}\footnote{For historical reasons, the radio relics referred to in this paper have variously been called radio {gischts} \citep{kempner2004}, {giant radio relics} \citep{pinzke2013}, and cluster radio {shocks} \citep{vanweeren2019}.}, respectively.

In the 1970s, it was recognised that radio relics could not be fit by traditional models \citep{wall1979}. Whilst their emission follows a power law, as is characteristic of synchrotron radiation, the spectral index $\alpha$ (where $S \propto \nu^{\,\alpha}$, for flux density $S$ at frequency $\nu$) varies from the outer to the inner edge, typically starting at $\alpha \lesssim -0.5$ and steepening strongly thereafter \citep{feretti2012}. Inactive radio galaxies were initially favoured as sources \citep[see e.g.][]{goss1987, harris1993}, however, by the end of the 1990s, it was clear that not every relic could be spatially associated with one \citep{feretti1996}. Moreover, relics show no apparent spectral cutoff \citep{komissarov1994, rajpurohit2020b}, which, combined with the relatively short cooling times of megahertz- and gigahertz-emitting electrons, implies an in-situ source \citep{ensslin2002, kang2012}.

It was eventually recognised that relics actually trace cosmological shocks \citep{ensslin1998}. In particular, they predominantly trace low Mach number shocks ($\mathcal{M} \lesssim 3-5$) driven by cluster mergers \citep[see e.g.][]{hoeft2007, markevitch2007}. The observation of such shocks in the X-ray band is now well established, as is their spatial association with radio relics \citep[see e.g.][and references therein]{brunetti2014, vanweeren2019}. This link has been strengthened by observed correlations between the orientation of the radio relic and the merger axis \citep{vanweeren2011}, and scalings of the radio relic power with both the X-ray luminosity \citep{bonafede2012} and mass of the host cluster \citep{degasperin2014}.

Two main mechanisms have been proposed to link such shocks with the resultant highly polarised, megaparsec-sized emission. However, whilst adiabatic compression scenarios have not been wholly ruled out\footnote{Note, adiabatic compression plays a major role in the related category of radio {phoenixes}, which are typically less linear in shape and can generate torus-like morphologies \citep[see e.g.][]{ensslin2002, pfrommer2011, raja2023}.} \citep{ensslin2001, button2020}, it is questionable whether they can replicate the observed scaling relations \citep{colafrancesco2017} and spectral slopes \citep[see e.g. evidence in][]{vanweeren2017}. Instead, the prevailing theory is that the merger shocks \mbox{(re-)accelerate} existing populations of electrons at the cluster outskirts via diffusive shock acceleration \citep[DSA; ][]{fermi1949, drury1983, blandford1987}. This theory is not without problems either, however, as such shocks are expected to have very low acceleration efficiencies ($\lesssim 1$\%) \citep{kang2005, kang2013, mou2023}. Acceleration from the thermal pool is consequently incompatible with the observed spectral intensities \citep{pinzke2013, vazza2014, botteon2020}. To solve this issue, a population of pre-existing, semi-relativistic `fossil' electrons are usually invoked \citep{markevitch2005, kang2011, pinzke2013, vazza2015, botteon2020b}, although some competing theories have emerged, such as turbulent re-acceleration \citep{fujita2015} and the multi-shock scenario \citep{inchingolo2022}.

Whilst first-order Fermi re-acceleration has emerged as the standard paradigm for explaining radio relics, our understanding of their origin remains far from complete. In particular, we highlight the following seven major problems: 

\begin{enumerate}
    \item {What is the origin of the seed electrons needed for re-acceleration?}
    
    \vspace{0.3em}
    \hspace{0.5cm} Four competing schools of thought currently exist: i) seed electrons are accelerated via diffusive shock acceleration during structure formation events and accretion shocks and therefore have an external origin \citep{pinzke2013, vazza2023}, ii) electrons are injected by radio galaxies at the extremities of a cluster \citep{ensslin1998,degasperin2015, johnston2017, botteon2020b, vazza2023}, iii) seed electrons originate from active galactic nuclei (AGN) lobes emitted from the cluster core, \citep{ensslin2001, shulevski2015, vazza2021, zuhone2021}, and iv) the electron population is common with that of radio haloes, and hence turbulent re-acceleration is the explanation \citep{beduzzi2024}. Whilst these options are not necessarily mutually exclusive, they each suffer from issues. For example, respectively, i) has yet to be shown to work in a fully cosmological MHD simulation, where the Fokker-Planck equation is solved explicitly\footnote{\citet{pinzke2013} were able to match observed luminosities using a Fokker-Planck solver. However, they used a hydrodynamic simulation, and hence made assumptions for the magnetic field. On the other hand, \citet{boess2024} applied their spectral solver to a cosmological MHD simulation but found that their mock radio relics had luminosities a factor of 10-100 below their observable counterparts (see e.g. their sect.~4.4).}, ii) radio galaxies are not always observationally evident, iii) it is unclear how AGN lobes survive thermal instabilities during buoyancy, and iv) it is unclear whether radio haloes can extend as far as some radio relics.

    \item {What is the origin of relic morphology?}
    
    \vspace{0.3em}
    \hspace{0.5cm} Radio relics exhibit a wide range of morphologies including irregular forms, such as in the case of Abell 2256 \citep{vanweeren2012b}, more regular, arc-like forms, such as the `Sausage' relic \citep{kocevski2007, vanweeren2010}, and even relatively linear forms, such as in the case of the `Toothbrush' relic \citep{vanweeren2012}. Moreover, with increasing resolution, it is clear that most relics have a filamentary nature \citep{rajpurohit2020, rajpurohit2022}. These are often attributed to magnetic fields \citep{rudnick2022} but shock geometry no doubt also plays a role. Numerical studies have already made some inroads towards answering this question, with simulations of relic morphology \citep{wittor2019, wittor2021b, lee2024, nuza2024}, investigations into substructure and relic `patchiness' \citep{dominguez-fernandez2021, dominguez-fernandez2024}, and a demonstration of a potential formation process for so-called `wrong way' relics \citep{boess2023} all being published recently. However, we are still far from being able to explain the morphology of individual relics.
    \vspace{0.3em}

    \item {Mach numbers inferred from radio measurements are typically larger than those inferred from X-ray data. How do we explain this discrepancy?}
    
    \vspace{0.3em}
    \hspace{0.5cm} Temperature and density jumps can be measured using X-ray observations, allowing Mach numbers to be inferred through the standard jump conditions. Meanwhile, the spectral slope of the radio emission can also be used to infer Mach numbers, assuming  DSA. However, while in principle they measure the same shock, the two measurements rarely agree \citep[see][and references therein]{wittor2021, lee2024b}. Potential explanations include the fact that X-ray measurements are subject to projection effects \citep{wittor2021} and that radio emission may be biased to higher values as a result of a Mach number distribution \citep{hoeft2011, skillman2013, hong2015, roh2019, wittor2019, dominguez-fernandez2021, wittor2021}. However, a simulation that shows a full {ab-initio} development of this effect is missing from the literature.
    \vspace{0.3em}
    
    \item {Observations appear to imply $\upmu$G magnetic fields in radio relics. How is this possible given that the surrounding intracluster medium (ICM) has strengths at least an order of magnitude lower?}
    
    \vspace{0.3em}
    \hspace{0.5cm} Magnetic field estimates have been made in relics based on inverse Compton emission studies \citep{chen2008}, constraints provided by the width of the radio relic \citep{markevitch2005, hoeft2007, vanweeren2010}, and constraints using Faraday rotation \citep{bonafede2013}. All of these models suggest magnetic field strengths on the order of $\upmu$G, albeit with some scatter in their estimates. Observations \citep{brunetti2001, govoni2017} and simulations \citep[see e.g.][]{skillman2013, dominguez-fernandez2019, nelson2024} of the surrounding ICM, meanwhile, typically imply magnetic fields that are only a fraction of this. This cannot be covered by standard shock compression alone \citep{donnert2016}, although there has been some evidence that radio observations bias values slightly high \citep{wittor2019}.

    \item {Why are standard cooling models unable to match spectral variations?}
    
    \vspace{0.3em}
    \hspace{0.5cm} Four cooling models are typically applied when trying to analyse radio relics \citep[see e.g.][]{vanweeren2012, rajpurohit2020}. The most simple of these assumes a one-time injection\footnote{In contrast to \citet{winner2019}, we use `injection' and `acceleration' as equivalent phrases in this paper.} with subsequent cooling that is dependent on \citep[`KP'][]{kardashev1962, pacholczyk1970} or independent of \citep[`JP'][]{jaffe1973} the cosmic ray (CR) electron pitch angle, respectively. Extensions include extended injection until the present time \citep[`CI'][]{pacholczyk1970} or until a fixed time in the past \citep[`KGJP'][]{komissarov1994}. Each of these mechanisms generates a characteristic spectral shape, the evolution of which can be probed using `colour-colour' diagrams \citep{katz-stone1993}. These are created by correlating spectral index maps, thereby producing a phase space diagram with $\alpha^{\nu_3}_{\nu_1}(x,y)$ as a function of $\alpha^{\nu_2}_{\nu_1}(x,y)$, where $\nu_1 < \nu_2 < \nu_3$, and $x$ and $y$ are spatial coordinates. Spectral features then translate to distinctive trajectories in this plane. However, despite fine-tuning, models are typically unable to replicate the tracks produced by observed radio relics.
    \vspace{0.3em}
    
    \item {Recent studies imply that CR electron acceleration is only efficient above a critical Mach number of $\mathcal{M}_\mathrm{crit} \approx 2.3$. How do we reconcile this with reports of radio relics at shocks apparently weaker than this?}
    
    \vspace{0.3em}
    \hspace{0.5cm} There are now several claims in the literature that suggest that efficient CR electron acceleration can only take place above $\mathcal{M}_\mathrm{crit} \approx 2.3$. This is typically because shocks below this value do not excite sufficiently strong plasma waves via CR electron-driven instabilities\footnote{A notable exception is the study by \citet{vink2014}, who arrive at a similar value through an energetics argument.}. These waves are necessary in order to confine the CR electrons during their acceleration to radio-emitting energies. This result has been observed in particle-in-cell (PIC) simulations at the pre-acceleration phase of CR electrons in both quasi-parallel \citep{shalaby2022, gupta2024} and quasi-perpendicular shocks \citep{kang2019, ha2021, boula2024}. However, efficient acceleration may also fail at quasi-perpendicular shocks due to a lack of CR proton generated plasma waves \citep{ha2018, ryu2019} or due to conditions on the firehose instability \citep{guo2014}. Whilst shocks derived from radio data are generally consistent with the above studies, shocks derived from X-rays are highly inconsistent. Indeed, there are now a substantial number of relics reported with X-ray-derived Mach numbers below $\mathcal{M} =  2.3$ \citep[see][and references therein]{wittor2021}. How can we explain this apparent inconsistency?
    \vspace{0.3em}

    \item {If electrons are most efficiently accelerated at quasi-parallel shocks, why does polarisation data appear to imply quasi-perpendicular acceleration?}
    
    \vspace{0.3em}
    \hspace{0.5cm} Both observations \citep{masters2013, liu2019} and simulations \citep{crumley2019, winner2020, shalaby2021, shalaby2022} imply that DSA is most effective for electrons at quasi-parallel shocks\footnote{It should be noted that (stochastic) shock drift acceleration (SDA) is most effective in quasi-{perpendicular} shocks \citep[see e.g.][]{amano2022}. However, whilst this process may aid the pre-acceleration phase, it is unclear how quasi-perpendicular shocks can facilitate electrons crossing the shock front more than a few times. This would prevent SDA accelerating electrons to Lorentz factors on the order of $10^4$, as is necessary for radio synchrotron emission. Diffuse shock acceleration, therefore, appears to be the more likely candidate for the production of radio relics at cluster shocks.} \citep[see theory introduced in][]{bykov1999}. This, it seems, is in contradiction with observations of radio relics; for example, observations of the `Sausage' relic appear to show the magnetic field vectors aligning almost perfectly with the shock surface \citep{vanweeren2010, digennaro2021}. This behaviour was replicated in part by \citet{wittor2019} and \citet{dominguez-fernandez2021b}, however neither included magnetic-obliquity-dependent shock acceleration in their simulations. It remains to be seen whether these results can still be replicated in such a scenario.

\end{enumerate}

This paper marks the first in a series where we attempt to tackle the problems listed above. In this study, we tackle problems 3 -- 5. Our strategy is to first run cosmological simulations in order to identify typical shock conditions, and then use these to inform significantly higher-resolution shock tube simulations. In addition, we implement upstream density turbulence, as inferred from both observations \citep{simionescu2011, eckhert2015, ghirardini2018} and previous simulations \citep{nagai2011, zhuravleva2013, battaglia2015, angelinelli2021}. By employing this method, we combine the strengths of previous numerical approaches, whilst also resolving relevant physics that is, as yet, out of the reach of current-generation cosmological simulations.

The paper is organised as follows: in Sect.~\ref{sec:methodology}, we introduce both the codes and the simulation setup used in this study. In Sect.~\ref{sec:cosmological-analysis}, we present an analysis of a cosmological galaxy cluster merger simulation. This is then used to inform a series of shock tube simulations, the results of which we present in Sect.~\ref{sec:shock-tube-analysis}. In this section, we show that upstream density turbulence results in: shock corrugation and the generation of downstream velocity turbulence (Sect.~\ref{sec:RT}), the formation of a Mach number distribution at the shock front (Sect.~\ref{subsec:mach-dist}), flatter CR electron spectra (Sect.~\ref{sec:spectra}), breaking of the assumption that distance from the shock front is a reliable indicator of time cooled for individual electrons (Sect.~\ref{sec:laminar-flow}), an amplification of the magnetic field to $\upmu$G levels (Sect.~\ref{sec:magnetic-field}), and synchrotron emission that is better able to replicate observed colour-colour diagrams (Sect.~\ref{sec:synchrotron-emission}). We finish the paper with a discussion of the caveats of our model (Sect.~\ref{sec:discussion}) and a summary of our main conclusions (Sect.~\ref{sec:conclusions}).

\section{Methodology}
\label{sec:methodology}

\subsection{AREPO}
\label{subsec:arepo}

All simulations presented in this paper were run with the moving-mesh code \textsc{Arepo} \citep{springel2010, pakmor2016, weinberger2020}. This code employs a set of unfixed mesh-generating points to construct a volume-filling Voronoi tessellation. Upon this, the equations for an ideal magnetohydrodynamic (MHD) fluid are solved using a second-order finite-volume Godunov scheme \citep{pakmor2011, pakmor2013} based on an HLLD Riemann solver \citep{miyoshi2005}. 

This method inherits the advantages of both Eulerian and Lagrangian codes. For example, cells are kept within a factor of two of a predetermined target mass (see following sections for values), being refined and merged as necessary. This means that spatial resolution increases naturally with structural complexity, allowing computational power to be concentrated where the system is most dynamic. Meanwhile, mesh-generating points follow the motion of the gas, leading to quasi-Lagrangian behaviour, which has the effect of substantially reducing numerical diffusion \citep[see e.g.][]{pfrommer2022}. The result is increased accuracy at the same resolution compared to competing codes; in particular, \textsc{Arepo} outperforms standard smoothed-particle hydrodynamics (SPH) \citep[see comparison studies in][]{vogelsberger2012, sijacki2012, keres2012, bauer2012}.

Especially germane to this study is the impact this has on the resolution of the turbulent cascade \citep{kolmogorov1941}. Several studies have shown that \textsc{Arepo} can reproduce this \citep[see e.g.][]{bauer2012, whittingham2021}, with strong evidence that it can also reproduce the closely related growth of the magnetic coherence scale in a fluctuating small-scale dynamo, given sufficiently high resolution\footnote{Milky Way-like galaxy simulations require approximately $10^7$ particles before a sufficiently high Reynolds number is resolved \citep{pakmor2017,whittingham2021}.} \citep{pakmor2017, whittingham2021, whittingham2023, pfrommer2022, pakmor2024}. The same studies have shown that the employed Powell 8-wave divergence cleaner \citep{powell1999} is highly robust even in strongly dynamic flows.

\subsection{Cosmological simulations}
\label{subsec:cosmological-setup}

To make sure that our shock tubes probe the most relevant physics, we first analysed the formation of shocks in a series of cluster mergers. We used the PICO-Clusters (Plasmas In COsmological Clusters) suite (Berlok et al., in prep) for this purpose, which comprises of 25 individual zoom-in simulations. The simulated clusters vary in mass\footnote{Cluster masses are given as $M_{200} = 4/3 \pi R_{200}^3$, where $R_{200}$ is the radius of a sphere centred on the cluster within which the average density is 200 times the critical density of the Universe, $\rho_\rmn{crit}$.} between $10^{14.9}-10^{15.5}\,\rmn{M}_\odot$ at $z=0$ . Cosmological volumes are periodic with a side-length of 1~co-moving Gpc $h^{-1}$ and use a Planck 2018 cosmology \citep{planck2020}; i.e. Hubble's constant is $H_0 = 100 \,h~\rmn{km~s}^{-1}~\rmn{Mpc}^{-1} = 67.3$ km s$^{-1}$ Mpc$^{-1}$ with the density parameters for matter, baryons, and a cosmological constant being $\Omega_\text{m} = 0.316$, $\Omega_\text{b} = 0.049$, and $\Omega_\Lambda = 0.684$, respectively.

Each zoom-in simulation was evolved from $z=127$ and has a dark matter resolution of $5.9\times10^7\,\rmn{M}_\odot$ (target gas mass of  $1.1\times10^7\,\rmn{M}_\odot$) in the highest resolution region, which extends without contamination of lower resolution particles to a minimum of $3\times R_{200}$ throughout the simulation. This resolution is roughly comparable to that of the TNG300 \citep{nelson2019}, TNG-Cluster \citep{nelson2024}, and MilleniumTNG \citep{pakmor2023} simulations. Like these simulations we also used the IllustrisTNG galaxy formation model \citep{weinberger2017, pillepich2018}. This includes a \citet{springel2003} ISM, stellar and AGN  feedback \citep{springel2005, weinberger2017}, radiative cooling and metal enrichment \citep{wiersma2009, vogelsberger2013}, magnetic fields \citep{pakmor2013, marinacci2018}, and the \citet{schaal2015} shock finder. 

Cosmic ray protons and electrons were not included in this model. This is not a problem, however, as we focus at this stage purely on the study of shocks in low-redshift massive mergers; i.e. those most expected to create radio relics. Such shocks are only minimally affected by CR protons in clusters \citep{pfrommer2017} and are negligibly affected by CR electrons, owing to their small energy content \citep{winner2019}.

The IllustrisTNG model has been shown to produce clusters that compare well with observations \citep{springel2018,pillepich2018b}. We note, in particular, that density and pressure profiles compare favourably \citep{pakmor2023}. This is important for this study, as it is these properties specifically that we probed in order to create higher resolution shock tube simulations.

\subsection{Shock tube simulations}
\label{subsec:shock-tube-setup}
\subsubsection{Resolution}
\label{subsec:shock-tube-setup-resolution}

We used the up- and downstream gas properties extracted from the aforementioned cosmological simulations to create initial conditions for a series of shock tubes. These shock tubes are three-dimensional with a periodic volume of $1800 \times 300 \times 300$ kpc$^3$. The volume can be further separated into four distinct regions, each 450 kpc long:

$\bullet$ Region I: Low-resolution piston region

$\bullet$ Region II: Piston region with increasing resolution

$\bullet$ Region III: High-resolution upstream region

$\bullet$ Region IV: Upstream region with decreasing resolution

\begin{figure}
\includegraphics[width=\columnwidth]{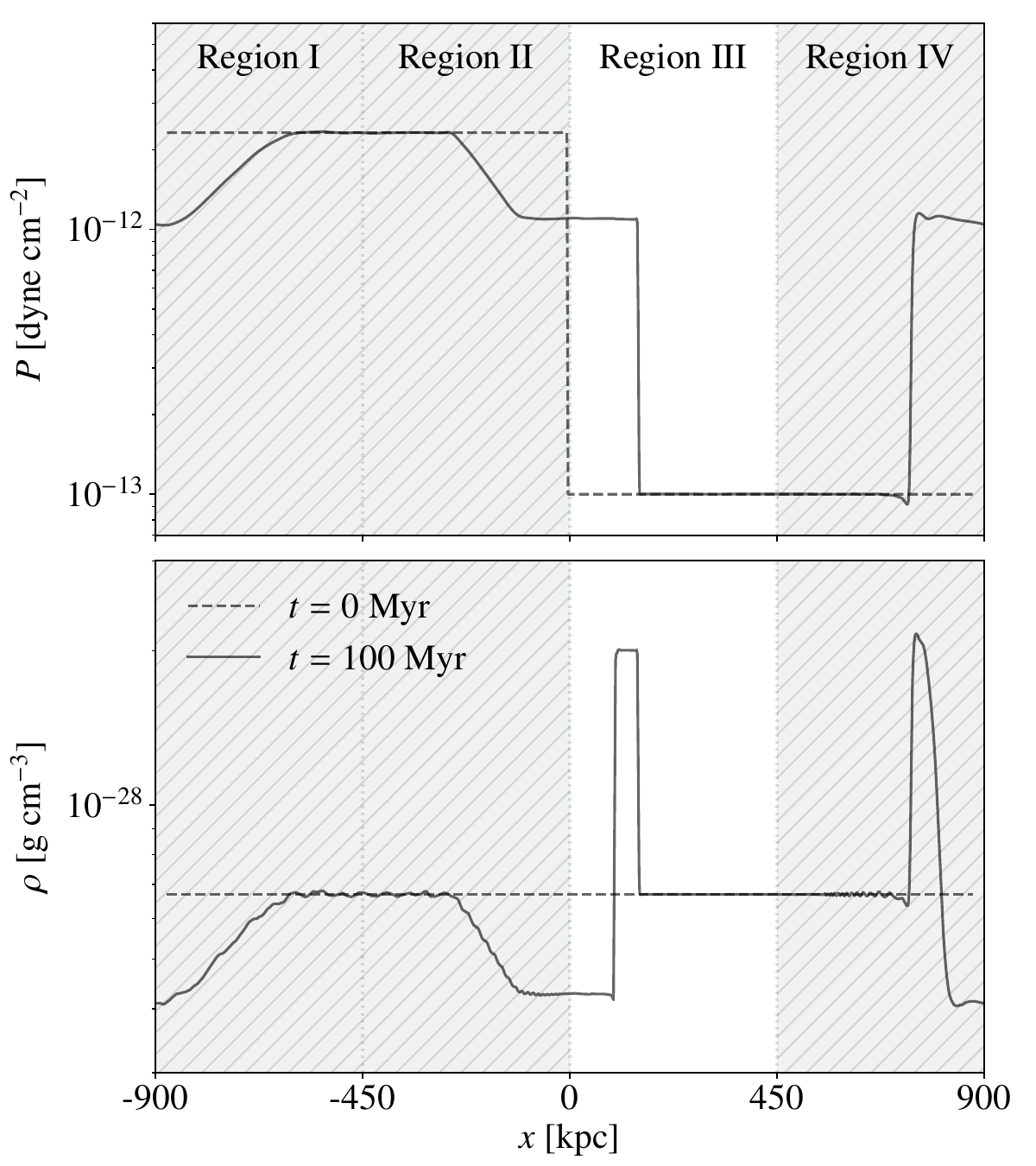}
\caption{{Top:} Mean pressure along the $x$ axis at $t=0$ Myr (dashed) and $t=100$ Myr (solid) in our Mach 3 `{Flat}' simulation (see Table~\ref{tab:sims}). Dotted lines indicate the edges of various regions in the shock tube (see Sect.~\ref{subsec:shock-tube-setup-resolution}). We greyed out panels that we do not analyse. {Bottom:} As above, but lines show the mean density. The initial pressure discontinuity causes a shock to propagate into region III. As a result of shock compression, a dense sheet moves rightwards, bounded by the density jumps at the contact discontinuity and the shock front. This feature is critical to the mechanism we analyse in this paper. }
\label{figure:1D-shock tube-profiles}
\end{figure}

\noindent We illustrate these regions along with the initial conditions for a Mach 3 shock in dashed lines in Fig.~\ref{figure:1D-shock tube-profiles}. Regions \text{I} and \text{IV} are necessary as our shock finder necessitates periodic boundary conditions. We must therefore allow for the reverse shock, as seen in Fig.~\ref{figure:1D-shock tube-profiles} in region \text{IV}. To mitigate for this, we varied the resolution across the volume, so that cells in region \text{I} have side-length 75 kpc, whilst cells in region \text{III} have a side-length of 2.5 kpc. By ramping the resolution up and down in this fashion, we aimed to focus our computational resources on region \text{III}, whilst also preventing the accumulation of truncation errors. 

We set the target gas mass resolution in the high-resolution region of our shock tubes to $m_\rmn{gas} \approx 1.5 \times 10^4 \,\rmn{M}_\odot$, with the exact value based on our initial density (see following section). We allowed the cells here to refine during the course of the simulation, which provides sub-kpc resolution after shock-compression. We fixed the resolution outside of region \text{III}, which leads to some minor variations from the expected density and pressure profiles to be evident in Fig.~\ref{figure:1D-shock tube-profiles}. These variations do not impact the accuracy of the dynamics in region III.

\subsubsection{Density and pressure values}
\label{subsec:density-and-pressure}

As we show in Sect.~\ref{sec:cosmological-analysis}, at the outskirts of clusters, merger shocks are able to generate dense, shock-compressed sheets. Indeed, this forms a critical part of the model for radio relics we introduce in this paper. This scenario can be replicated by setting the mean density of all four regions to be the same and driving the shock purely by setting a jump in pressure. Indeed, this turns out to be a good approximation to the scenario observed in our cosmological simulations. Following these simulations, we set the density in our shock tube to $n_\mathrm{e} = 3.5 \times 10^{-5}$ cm$^{-3}$ (or, equivalently, $\rho \approx 6.7 \times 10^{-29}$ g cm$^{-3}$) and the initial upstream pressure to $P_1 = 1 \times 10^{-13}$ dyne cm$^{-2}$. Regions I and II then form the high-pressure piston, whilst regions III and IV form the low-pressure upstream region into which the shock runs. These initial conditions are shown as dashed lines in Fig.~\ref{figure:1D-shock tube-profiles}. 

The Mach number of the shock is determined entirely in our case by the choice of pressure in the piston. Note, this differs from the classic \citet{sod1978} shock tube setup, in which density also plays a role. We chose to investigate the impact of $\mathcal{M}=2$ and $\mathcal{M}=3$ shocks, as these are typical values reported in radio relics \citep[see e.g.][]{vanweeren2017}. This fixed the initial downstream pressure to $P_2 = 1 \times 10^{-12}$ dyne cm$^{-2}$ and $P_2 = 2.32 \times 10^{-12}$ dyne cm$^{-2}$, respectively, with exact values being calculated using a Riemann solver. This setup generates a narrow, shock-compressed region as desired (see solid lines in Fig.~\ref{figure:1D-shock tube-profiles}). 

By initialising the shock in this way, we may test a variety of upstream conditions. Indeed, we examine a full range of variables in a companion paper (Whittingham et al., in prep.). In this current paper, however, we restrict ourselves simply to the impact of including density and magnetic turbulence at all.

\subsubsection{Simulation variations}
\label{subsec:sim-vars}

To unpin the impact of magnetic and density turbulence we ran four separate simulations. These are listed in Table~\ref{tab:sims} and cover the various possible permutations. Density turbulence was added only to region III, with the other regions given a constant density (see Sect.~\ref{subsec:density-and-pressure} for values). A short buffer region without density turbulence was also provided at the start of region III to allow for the initial formation of a planar shock.

In simulations with magnetic turbulence, on the other hand, this was added to all regions in order to keep magnetic divergence to a minimum. In simulations without magnetic turbulence, the magnetic field was oriented in the $x$-direction and given a fixed strength equal to the root-mean-square (RMS) value in the turbulent runs. We did not correlate the magnetic field and density fluctuations in this initial study, leaving a study of this improvement to future work.

\renewcommand{\arraystretch}{1.2}
\begin{table}
\caption{Shock tube simulations presented in this paper.}
\begin{tabular}{|l||l|l|}
\hline
{Name} & \makecell{{Turbulent density} \\ {fluctuations?}} & \makecell{{Turbulent magnetic} \\ {fluctuations?}} \\ \hline
{Flat-ConstB} & No & No \\ \hline
{Flat} & No & Yes \\ \hline
{Turb-ConstB} & Yes & No \\ \hline
{Turb} & Yes & Yes \\ \hline
\end{tabular}
\tablefoot{Our fiducial simulation is {Turb}. Each simulation was run twice: once in a Mach 2 and once in a Mach 3 variation (see Sect.~\ref{subsec:density-and-pressure}).}
\label{tab:sims}
\end{table}

\subsubsection{Generating turbulence}
\label{subsec:generating_turb}

We generated turbulence using the method given in \citet{ruszkowski2007} and \citet{ehlert2018}. In this method, turbulence is first created in $k$-space before being converted to configuration space using a fast Fourier transform. For density turbulence, we applied \citet{kolmogorov1941} turbulence ($P(k)\propto k^{-5/3}$) below an injection scale of 150 kpc, with white noise imposed above this. This is similar in magnitude to the scale used in previous shock tube studies of relics \citep[see e.g.][]{dominguez-fernandez2021}. We additionally implemented log-normal variance, as inferred from both simulations and observations \citep[see e.g.][]{kawahara2008}, using a Box-Muller random variate method. Following \citet{zhuravleva2013}, we set the relative variance of the distribution to $\sigma_\rho/\mu_\rho = 0.4$, where $\sigma_\rho$ is the standard deviation and $\mu_\rho$ is the mean of the density distribution, respectively.

We expect the peak of the magnetic power spectra to lag behind that of the kinetic power spectra by a factor of a few \citep[see, e.g.][]{tevlin2024}, and so chose an injection scale here of 40 kpc. As before, we applied Kolmogorov turbulence below this scale and white noise above it. We set the RMS strength, such that in the upstream $\langle P_\mathrm{th} / P_B\rangle = 100$, where $P_\mathrm{th}$ and $P_B$ are the thermal and magnetic pressures, respectively. We find this is typical of values in the ICM value in our cosmological simulations \citep{tevlin2024}. The result is an RMS field strength of approximately 0.16~$\upmu$G. Following, \citet{ehlert2018}, we also applied Gaussian variance, with a standard variation, $\sigma_B$, determined by Parseval's formula:
\begin{equation}
    \sigma_B^2 = \frac{1}{N^2} \sum_\bfit{k} | \bfit{B}_{i, \bfit{k}} |^2,
\end{equation}
where $i$ covers the three spatial components, \textit{N} is the number of cells in the initial conditions, and the sum runs over all \textit{k}-values. Magnetic divergence was cleaned by projecting it out in $k$-space, i.e. by subtracting $\hat{\bfit{k}} (\hat{\bfit{k}} \bm{\cdot} \bfit{B}_\bfit{k})$.

Note that our method of creating turbulence differs fundamentally from the method used in the shock tubes of \citet{dominguez-fernandez2021, dominguez-fernandez2021b, dominguez-fernandez2024}, who first drive turbulence, before letting it then decay over time. In the shock tube simulations presented in this study, all cells are initially at rest; i.e. there is no initial velocity turbulence. We discuss the advantages and disadvantages of our method in Sect.~\ref{sec:discussion}.

\subsubsection{CR physics and tracer particles}
\label{subsec:tracers}

We used the standard \textsc{Arepo} code (see Sect.~\ref{subsec:arepo}) in our shock tubes, augmented with the \citet{pfrommer2017} CR proton module, in which CR protons are treated as a relativistic fluid with effective adiabatic index, $\gamma_\rmn{a}= 4/3$. We injected CR protons at shocks according to DSA theory (see Sect.~\ref{subsec:shock finder}) and advected them thereafter with the gas. We did not account for streaming and diffusion, as strong turbulence in the vicinity of the shock is believed to decrease the diffusion coefficient to values approaching Bohm diffusion \citep{caprioli2014}. This leaves advection the dominant transport process. Cosmic ray protons are dynamically unimportant in our simulations, but are intrinsically linked to the modelling of CR electrons (see Sect.~\ref{subsec:crest}). 

As \textsc{Arepo} is only a quasi-Lagrangian code, we also employed the use of tracer particles \citep{genel2013} for CR electron modelling. One tracer particle was placed in each upstream cell, with additional tracers placed at the same resolution at the end of region II, just behind the initial pressure discontinuity. This allows us to take advantage of the contact discontinuity that forms (see Fig.~\ref{figure:1D-shock tube-profiles}), helping us to define the volume of cells in the injected region. In total, this results in approximately $2.9\times10^6$ tracer particles. The tracer particle functionality has been adapted to save values on-the-fly in order to be used by \textsc{Crest} \citep{winner2019}. Of particular importance is the recording of up- and downstream shock properties. For this, we used a numerical shock finder.

\subsection{Shock finder}
\label{subsec:shock finder}

In both cosmological and shock tube simulations, we used the \citet{schaal2015} shock finder, with extensions for CRs, as presented in \citet{pfrommer2017}. This is based on the Rankine-Hugoniot jump conditions with modifications for the case of additional CR pressure and injection. These produce the formula

\begin{equation}
\label{eq:mach-number}
\mathcal{M}^2 = \left(\frac{P_2}{P_1} - 1 \right) \frac{x_\rmn{s}}{\gamma_\mathrm{a, eff}(x_\rmn{s} - 1)},
\end{equation}
where $\gamma_\mathrm{a, eff}$ is the effective adiabatic index, $P_1$ and $P_2$ denote the total pre- and post-shock pressures, and $x_\rmn{s}$ is the density jump at the shock.

The dissipated energy flux at the shock can be expressed as
\begin{equation}
    \dot{E}_\rmn{diss} = \varepsilon_\rmn{diss}A_\rmn{shock}\frac{\mathcal{M}c_\mathrm{s,1}}{x_\rmn{s}},
    \label{eq:shock-dissipated-energy}
\end{equation}
where $\varepsilon_\rmn{diss}$ is the post-shock energy density minus the adiabatically compressed pre-shock energy density, $A_\rmn{shock}$ is the cell's shock surface, and $c_{\rmn{s},1}$ is the upstream sound speed \citep[see][for details]{pfrommer2017}.

We guarded against spurious shocks by further ensuring the following conditions:

\begin{enumerate}[label=(\roman*)]
    \item $\bm{\nabla} \bm{\cdot} \bm{\bupsilon} < 0$
    \item $\bm{\nabla}T \bm{\cdot} \bm{\nabla}\rho > 0$
    \item $\mathcal{M} > \mathcal{M}_\mathrm{min}$
    \item $x_\rmn{s} > x_\mathrm{s,\,min,}$  
\end{enumerate}
where $\bm{\bupsilon}$, $T$, and $\rho$, are the gas velocity, temperature, and density respectively. The guards act to: i) ensure converging velocity flows, ii) filter against tangential and contact discontinuities, iii) ensure a minimum Mach number (here, $\mathcal{M}_\mathrm{min} = 1.3$), and iv) ensure a minimum density jump, respectively. This last criterion is especially important for ensuring the stability of Eq.~\eqref{eq:mach-number} at weak shocks and is calculated using the hydrodynamic jump condition with $\mathcal{M}_\mathrm{min}$.

Shocks are numerically broadened in our simulations; typically with a thickness of two to three cells. The cell with the minimum velocity divergence (i.e. maximum compression) is labelled the `{shock surface}' cell, whilst the cell directly outside the shock zone is labelled the `{post-shock}' cell. We ran the shock finder in the before-snapshot mode in the cosmological simulation and in the on-the-fly mode in our shock tubes. 

\subsection{CREST}
\label{subsec:crest}

Cosmic ray electron modelling in this study was performed using the post-processing code \textsc{Crest} \citep{winner2019}. This code uses the previously discussed tracer functionality in \textsc{Arepo} to store relevant data on the MHD timestep, allowing us to evolve the Fokker-Planck equation in the Lagrangian frame. In particular, we accounted for adiabatic changes; cooling via  Coulomb, bremsstrahlung, inverse Compton, and synchrotron losses; and DSA with magnetic obliquity \citep[see formulae in][]{pais2018, winner2020}. In principle, \textsc{Crest} is also capable of Fermi I re-acceleration, Fermi II momentum diffusion, and Bell amplification, but we did not use these features in the following work.

On each tracer particle, \textsc{Crest} samples the underlying CR electron spectral density field. In practice, this means that all points closest to a given tracer are represented by that tracer's spectrum. To compute a thermodynamically extensive quantity such as the CR electron energy, we therefore constructed the volume associated with a given tracer by using a Voronoi tessellation of space with the tracer position as the mesh-generating point. This gives the tracer an evolving volume, $V_\rmn{cell}$, where $E_\rmn{e} = \varepsilon_\rmn{e} V_\rmn{cell}$ is the CR electron energy in that volume, and $\varepsilon_\rmn{e}$ is the CR electron energy density obtained by taking the second moment of the CR electron spectrum.

Tracers were initially assigned a purely thermal spectrum. This is acceptable, as whilst a pre-existing non-relativistic population is likely required to reach radio relic luminosities (see Sect.~\ref{sec:intro}), the addition of this feature should only affect the normalisation of the spectra, not its shape at radio-emitting frequencies \citep[see e.g.][]{pinzke2013, winner2019}. Modelling the impact of such a population is outside the scope of this work, and hence left to a future study.

For our cooling modules, we assumed that the gas is primordial and hence that the mass fraction of hydrogen is $X_\rmn{H} = 0.76$. We also assumed that the gas is fully ionised, resulting in a mean molecular weight of $\mu = 0.588$ and an ionisation fraction of $x_\rmn{e} = 1.157$. To aid comparison with observations we assumed a redshift of $z=0.2$, which is the approximate redshift of the Toothbrush and Sausage radio relics \citep{vanweeren2010, vanweeren2012}. The energy density of the cosmic microwave background (CMB) was therefore set to $\epsilon_\rmn{CMB} = 8.65 \times 10^{-13}$ erg cm$^{-3}$, or equivalently a magnetic field strength of $B_\rmn{CMB} \approx 4.7$ $\upmu$G.

Throughout this work, we use dimensionless momentum, $p = \tilde{p} / (m_\rmn{e} c)$, where $\tilde{p}$ is the dimensional momentum, $m_\rmn{e}$ is the electron rest mass, and $c$ is the speed of light. We also convert the transport equation \citep[see][]{winner2019} into one-dimensional momentum space, such that $f^\rmn{1D} = 4 \pi p^2 f^\rmn{3D}$, where $f^\rmn{1D}$ and $f^\rmn{3D}$ are the one- and three-dimensional distribution functions, respectively. We let $p$ range in all simulations from $10^{-1}$ to $10^{8}$ and used 10 logarithmically spaced bins per decade. We find this is sufficient to produce converged spectra. In the remainder of this section, we summarise the DSA implementation, having the greatest impact on our study. We direct the reader, however, to \citet{winner2019} for a more thorough discussion of all the physics represented in the code.

Cosmic ray electrons are accelerated in \textsc{Crest} in shock surface and post-shock cells by attaching to the one-dimensional distribution function a power-law slope\footnote{Note, we define $\alpha < 0$ here, in contrast to the notation used in \citet{winner2019}.} with
\begin{equation}
\alpha = -\frac{x_\rmn{s} + 2}{x_\rmn{s}-1}
\label{spectral-slope}
.\end{equation}
This slope was limited to a maximum of -2.2, following \citet{caprioli2020}. This has a very minor impact on our results, however, given the weak shocks we simulate and the correspondingly steep spectral slopes. To avoid spurious heating before the shock due to numerical broadening, we stopped updating the recorded density on encountering a shocked cell. When the tracer reaches a shock surface or post-shock cell, the density was then updated to its post-shock value. This produces a discrete jump at the shock. 

Following \citet{webb1984} and \citet{pinzke2010}, we set the maximum momentum up to which we accelerate to
\begin{equation}
    p_\mathrm{max} = \frac{\bupsilon_\mathrm{post}}{c}\sqrt{\frac{6 \pi e}{\sigma_\mathrm{T}} \frac{B}{(B^2 + B_\mathrm{CMB}^2)} \frac{x_\rmn{s} (x_\rmn{s}-1)}{(x_\rmn{s}+1)}},
\end{equation}
where $\bupsilon_\mathrm{post}$ is the post-shock velocity in the shock rest-frame, $e$ is the elementary charge, $\sigma_\mathrm{T}$ is the Thomson cross section, and $B$ is the strength of the magnetic field.

We then required that $p_\mathrm{min}$, the minimum acceleration momentum, fulfils
\begin{equation}
\int\limits_{p_\mathrm{min}}^{p_\mathrm{max}} f_\mathrm{th}(p_\mathrm{min})p_\mathrm{min}^{-\alpha} p^{\alpha} E_\mathrm{e,\,kin}(p)\mathrm{d}p = \Delta\varepsilon_\mathrm{e},
\label{eq:p_min}
\end{equation}
where $f_\rmn{th}(p)$ is the thermal Maxwellian\footnote{See eq.~31 of \citet{winner2019}.}, $E_\mathrm{e,\,kin}(p) = (\sqrt{1+p^2} - 1) m_\mathrm{e} c^2 $ is the electron kinetic energy, and $\Delta\varepsilon_\mathrm{e}$ is the increase in energy density of the CR electron population due to acceleration. We used 0.1\% of the liberated thermal energy at the shock, converting this to an energy density by dividing by the cell volume \citep[see][]{winner2019}.

The normalisation of the injected spectrum is subsequently
\begin{equation}
C = f_\mathrm{th}(p_\mathrm{min}) p_\mathrm{min}^{-\alpha}.
\end{equation}
At the high momentum end of the spectrum, we imposed a super-exponential cutoff \citep[see][]{ensslin1998, zirakashvili2007, pinzke2010}, which modifies the spectrum thusly:
\begin{equation}
\tilde{f}(p) = f(p)\left[1 + 0.66\left(\frac{p}{p_\mathrm{max}}\right)^{2.5} \right]^{1.8} \exp{\left[ -\left(\frac{p}{p_\mathrm{max}}\right)^2 \right]}
.\end{equation}

As explained in Sect.~\ref{sec:intro}, there are now several studies that suggest that CR electron acceleration can only take place above a critical Mach number of $\mathcal{M}_\mathrm{crit} = 2.3$. Subsequently, we did not accelerate electrons below this Mach number in the simulations that include upstream density turbulence\footnote{We did not apply this condition to simulations without upstream density turbulence, as the Mach number distribution in this case is insufficiently broad.}. We show in Whittingham et al. (in prep.) that this has remarkably little impact on the subsequent emission, even when the Mach number distribution peaks at $\mathcal{M} = 2$. This is due to the fact that emission is dominated by the tail of the Mach number distribution. 

\subsection{CRAYON+}
\label{subsec:crayon}

In the final step of our methodology, we post-processed the resultant \textsc{Crest} output with the \textsc{Crayon+} code \citep{werhahn2021}, which is able to convert spectra into instantaneous emission for a range of CR electron processes. Naturally, we are most interested in the radio synchrotron emission. We summarise the salient formulae here, but encourage the interested reader to see section~2.4 and associated appendices of \citet{werhahn2021b} for a more comprehensive overview.

Following \citet{rybicki1986}, the synchrotron emissivity $j_\nu$ of a tracer at frequency $\nu$ is
\begin{equation}
j_\nu = \frac{\sqrt{3}e^3 B_\perp}{m_\rmn{e} c^2} \int\limits_{0}^{\infty} f(p) F(\nu / \nu_c) \mathrm{d}p \;\,\propto\; B_\perp^{1-\alpha_\nu}\nu^{\alpha_\nu},
\label{eq:synchrotron-emissivity}
\end{equation}
where $B_\perp$ is the projection of the magnetic field onto the plane perpendicular to the line of sight, $\alpha_\nu$ is the radio spectral index, and $F(\nu/\nu_\rmn{c})$ is the dimensionless synchrotron kernel, with $\nu_\rmn{c}$ the critical frequency. These last two variables are, in turn, defined as
\begin{equation}
\nu_\rmn{c} = \frac{3 e B_\perp \gamma^2}{4 \pi m_\rmn{e} c},
\label{eq:critical-frequency}
\end{equation}
where $\gamma = \sqrt{1 + p^2}$ is the Lorentz factor and
\begin{equation}
F(x) = x \int\limits_{x}^{\infty} K_{5/3} (\xi)\mathrm{d}\xi, 
\end{equation}
where $K_{5/3}$ is the modified Bessel function of order 5/3, with $x = \nu/\nu_c$.  To get the right-hand side of Eq.~\eqref{eq:synchrotron-emissivity}, we have assumed a power-law spectrum over an appropriate range of frequencies. 

The specific radio synchrotron intensity, $I_\nu$, at frequency $\nu$ is then obtained by integrating $j_\nu$ (with units of erg s$^{-1}$ Hz$^{-1}$ cm$^{-3}$) along the line of sight, $s$:
\begin{equation}
I_\nu = \frac{1}{4\pi}\int\limits_0^{\infty} j_\nu ds
\label{eq:intensity}
.\end{equation}
This gives $I_\nu$ in units\footnote{This may be converted to the more typical observational units of Jy~arcsec$^{-2}$ with the transformation 1 Jy arcsec$^{-2}$ = $4.25 \times 10^{-13}$ erg cm$^{-2}$ s$^{-1}$ Hz$^{-1}$ sterad$^{-1}$.} of erg cm$^{-2}$ s$^{-1}$ Hz$^{-1}$ sterad$^{-1}$. In the remainder of the paper, where projections are shown we have set the upper limit of the integral in Eq.~\eqref{eq:intensity} to be the depth of the box, equal to 300 kpc.

\section{Cosmological analysis}
\label{sec:cosmological-analysis}

\begin{figure*}
    \centering
    \includegraphics[width=2.0\columnwidth]{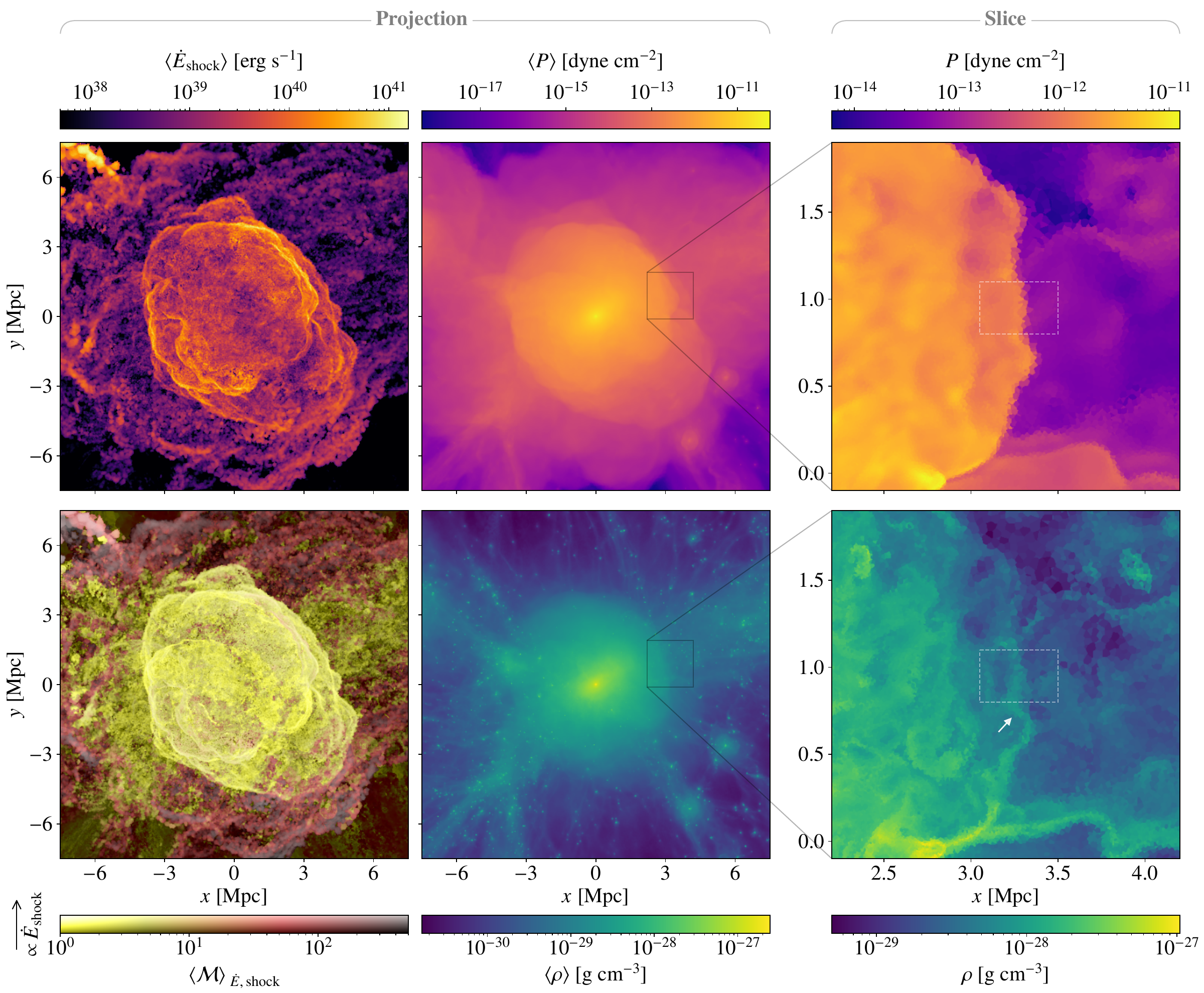}
    \caption{Cosmological simulation of a galaxy cluster undergoing a major merger at $z=0.14$. {Four left-most panels, clockwise from top left:} Projected shock-dissipated energy rate, gas pressure, gas density, and dissipation-weighted Mach number, respectively. Projections have a depth of $\pm 7.5$ Mpc from the cluster centre. {Right-most panels:} enlarged cut-outs showing slices through the midplane of the projection. The white, dashed box indicates the size of the upstream region in our idealised shock tube simulations. The merger drives a shock wave out to the cluster outskirts. The collision of this shock wave with an accretion shock leads to the production of a thin shell of compressed gas (see white arrow). A movie of this process can be found \href{https://www.youtube.com/watch?v=jKhqhaCTCL4}{here}.}
    \label{figure:cosmological}
\end{figure*}

As explained in Sects.~\ref{sec:intro} and~\ref{sec:methodology}, the first step in our strategy is to analyse shock conditions in cosmological simulations. To this end, we present here a case study. We use a variation\footnote{The simulation presented here has been run with Arepo-1, whilst the PICO-Cluster suite was run with Arepo-2.} of {Halo 0003} from the PICO-Cluster suite (Berlok et al., in prep) as it features a particularly energetic merger, making it especially likely that the generated shock wave will form a radio relic \citep[see similar shocks in e.g.][]{boess2023, lee2024}. However, our analysis generalises to all the cluster mergers we investigated.

The major merger in {Halo 0003} takes place between $z\approx0.3$ and $z\approx0.1$. At the start, the progenitors have $M_{200}$ masses of $10^{14.8}\,\rmn{M}_\odot$ and $10^{15.2}\,\rmn{M}_\odot$, giving a mass ratio of roughly 1:2.5. We show gas conditions during the merger in Fig.~\ref{figure:cosmological}, focusing on a period when the energy dissipation in the shock is at its most intense. At this time, the shock has become detached from the merging cluster that drove it, known as the `runaway' phase \citep[see definitions in][]{zhang2019}. It has reached a distance of a little under $1.8 \times R_{500}$, which is within the range expected for radio relics \citep[see, e.g.][]{bagchi2011, erler2015}.

In the top left panel of Fig.~\ref{figure:cosmological}, we show the projected shock-dissipated energy, where this has a depth of 15 Mpc. The merger has led to the formation of two megaparsec-sized, arc-like shocks. These align with the merger axis and are relatively symmetric, as expected when the initial cluster masses are similar \citep{vanweeren2011b, lee2024b}. The shocks from the major merger are superimposed on a background of less energetic shocks. Some of these, such as those towards the lower-right are from ongoing minor mergers, whilst the majority result from accretion.

The bottom-left panel has a two-dimensional colourbar, where the colour indicates the dissipated energy-weighted Mach number, with saturation being set proportional to this energy. It can be seen that the merger shock is made up predominantly of low Mach numbers ($\mathcal{M}\lesssim5$). Shocks at the edge of the cluster, meanwhile, are much higher ($\mathcal{M}\gtrsim100$)\footnote{In the terminology of \citet{ryu2003}, these are {external} shocks.}. These shocks define the filamentary structure, within which the cluster sits. Indeed, the merging cluster has arrived from the filament evident at the top right of the panel (see \href{https://youtu.be/Ka-Odrwwamo}{movie}).

In the middle column, we show the projected pressure (top) and density (bottom) respectively. The brightest central galaxy (BCG) is located at the peak value in both panels. Infalling galaxies can also be seen as bright spots at the edge of the cluster. The merger has led to a region of increased pressure centred on the cluster. This region is well-delineated, in contrast to the more diffuse density distribution. It is also clear that the region of high pressure is bounded by the merger shocks.

\begin{figure*}
    \centering
    \includegraphics[width=2.0\columnwidth]{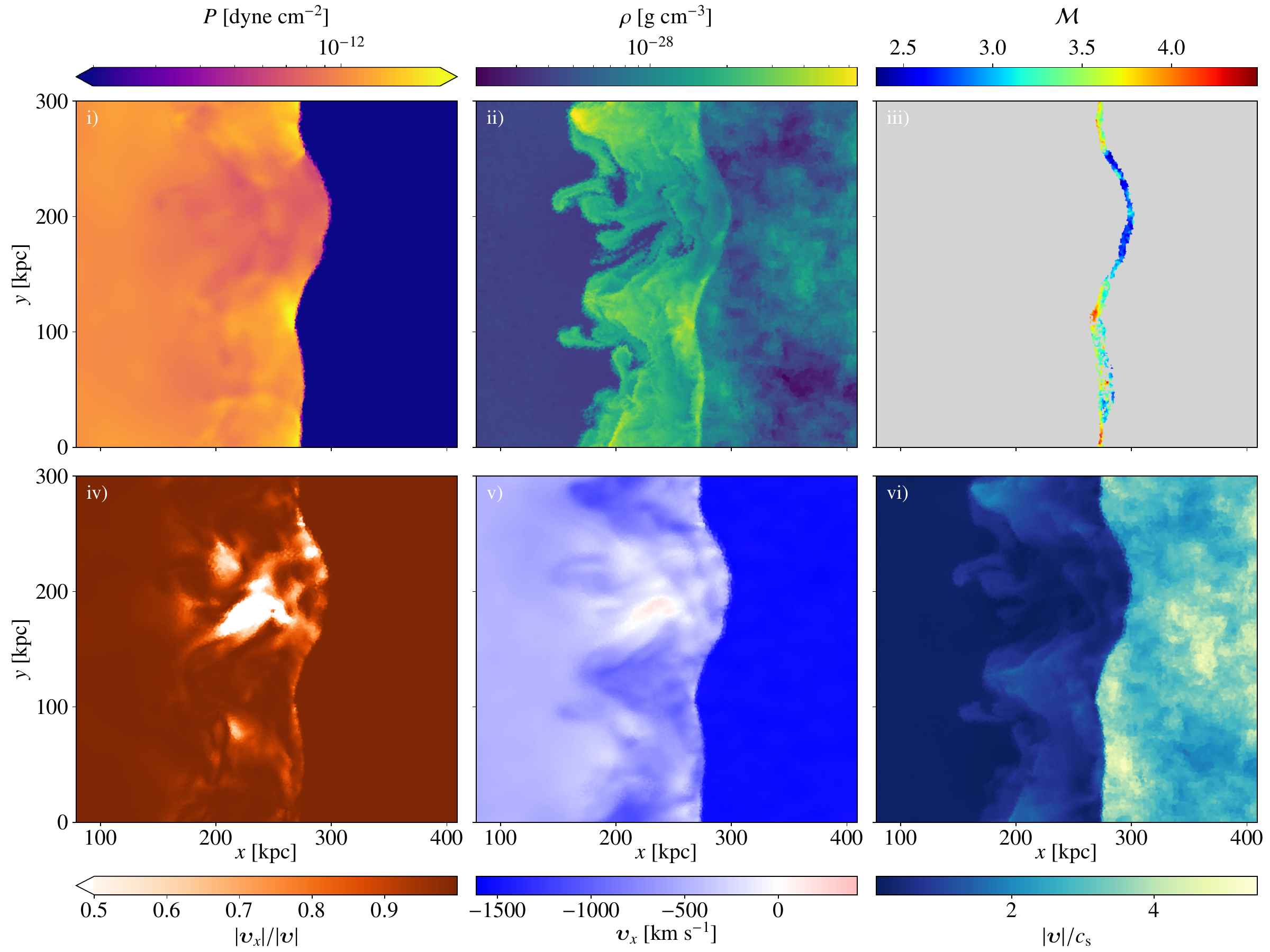}
    \caption{Mach 3 shock is driven into a turbulent upstream density field. The initial pressure and density values for the shock tube correspond to those shown in Fig.~\ref{figure:cosmological}. {Panels:} i) gas pressure, ii) gas density, iii) dissipated-energy-weighted Mach number, iv) the fraction of the gas speed in the $x$-component, v) the $x$-component of gas velocity, vi) gas speed divided by sound speed. Data are shown in slices except for iii), which is a thin projection of 35 kpc. All values are shown in the shock rest frame at $t=180$ Myr. Upstream density turbulence directly leads to shock corrugation, a distribution of Mach numbers at the shock front, and downstream velocity turbulence. An animated version of this figure can be found \href{https://www.youtube.com/watch?v=ERBftXpMqgs}{here}.}
    \label{figure:shock-tube}
\end{figure*}

To remove the effects of projection, in the right-hand column we show slices of the same quantities. We zoom in on a region of size $2\times2$~Mpc$^2$ in order to better explore the shock details. At this time and position, the outwards-moving merger shock encounters an inwards-moving accretion shock\footnote{In the terminology of \citet{ryu2003}, these are {internal} shocks.}{$^{,}$}\footnote{
We leave a full investigation of the conditions under which this scenario takes place, as well as an estimate of the frequency of such mergers, to future work. In the current study, we only show that it is a viable solution to the aforementioned radio relic problems.}. These are evident as ridges in the density and pressure slices in the corresponding \href{https://youtu.be/Ka-Odrwwamo}{movie}. The time at which the two shocks meet can be considered a Riemann problem with density, pressure and velocity specified at both sides of the discontinuity. The result of this is two-fold: firstly, an inwards travelling rarefaction wave is generated, and secondly, a thin, shock-compressed density shell forms behind the merger shock. This feature is marked by the white arrow in the bottom right panel, and is consistent with previous studies \citep{zhang2019, zhang2020}. 

It can be seen that the pressure downstream of this feature is approximately constant, whilst the density jumps. This indicates that the edge closest to the arrow is actually a contact discontinuity. The jump itself is exacerbated by the rarefaction wave, which lowers the density downstream of this discontinuity. Such waves have also been observed in previous studies of cluster mergers \citep{shi2020}. This feature is key to several results shown in the remainder of the paper.

The subsequent evolution of the shock wave can be well-approximated as a pressure-driven shock wave on a flat density background (see Sect.~\ref{subsec:shock-tube-setup}). This lends itself to shock tube simulations, where we can approach the problem with significantly higher resolution. This is necessary, as whilst the overall pressure and density profiles of our simulated clusters are converged \citep[Berlok et al., in prep;][]{tevlin2024}, density turbulence is not sufficiently resolved at these radii. In both of the right-most panels, we have added a white, dashed box with dimensions 450 kpc $\times$ 300 kpc, which represents the size of the high-resolution region in our shock tube initial conditions. This region is traversed by only $\mathcal{O}(10)$ cells, which is clearly insufficient. The shock tube method also allows us to precisely define the various turbulent parameters, allowing for a clearer investigation into their impact on downstream properties.

\section{Shock tube analysis}
\label{sec:shock-tube-analysis}

Our shock tube setup, including initial pressure and density values, is informed by the above cosmological analysis, with values given in Sect.~\ref{subsec:shock-tube-setup}. We note that the pressure jump dominates the white, dashed box shown in Fig.~\ref{figure:cosmological}, and so chose to make the shock exclusively pressure-driven, i.e. the downstream density is set equal to the mean upstream density. The mass resolution in the high-resolution upstream region of these simulations (i.e. region III) is approximately 730 times greater than in the cosmological simulation, allowing for the resolution of density turbulence.  

As previously mentioned, multiple studies have shown that the ICM is clumpy. We implemented this in our shock tube simulations using the method described in Sect.~\ref{subsec:generating_turb}. The injection scale of the density turbulence was set to half the smallest box dimension, i.e. 150 kpc, and the amplitude of the fluctuations was taken from \citet{zhuravleva2013}. We remind the reader that there is no initial velocity turbulence; that is, the density fluctuations are initially at rest.

We show the state of our fiducial Mach 3 simulation at $t=180$~Myr in Fig.~\ref{figure:shock-tube}, where the shock front is travelling from left to right. Each panel shows a 2D slice (except for panel {iii}, which is a thin projection) and $x=0$ marks the beginning of `region III' (see Sect.~\ref{subsec:shock-tube-setup}). The time was chosen such that several characteristics of the shock are on display simultaneously. In particular, we draw attention to the following three consequences of including upstream density turbulence: i) the formation of downstream turbulence, ii) shock corrugation, iii) the formation of a Mach number distribution.

\subsection{Downstream turbulence and shock corrugation}
\label{sec:RT}

Perhaps the most striking of these three is the generation of downstream turbulence\footnote{We show that our resolution of this is numerically stable in Fig.~\ref{figure:low-res-density-slice}.}. The impact of this can be seen in almost all panels, but we start with panel {iv)}. Here we show the fraction of the shock-frame gas velocity in the $x$-component. To calculate the shock-frame, we took the median $x$-coordinate for all shock surface cells at each snapshot. We then converted this to an average shock velocity by dividing by the snapshot cadence. Finally, we subtracted the subsequent value from the lab-frame gas velocities. In a purely laminar flow aligned with the shock normal, there would be no deviations from $|\bupsilon_x|/|\mathbf{\bupsilon}| = 1$, as is the case in the simulations without density turbulence.

This panel is supported by panel {v)}, which shows the $x$-component of the velocity. Blue shades indicate gas moving from right to left, whilst red shades indicate gas moving the opposite way. These values are, once again, both given in the frame of the shock. The reduction in speed after crossing the shock is a standard theoretical result, however, panel {v)} shows that including density turbulence causes gas to move away from the shock front at varying speeds. Indeed, in some cases, gas even moves towards the shock front from the left. Whilst we directly model CR electrons later in the paper, this already shows us that distance from the shock front is not a reliable indicator of time since injection.

The turbulence results predominantly from a Rayleigh-Taylor instability, which leads to the formation of density `fingers'. These are visible, for example, in panel {ii)} at  $y\approx120$ kpc and $y\approx290$ kpc. This spacing is no coincidence, as such fingers form at positions where the shock reaches higher density clumps, which in turn is set by the injection scale. In the non-linear regime of this instability, the resulting velocity shear causes the Kelvin-Helmholtz instability to form as a parasitic instability. The effect of this can be seen particularly clearly in the upper half of panel {ii)} and the associated \href{https://www.youtube.com/watch?v=ERBftXpMqgs}{movie}.

We explain the emergence of the Rayleigh-Taylor instability as follows: density fluctuations result in the shock speed varying along its surface, with the shock accelerating into lower density regions. An initially planar shock will subsequently warp. This leads to the pressure dropping behind the most advanced part of the shock, as can be seen in panel {i)}. The consequence of this is that the pressure gradient starts to point towards the far downstream. The density jump from the downstream behind the contact discontinuity towards the shock-compressed region, however, provides a gradient pointing the opposite way. What is more, as higher density fluctuations are less easily accelerated by the shock, the back of the compressed region also begins to warp. Indeed, as seen in Fig.~\ref{figure:shock-tube}, this happens to a greater extent than at the shock front.

\begin{figure}
    \centering
    \includegraphics[width=0.8\columnwidth]{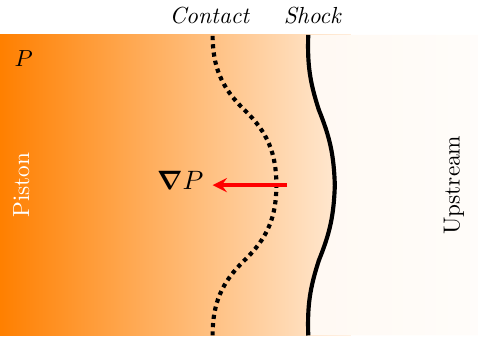}
    \includegraphics[width=0.8\columnwidth]{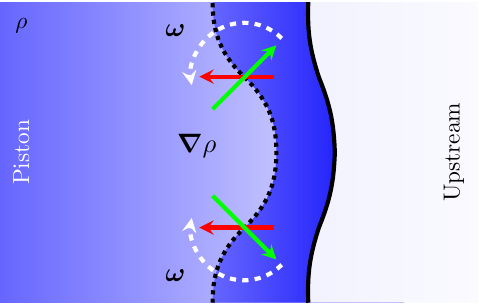}
    \caption{Schematic showing how upstream density turbulence causes velocity turbulence to be generated downstream. The corrugation of the shock front naturally leads to a misalignment of pressure and density gradients (shown here by red and green arrows, respectively). This results in a baroclinic term, which in turn induces vorticity, causing the contact discontinuity to become Rayleigh-Taylor unstable.}
    \label{figure:RTI-schematic}
\end{figure}

We show a schematic that illustrates this scenario in Fig.~\ref{figure:RTI-schematic}, with orange colours representing pressure and blue colours representing density, respectively. Pressure and density gradients have become misaligned, which, due to the inviscid vorticity equation, results in a baroclinic torque, as
\begin{equation}
    \frac{\rmn{D}\bm{\omega}}{\rmn{D}t} = \frac{1}{\rho^2}\bm{{\nabla}}{\rho}\bm{\times}\bm{{\nabla}}{P},
\label{eq:inviscid-vorticity}
\end{equation}
where the left-hand term is the Lagrangian derivative of vorticity $\bm{\omega}$. This mechanism rapidly generates downstream velocity turbulence, and leads to a reinforcement of the finger-like structures, with eddies directed as shown by the white, dashed arrows.

Upstream density fluctuations are critical for the formation of this scenario. Without them, there is only a very limited pressure gradient and no corrugation at the back of the shock-compressed region. The later condition in particular would mean that $\bm{{\nabla}}{\rho}$ and $\bm{{\nabla}}{P}$ were approximately anti-parallel, leading to a vanishing cross-product in Eq.~\eqref{eq:inviscid-vorticity} and hence little to no growth of the instability. We show this to be true in our simulations in Appendix~\ref{appendix:RTI-without-density-perturbations}.

We note that \citet{hu2022} identified the related Richtmyer-Meshkov instability as causing turbulence in their shock tube simulations. We may discount this here as the Richtmyer-Meshkov instability takes place at the shock front, whilst the instability we observe takes place at the contact discontinuity. Vorticity may also be generated by curved shock surfaces according to the theorem by \citet{crocco1937}. We discount this as the primary driver in this case, however, as this mechanism would produce vorticity oriented in the opposite direction. This would reduce the size of the fingers rather than increase them.

\subsection{Mach number distribution}
\label{subsec:mach-dist}
\subsubsection{Form}
\label{subsec:mach-form}

We now turn our attention to the third consequence of including density fluctuations: the formation of a Mach number distribution. It has long been known that shocks in clusters vary in strength \citep{ryu2003, kang2005, pfrommer2006, hoefft2008, skillman2008, vazza2009}. It is also inferred from synchrotron fluctuations that this is true within radio relics as well \citep{rajpurohit2020, rajpurohit2021}. This was supported by the shock tube simulations of \citet{dominguez-fernandez2021}. Like them, we find that an initially well-defined Mach number decomposes into a distribution upon encountering turbulence. This can be seen in the top right panel of Fig.~\ref{figure:shock-tube}, which shows a thin projection\footnote{Numerical broadening occasionally leads to cells not being recognised as part of the shock front. Such cells are captured at the next hydro-timestep, however, meaning that no gas cells downstream remain unshocked as a result of the numerical implementation. Projecting therefore gives a more faithful indication of the coherence of the shock front.} of depth 35 kpc, where colours indicate the Mach number at the shock front.

\begin{figure}
    \centering
    \includegraphics[width=1.0\columnwidth]{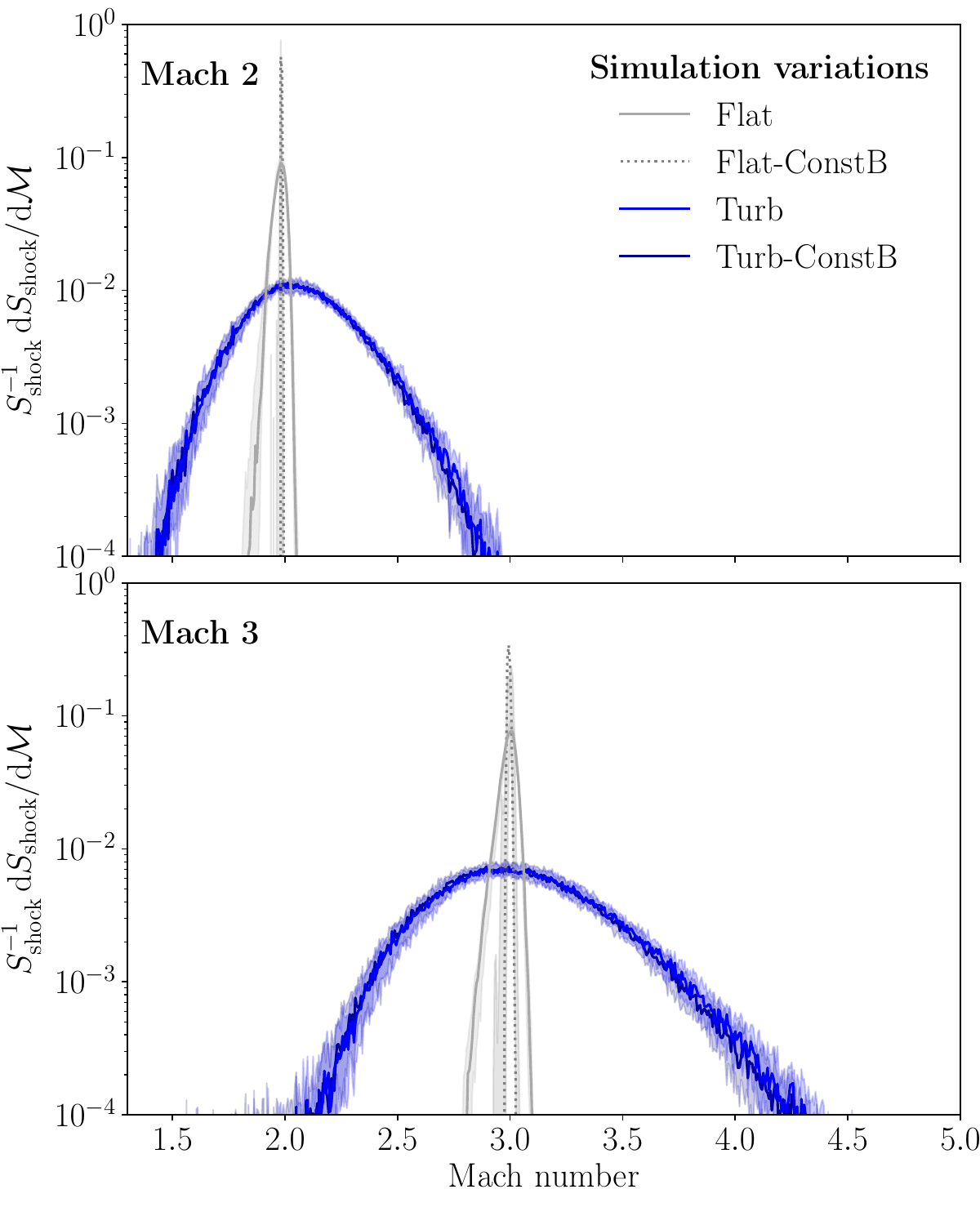}
    \caption{Mach number distributions for all models, where each cell has been  weighted by its normalised contribution to the shock surface. Lines indicate the median taken over all snapshots, whilst the shaded values indicate the interquartile range. The addition of magnetic turbulence ({Flat}) to a homogeneous density distribution ({Flat-ConstB}) broadens the distribution only very mildly. By contrast, the addition of upstream density fluctuations in the simulation ({Turb} and {Turb-ConstB}) turns an extremely narrow distribution into a much broader one. The width of the distribution is proportional to the peak Mach number.}
    \label{figure:mach-no-pdf}
\end{figure}

The shock in Fig.~\ref{figure:shock-tube} was initially set up with $\mathcal{M}=3$. However, whilst many cells continue to show similar values, it is clear that that there is now a distribution, with values lying approximately between $\mathcal{M}=1.9$ and $\mathcal{M}=4.8$. We quantify the development of this distribution in Fig.~\ref{figure:mach-no-pdf}. For this, we calculated the distribution at each snapshot, where each cell has been weighted by its contribution to the overall shock surface. We show the median of these distributions in Fig.~\ref{figure:mach-no-pdf} as solid lines and provide the 25$-$75\% percentile range as shaded values.

The dotted grey line indicates the values for simulation {Flat-ConstB}, which has neither density nor magnetic turbulence in the initial conditions. It can be seen that the distribution forms an approximate delta function. This indicates the base-line accuracy of our shock finder. When we add magnetic fluctuations, as in {Flat}, the distribution broadens slightly. This is due to the additional pressure fluctuations caused by the magnetic field (see Appendix~\ref{appendix:beta}). However, this effect is totally dominated by the impact of including density fluctuations. Indeed, there is essentially no difference between the Mach number distributions in {Turb-ConstB} and {Turb}. The initial conditions for these simulations differ only by their inclusion of a uniform or turbulent magnetic field, respectively.

The peak of the distribution meanwhile remains at the initial Mach number ($\mathcal{M}=2$ in the top panel and $\mathcal{M}=3$ in the bottom). Finally, we note that the distribution remains remarkably stable over time; the difference between the 25th and 75th percentile is very low. This is in contrast to \citet{dominguez-fernandez2021} who, whilst also finding a tail towards higher Mach numbers, do not see such stability (see their fig. B2). We attribute this to our different methods of generating turbulence (see Sect.~\ref{sec:discussion}).

\subsubsection{Origin}
\label{sec:mach-no-dist-origin}

\begin{figure*}
    \centering
    \includegraphics[width=2.0\columnwidth]{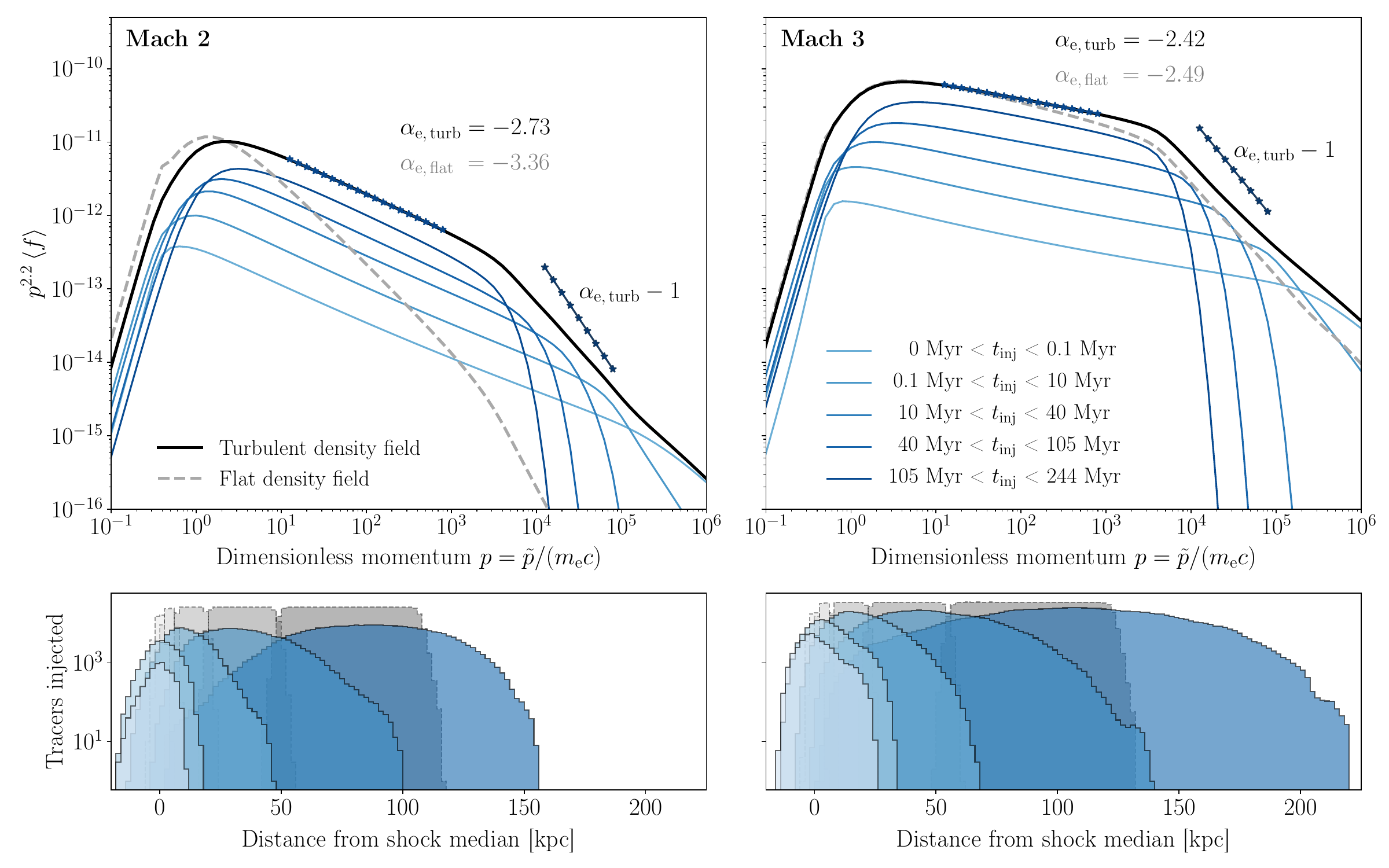}
    \caption{{Top row:} Volume-weighted non-thermal electron spectra generated from our {Turb} (solid black) and {Flat} (dashed grey) simulations at $t=250$ Myr. Blue lines show contributions to the {Turb} spectrum, where tracers have been binned by time since injection. {Bottom row:} Histograms indicating the total number of injected tracers relative to the shock front, where bins have a width of 2~kpc. Blue and grey colours represent our {Turb} and {Flat} simulations respectively, whilst colour saturations indicate the same time bins as above. The broadened Mach number distribution seen in Fig.~\ref{figure:mach-no-pdf} causes a shallower slope, $\alpha_\rmn{e}$, in the Turb runs compared to the theoretical expectation. This is especially noticeable for low Mach number shocks. In the case of upstream density fluctuations, substantial mixing takes place downstream, such that distance from the shock front is no longer a good indication of cooling time.}
    \label{figure:spectra-binned-by-tinj}
\end{figure*}

Returning to  Fig.~\ref{figure:shock-tube}, it can be seen from panel {iii)} that the higher Mach numbers are typically found towards the back of the shock front, whilst lower Mach numbers are found at its most advanced position. An intuitive understanding of this can be gained by inspecting panel {vi)}. Here we show a slice, where colours indicate the shock-frame gas speed as a fraction of the sound speed. We expect this quantity to provide a fair approximation to the more rigorous result given by Eq.~\eqref{eq:mach-number}. We remind the reader that we have defined the shock-frame based on the median of the local shock velocities. This is reasonable as the local velocity fluctuations along the shock front are small compared to its median shock speed.

Gas is stationary in the lab-frame and so, under our approximation, approaches the shock at the same speed, as can be seen in the right-hand side of panel {v)}. Additionally, the gas is in approximate pressure equilibrium, as can be seen by inspecting panel {i)}. Consequently, $|\mathbf{\bupsilon}|/c_\rmn{s}$ in the upstream is determined purely by the density fluctuations, as $c_\rmn{s}= \sqrt{\gamma_\rmn{a} P/\rho}$, where the adiabatic index is set to $\gamma_\rmn{a} = 5/3$. This means that denser (more rarefied) gas has a lower (higher) sound speed, respectively. The consequence of this is that we expect denser gas to produce higher Mach numbers, and more rarefied gas to produce lower Mach numbers. This is indeed the case, as can be seen by comparing panels {ii)} and {iii)}. As discussed in Sect.~\ref{sec:RT}, lower density gas also causes the shock front to accelerate locally. As a result, lower Mach numbers are found further forward. We show that this is generically true across the shock front in Appendix~\ref{appendix:mach-in-projection}.

\subsection{Radio versus X-ray Mach number discrepancy}
\label{sec:spectra}

It has been argued previously that the highest Mach numbers in a distribution are responsible for the radio emission \citep[see e.g.][]{hoeft2011, wittor2021, dominguez-fernandez2021}. This is currently the leading theory for relieving the tension in the so-called $\mathcal{M}_\rmn{radio} - \mathcal{M}_\rmn{X-ray}$ discrepancy, where Mach numbers derived from radio observations are found to be higher than those derived from X-ray observations. The effect has yet to be modelled end-to-end, however, using a full Fokker-Planck solver. We resolve this issue here, using the tracers discussed in Sect.~\ref{subsec:tracers} and the CR electron post-processing code \textsc{Crest}, introduced in Sect.~\ref{subsec:crest}.

We show the resulting volume-weighted non-thermal electron spectra at $t=250$ Myr in the upper panels of Fig.~\ref{figure:spectra-binned-by-tinj}. At this time, the shock in the Mach 3 simulations has reached the end of the high-resolution upstream region (region III). We remind the reader that we present dimensionless momenta $p= \tilde{p} /(m_\rmn{e} c)$ and that we use the 1D distribution function, $f^\rmn{1D} = 4 \pi p^2 f^\rmn{3D}$. We have further multiplied the spectra by $p^{2.2}$, which is the maximum slope of injection we allow (see Sect.~\ref{subsec:crest}). Using this convention, such a slope would appear as a horizontal line.

The grey, dashed lines show spectra from our {Flat} simulations, which do not include density fluctuations. We have calculated their spectral slope, $a_\rmn{e, flat}$, in each panel by fitting a straight line between $10 < p < 10^3$ using the method of least squares. This is approximately the region unaffected by cooling. 

If we assume a single Mach number is responsible for the CR electron spectra slope, $\alpha_\rmn{e}$, we may derive the following formula from the jump conditions \citep{ensslin1998, clarke2011}:

\begin{equation}
    \mathcal{M} = \sqrt{\frac{2 (2-\alpha_\rmn{e})}{1 - 2\alpha_\rmn{e} - 3\gamma_\rmn{a}}},
    \label{eq:mach-no-slope}
\end{equation}
or equivalently
\begin{equation}
    \alpha_\rmn{e} = - \frac{2 (\mathcal{M}^2 + 1)}{\mathcal{M}^2 - 1},
    \label{eq:mach-no-slope-inverse}
\end{equation}
where we set the adiabatic index $\gamma_\rmn{a}=5/3$. These formulae predict that $\mathcal{M}=2$ and $\mathcal{M}=3$ shocks should produce slopes of $\alpha_\rmn{e}=-3.33$ and $\alpha_\rmn{e}=-2.5$, respectively. The actual values of $\alpha_\rmn{e, flat}=-3.36$ and $\alpha_\rmn{e, flat}=-2.49$, on the other hand, imply $\mathcal{M}=1.98$ and $\mathcal{M}=3.03$. Whilst some of this variation is due to the distribution of Mach numbers shown in Fig.~\ref{figure:mach-no-pdf}, at higher Mach numbers the error arises predominantly due to the fact that Eq.~\eqref{eq:mach-no-slope} has a pole at $\alpha_\rmn{e}=-2$ and consequently shallow slopes must be measured with exponentially more accuracy. With this in mind, we consider the {Flat} runs to be good approximations of single Mach number shocks. That is, their slopes are what would be expected if the radio-derived Mach number matched that derived from X-ray observations, which tend to average over upstream and downstream properties \citep{wittor2021}.

The solid black lines in Fig.~\ref{figure:spectra-binned-by-tinj} show spectra from our {Turb} runs, which include density fluctuations. It can be seen that the resultant spectral slopes, labelled $\alpha_\rmn{turb}$, are systematically shallower than in the {Flat} case. Indeed, Mach numbers derived from these slopes would imply shocks with strength $\mathcal{M} = 2.54$ and $\mathcal{M} = 3.25$ for our two simulations with initial Mach numbers of $\mathcal{M} = 2$ and 3, respectively\footnote{Without the use of $\mathcal{M}_\rmn{crit} = 2.3$ (see Sect.~\ref{subsec:crest}), the spectral slope for the Mach 2 run would be $\alpha_\rmn{e, turb}=-2.79$, implying $\mathcal{M} = 2.45$.}. This is despite the fact that the shock strength continues to peak at $\mathcal{M} = 2$ and $\mathcal{M} = 3$, respectively, as seen in Fig.~\ref{figure:mach-no-pdf}.

It is notable that the discrepancy is greater in weaker shocks. This might seem to be in contrast to our finding in Fig.~\ref{figure:mach-no-pdf}, which shows that the tail of the Mach number distribution actually extends further in {stronger} shocks. This effect is outweighed, however, by the strong inverse dependence of $\alpha_\rmn{e}$ on $\mathcal{M}$, with $\alpha_\rmn{e}$ having a strong dependence on $\mathcal{M}$ at $\mathcal{M}=2$, but a substantially weaker dependence at $\mathcal{M}=3$. For example, from Eq.~\eqref{eq:mach-no-slope-inverse}, it can be seen that a shock with $\mathcal{M}=2.5$ would produce a spectral slope of $\alpha_\rmn{e}=-2.76$, which is 0.57 shallower than that of a $\mathcal{M}=2$ shock. On the other hand, a $\mathcal{M}=3.5$ shock will produce a spectral slope of $\alpha_\rmn{e}=-2.36$, which is only 0.14 shallower than an $\mathcal{M}=3$ shock. The tail of the Mach number distribution is therefore more influential in weaker shocks, as the shallower slopes from these fluctuations can more easily affect the integrated spectra at higher frequencies, thus increasing the discrepancy between the expected and inferred Mach number.

Of course, in radio observations, only the momenta range $10^4 \lesssim p \lesssim 10^5$ is available for observation, as it is this range that produces megahertz- and gigahertz-emitting electrons. Standard theory dictates that the slope of the CR electron spectrum in this range should be $a_\rmn{e} - 1$ due to cooling\footnote{Equivalently, the spectral index of the emission will steepen by 1/2.} \citep{condon2016}. Indeed, it is this fact that is used to infer the spectral slope in radio relic observations \citep[see, e.g.][]{vanweeren2012, rajpurohit2020}. This result relies, however, upon the spectrum being produced by a single Mach number shock. We have added the theoretical slope to Fig.~\ref{figure:spectra-binned-by-tinj} for the {Turb} runs to guide the eye. It can be seen that, when the shock is made up of a distribution of Mach numbers, the slope in this region is actually slightly shallower than $a_\rmn{e} - 1$. Once again, this is especially evident in the $\mathcal{M}=2$ run. Indeed, if we measure the slope in this range for the Mach 2 {Turb} run, we would infer a spectral slope of $\alpha_\rmn{e, turb}=-2.5$. This in turn produces an estimated Mach number of $\mathcal{M}=3.01$, which is $\sim$50\% higher compared to the true peak value. For comparison, we show the slope measurements and inferred Mach numbers for all of our runs in Table~\ref{tab:inferred_mach_no}.

\renewcommand{\arraystretch}{2}
\begin{table}
\caption{Spectral slope and the inferred Mach number for our different shock tube models.}
\begin{tabular}{|c|c||c|c|c|c|}
\cline{3-6}
\multicolumn{2}{c|}{}                 & {Flat} & {Turb} & {Flat} & {Turb} \\ \cline{3-6} \noalign{\vskip\doublerulesep
         \vskip-\arrayrulewidth\vskip\doublerulesep
         \vskip-\arrayrulewidth} \cline{2-6} 
\multicolumn{1}{c|}{\parbox[t]{-1mm}{\multirow{2}{*}{\rotatebox[origin=c]{90}{{Theory}}}}}
  & {Initial ${\mathcal{M}}$} & 2.00          &  2.00          &  3.00          &  3.00          \\ \cline{2-6} 
\multicolumn{1}{c|}{} & {Expected slope}             & -3.33         & -3.33         & -2.5          & -2.5         \\ \cline{2-6} \noalign{\vskip\doublerulesep
         \vskip-\arrayrulewidth\vskip\doublerulesep
         \vskip-\arrayrulewidth} \cline{2-6} 
\multicolumn{1}{c|}{\parbox[t]{8mm}{\multirow{2}{*}{\rotatebox[origin=c]{90}{\makecell{{Low} \\ {momenta}}}}}}
 & {Measured slope}             & -3.36         & -2.73         & -2.49         & -2.42         \\ \cline{2-6} 
\multicolumn{1}{c|}{} & {Implied ${\mathcal{M}}$} &  1.98          &  2.54          &  3.03          &  3.25          \\ \cline{2-6} \noalign{\vskip\doublerulesep
         \vskip-\arrayrulewidth\vskip\doublerulesep
         \vskip-\arrayrulewidth} \cline{2-6} 
\multicolumn{1}{c|}{\parbox[t]{7mm}{\multirow{2}{*}{\rotatebox[origin=c]{90}{\makecell{{High} \\ {momenta}}}}}
\hspace{3mm}} & {Measured slope}               & -4.28         & -3.50         & -3.45         & -3.35       \\ \cline{2-6} 
\multicolumn{1}{c|}{\hspace{3mm}} & {Implied ${\mathcal{M}}$} &  2.03          &  3.01          &  3.15          &  3.54          \\ \cline{2-6} 
\end{tabular}
\vspace{2mm}
\tablefoot{Top two rows: Initial Mach number for each simulation and the expected slope calculated using Eq.~\eqref{eq:mach-no-slope-inverse}. Middle two rows: Measured slope at low ($10 < p < 10^3$) momenta and the Mach number that one would infer using Eq.~\eqref{eq:mach-no-slope}. Bottom two rows: As previous but for high momenta ($10^4 < p < 10^5$) and making the typical assumption that this slope is $\alpha_\rmn{e} - 1$. The tail of the Mach number distribution produces shallower slopes and breaks the $\alpha_\rmn{e} - 1$ assumption. These effects both lead to an overestimation of the peak Mach number in the shock.}
\label{tab:inferred_mach_no}
\end{table}

\begin{figure*}
    \centering
    \includegraphics[width=2.0\columnwidth]{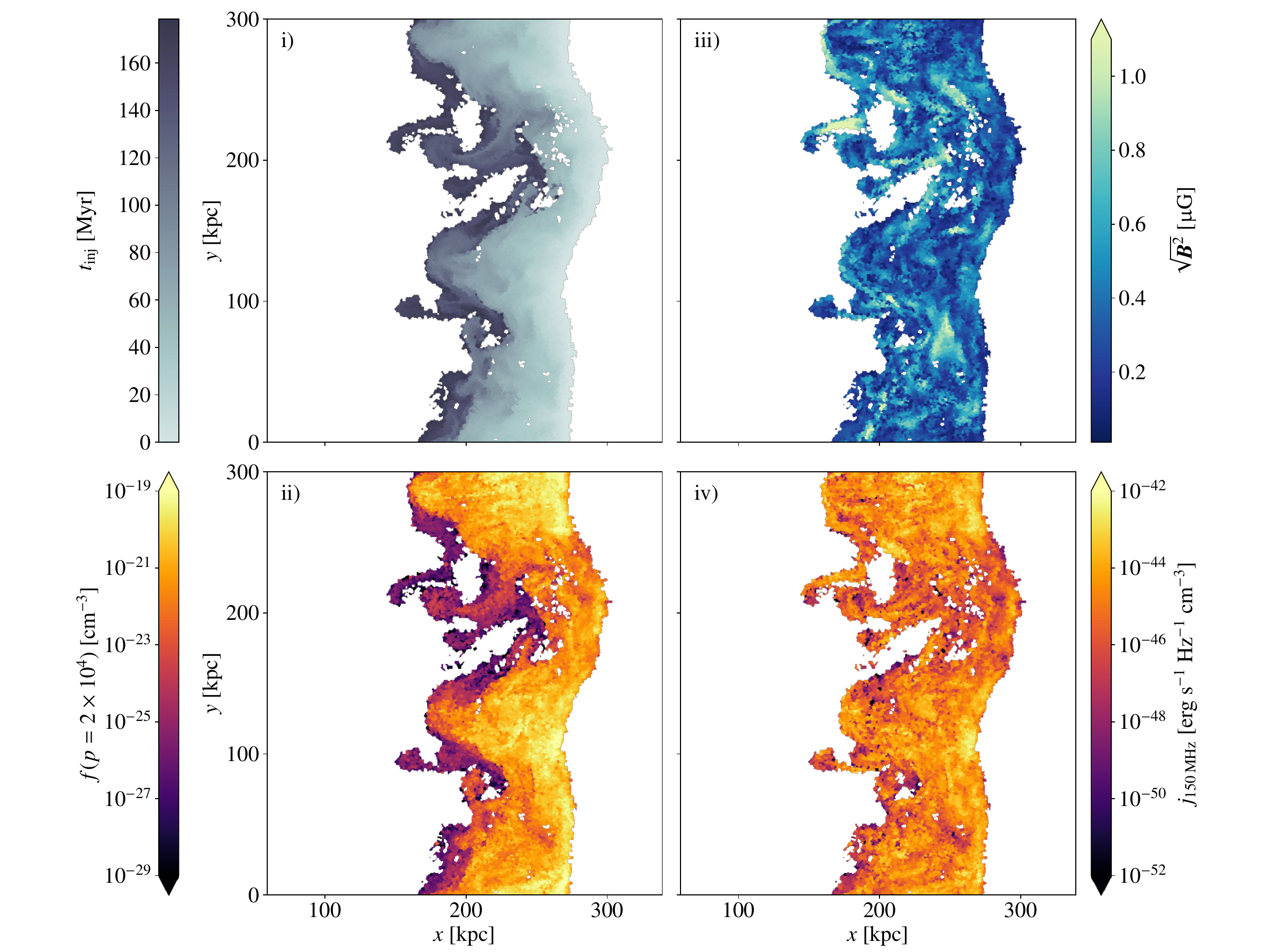}
    \caption{Slices through our fiducial Mach 3 simulation at $t=180$ Myr showing: i) time since injection, ii) CR electron number density at $p=2\times 10^4$ (see text), iii) magnetic field strength, and iv) synchrotron emissivity at 150 MHz. Only tracers that have undergone DSA are shown. A Rayleigh-Taylor instability induces a counter-streaming plume (in the shock rest frame) that brings aged, and therefore cooled, electrons close to the shock front. However, these electrons are still relatively bright at 150 MHz due to magnetic field amplification up to $\upmu$G levels. An animated version of this figure can be found \href{https://youtu.be/WjnCC3Ul0s0}{here}.}
    \label{figure:slice-downstream}
\end{figure*}

As previously, this effect originates due to the shallower slopes of higher Mach number shocks. At the higher momenta end, where $f(p)$ is lower, individual tracers can have a larger impact on the volume-weighted mean. The effect can be more clearly seen if we plot the spectra binned by the time elapsed since their injection at the shock front. These bins are given in the legend in the top right panel of Fig.~\ref{figure:spectra-binned-by-tinj}. We have chosen the time bins such that the number of tracers in them approximately doubles each time, which produces an approximate doubling of the amplitude of the binned spectra in regions where the cooling time exceeds the simulation run-time. The maximum time since injection is 244 Myr, as we ignore tracers injected during the initial buffer region (see Sect.~\ref{subsec:sim-vars}). As the electrons cool rapidly at $p \gtrsim 10^4$, binning in this fashion allows us to pick out regions of momenta which are dominated by a specific age. Indeed, it is this layering that produces the overall volume-weighted slope. It can be seen that the binned spectra bend slightly, becoming slightly shallower at higher momenta. This results from the layering of different injected slopes as tracers sample the Mach number distribution. It is this bending effect that ultimately produces a slope of less than $a_\rmn{e} - 1$.

\subsection{Breaking of the laminar flow assumption}
\label{sec:laminar-flow}

It is frequently assumed that distance from the shock front is a reliable measure for time cooled \citep[see e.g.][]{vanweeren2012, rajpurohit2020}. We now investigate whether this assumption is valid, particularly in light of the turbulence we identified in Sect.~\ref{sec:RT}. For this purpose, we reuse the time bins just discussed, making histograms of the distance from the shock median for each binned population. These histograms are presented in the bottom row of Fig.~\ref{figure:spectra-binned-by-tinj}. The grey shaded regions indicate data from the {Flat} simulations. It can be seen that whilst there is some overlap between time bins for the Flat simulation, owing to the aforementioned magnetic pressure fluctuations, the amount is relatively minimal. This indicates that the flow is predominantly laminar. In contrast, tracers from the {Turb} simulations, represented by the blue shaded regions, overlap strongly. Indeed, there are tracers in the last time bin that are still close to the shock front. 

Two further aspects further distinguish the {Turb} simulations from their {Flat} counterparts: firstly, there are well-populated regions of negative distance due to the corrugation of the shock front (see Sect.~\ref{sec:RT}), and secondly, the maximum distance from the shock is substantially larger in the {Turb} runs. Indeed, the relative maximum distance increases proportionally with the strength of the shock: tracers reach 32\% and 67\% further in Mach 2 and Mach 3 shocks, respectively. This comes about primarily due to the inertia of the high-density fluctuations; higher density clumps experience less acceleration and can therefore better penetrate downstream. Higher Mach number shocks, meanwhile, have faster speeds, resulting in a greater distance opening up between the shock front and such clumps. This effect is further reinforced by the Rayleigh-Taylor instability as analysed in Sect.~\ref{sec:RT}, which itself is more effective at faster shock speeds, when the downstream is more dynamic. 

The consequence of all of these effects is that, once turbulent density fluctuations are taken into account, distance from the shock front is no longer a good indicator for time since injection for individual electrons. Indeed, as we show in App.~\ref{appendix:spectra_by_dist}, it is not possible to make an analogous figure to Fig.~\ref{figure:spectra-binned-by-tinj} where the tracers have been binned by distance.

These effects can be seen particularly well in Fig.~\ref{figure:slice-downstream}, where we show slices through the injected region in our Mach 3 {Turb} simulation at $t=180$ Myr. The slices presented here may be directly compared with those presented in Fig.~\ref{figure:shock-tube}; data have simply been taken from the tracers here rather than the gas cells. In panel {i)} of Fig.~\ref{figure:slice-downstream}, we show the time since injection, with lighter colours indicating more recent injection. Regions without colour indicate a lack of injection. Within the shock-compressed region, this results from the shock front falling below the critical Mach number (see discussion in Whittingham et al., in prep.). The impact of the range of post-shock gas speeds, as already shown in panel {v)} of Fig.~\ref{figure:shock-tube}, can clearly be seen, with relatively freshly injected CR electrons evident at varying distances from the shock front. With this said, the overall distribution of CR electron ages mostly traces the shape of the Rayleigh-Taylor fingers, with extra perturbations formed by the additionally generated turbulence. As this instability acts at the contact discontinuity, the resultant eddies are generally populated by older material. This effect is also responsible, however, for bringing the same aged CR electrons closer to the shock front.

In panel {ii)} of Fig.~\ref{figure:slice-downstream}, we show the CR electron distribution function $f(p)$, where we have set $p = 2 \times 10^4$. We choose this momentum in particular as it is the one which contributes most to the 150 MHz emission\footnote{To calculate this, we inverted the critical frequency formula given in Eq.~\eqref{eq:critical-frequency} to find the momentum primarily responsible for the 150 MHz emission in each tracer (assuming $\nu_\rmn{syn} = 2 \nu_\rmn{c}$, following \citealt{werhahn2021}). We then took the emission-weighted mean of the resulting distribution of momentum values.}. Older electrons have had more time to cool, and so $f(p)$ at a fixed momentum generally reflects the distribution shown in panel {i)}. Some noticeable differences are evident at the shock front, however. Here it can be seen that the most advanced part of the shock has lower $f(p)$ values compared to neighbouring regions. This is due to the weaker shocks that take place here, as may be seen by comparing with panel {iii)} in Fig.~\ref{figure:shock-tube}. Such shocks produce spectra with steeper slopes and lower normalisations\footnote{We do not include Mach-dependent acceleration efficiencies in our simulations, but lower Mach number shocks naturally result in less dissipated energy through Eq.~\eqref{eq:shock-dissipated-energy} and hence lower normalisations through Eq.~\eqref{eq:p_min}.}, which act to reduce $f(p)$ values. Both of these effects are also evident in Fig.~\ref{figure:spectra-binned-by-tinj}. 

Observationally, $f(p)$ is not available to us; instead, we must use synchrotron emission to infer it. However, this is modulated by the magnetic field strength of the emitting region, as can be seen by inspection of Eq.~\eqref{eq:synchrotron-emissivity}. We therefore focus now on the magnetic field downstream.

\subsection{Magnetic field amplification}
\label{sec:magnetic-field}
\subsubsection{Strength}
\label{sec:magnetic-field-strength}

\begin{figure}
    \centering
    \includegraphics[width=1.0\columnwidth]{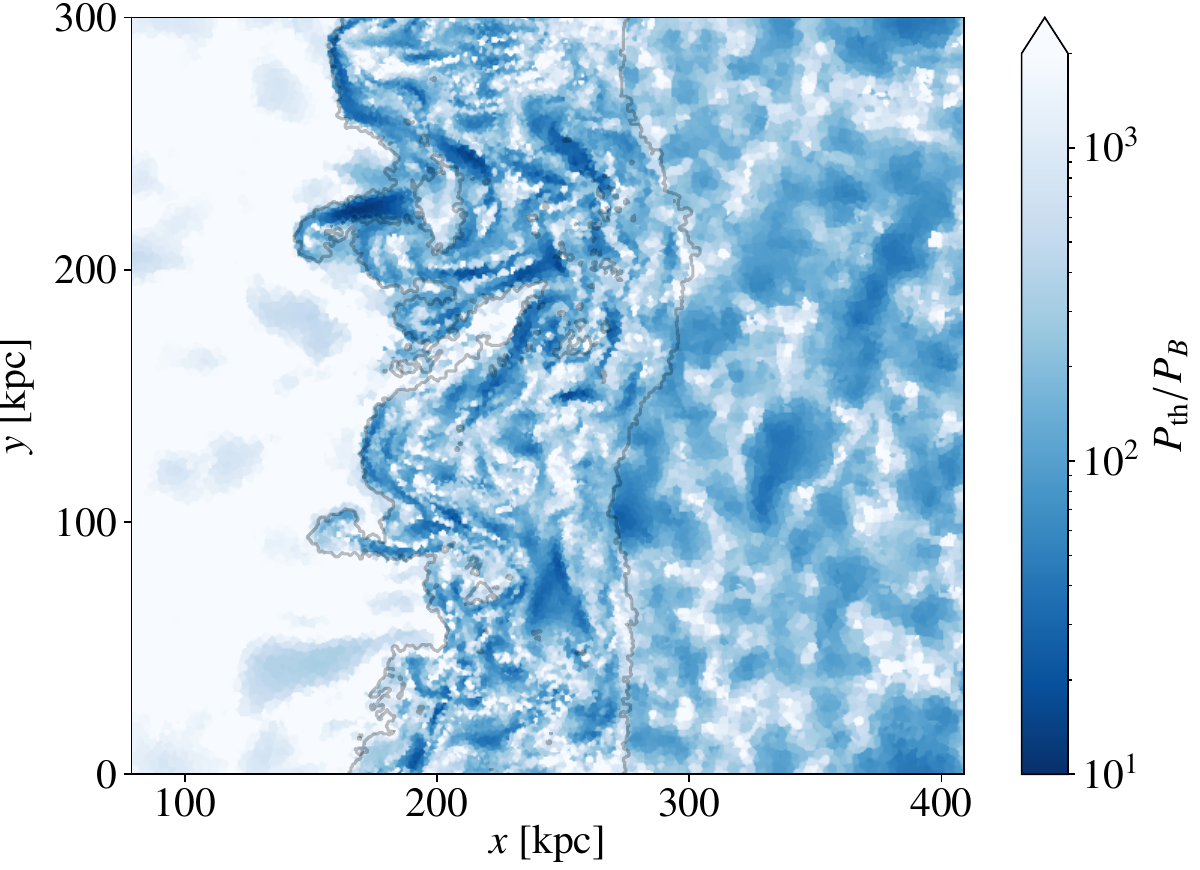}
    \caption{Slices through our fiducial Mach 3 simulation at $t=180$ Myr showing plasma beta values. The grey lines mark the corrugated shock surface and the contact discontinuity (i.e. the region within which tracers have been injected). Despite significant amplification, beta values are typically well above 10, limiting the ability of the magnetic field to affect dynamics.}
    \label{figure:plasma-beta-slice}
\end{figure}

As discussed in Sect.~\ref{subsec:generating_turb}, the upstream turbulent magnetic field in our simulations has an RMS field strength of $B_\rmn{1} \approx 0.16$~$\upmu$G. In the $\mathcal{M}=3$ case without density fluctuations we expect a shock compression ratio of $x_\rmn{s}=3$. Assuming that the magnetic components perpendicular to the direction of shock propagation are amplified by this same ratio, for an isotropic magnetic field we should expect the RMS strength immediately behind the shock to be 
\begin{align}
\label{eq:B_amplification}
    B_\rmn{2} = B_\rmn{1} \sqrt{\frac{1 + 2 x_\rmn{s}^2}{3}},
\end{align}
and hence $B_\rmn{2}\approx 2.5 B_1 = 0.4$~$\upmu$G. This is problematic, as multiple studies now indicate that magnetic fields in radio relics are at least a factor 3 higher than this (see citations given in Sect.~\ref{sec:intro}).

In panel {iii)} of Fig.~\ref{figure:slice-downstream}, we show the magnetic field strength in the injected region. It is evident that values indeed start around $B = 0.4$~$\upmu$G. However, many regions of the downstream actually exceed $\upmu$G strengths, in accordance with observed radio relics. Even at its highest strengths, however, the magnetic field remains dynamically subdominant. As can be seen in Fig.~\ref{figure:plasma-beta-slice}, minimum values reach $P_\rmn{th} / P_{B} = 10$, with the majority having $P_\rmn{th} / P_{B} > 100$\footnote{We provide distributions of the plasma beta values before and after injection for the whole simulation in Appendix~\ref{appendix:beta}, which allows us to quantify this statement further.}. Moreover, peak magnetic field strengths also remain below the equivalent value for the CMB magnetic field strength at redshift $z=0.2$, $B_\rmn{CMB} \approx 4.7$~$\upmu$G. This means that inverse Compton cooling still dominates over synchrotron cooling in the downstream.

\begin{figure}
    \centering
    \includegraphics[width=1.0\columnwidth]{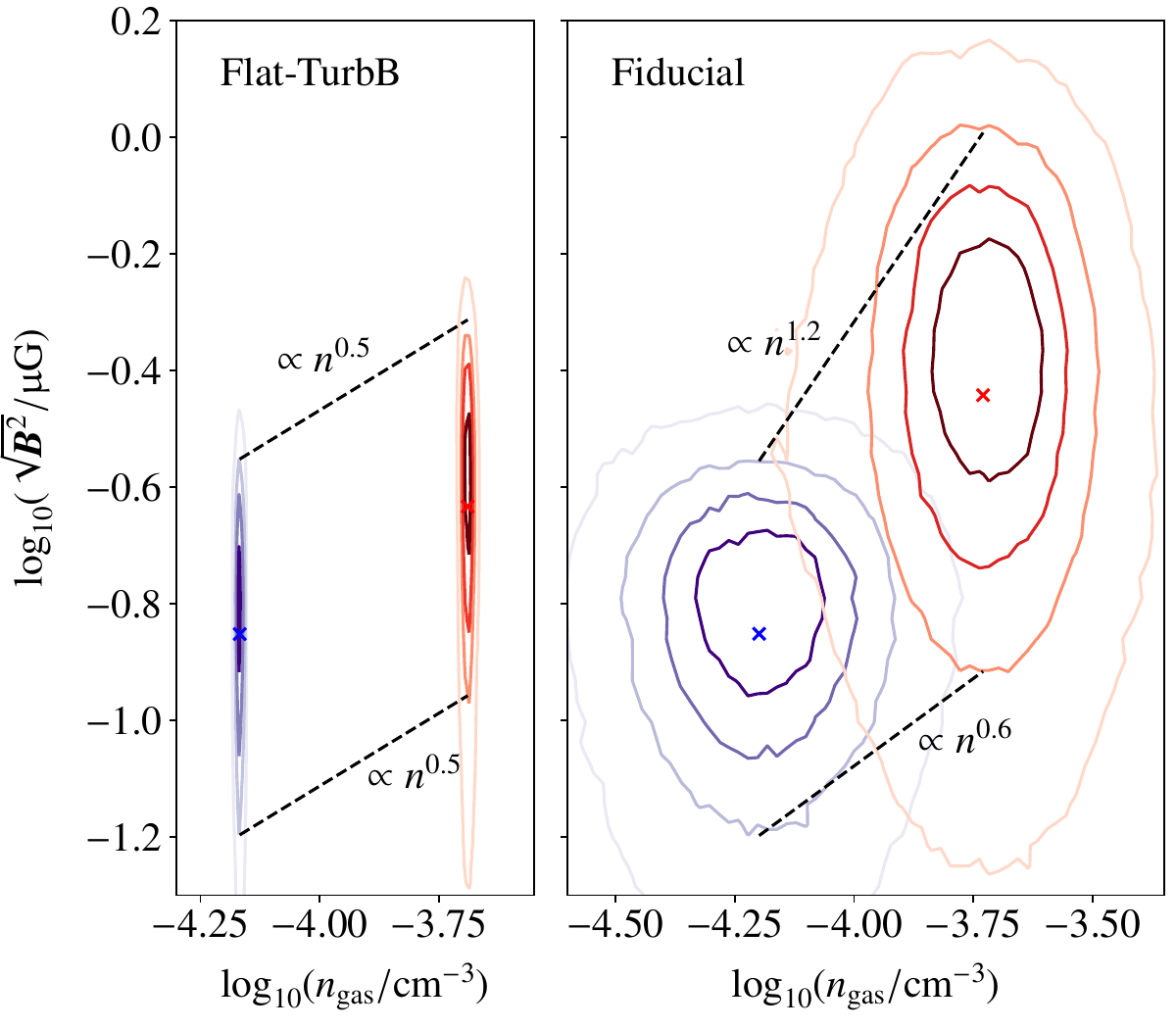}
    \caption{{Left:} Phase space diagram of magnetic field strength versus gas number density for our Mach 3 {Flat-TurbB} simulation. Contours cover 25\%, 50\%, 75\%, and 95\% of the probability density (represented by darker to lighter shades, respectively). Blue colours indicate the initial upstream distribution, whilst red colours show the final state in the injected region at $t = 250$~Myr. A cross marks the median of the distribution. {Right:} As previous, except data are taken from the Mach 3 fiducial simulation. Dashed lines indicate the empirical scaling of the 75\% boundaries. Amplification is weaker than the naive expectation for {Flat-TurbB} due to magnetic decay. Meanwhile, for the fiducial simulation, amplification is significantly stronger than that possible from compression alone.}
    \label{figure:rho-B-phase-diagram}
\end{figure}

We quantify the range of magnetic field strengths reached in Fig.~\ref{figure:rho-B-phase-diagram}, where we show the field strength as a function of density for the initial upstream and final downstream regions. To do this we take all gas cells in the high-resolution upstream region (region III) at $t=0$ and all gas cells at $t=250$ Myr that are in the shock-compressed region. Contours of these distributions are shown in blue and red, respectively.

It can be seen that the fiducial simulation, {Turb}, indeed regularly reaches $\upmu$G strengths and generally significantly exceeds those achieved in the {Flat-TurbB} simulation. Moreover, the {Flat-TurbB} simulation fails to match the naively expected amplification: with $x_\rmn{s} = 3$ and an amplification of the magnetic field by a factor of 2.5 according to Eq.~\eqref{eq:B_amplification}, we should expect a scaling of $B\propto n^{\alpha_B}$, with $\alpha_B=\log(2.5)/\log(3)\approx0.83$. Empirically, however, we recover $B\propto n^{0.5}$. This is a direct result of turbulent magnetic decay caused by magnetic tension forces. Indeed, we show in App.~\ref{appendix:decay} that the $B\propto n^{0.5}$ scaling can even be predicted, once this process is taken into account.

The decay of the magnetic field significantly exacerbates the problem of $\upmu$G magnetic fields in radio relics; without subsequent amplification in the downstream, $0.4$~$\upmu$G becomes a {maximum} value for our simulation\footnote{We note that \citet{donnert2016} model exponential decay of the field but assume that adiabatic expansion is the primary driver. \citet{kang2015}, meanwhile, show that a declining magnetic field strength can produce curved spectra, but built this model on a declining gas pressure. Here, however, we show that magnetic fields will decay even if the gas density and pressure remain constant.}. This makes the empirical scalings of the fiducial simulation all the more remarkable; compression alone cannot produce scalings greater than $B \propto n$, and magnetic decay acts to reduce the field strengths over time as well.

\subsubsection{Origin}
\label{subsec:mag-field-origin}

\begin{figure}
    \centering
    \vspace{-0.2cm}
    \includegraphics[width=1.0\columnwidth]{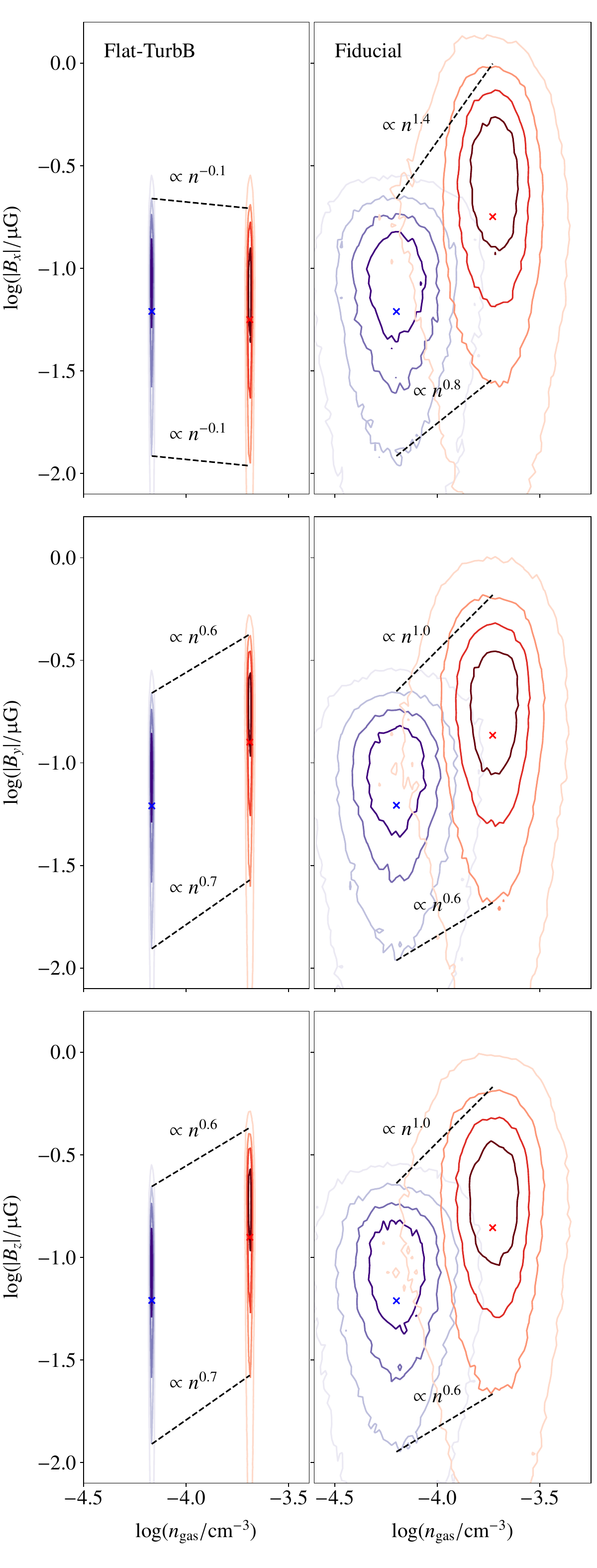}
    \caption{{Left:} As Fig.~\ref{figure:rho-B-phase-diagram} but now showing individual components of the magnetic field. All components are more highly amplified in the fiducial simulation owing to time-varying compression. It is particularly notable, however, that the $x$ component is the most highly amplified. This is indicative of amplification by shearing.}
    \label{figure:rho-B-phase-diagram--components}
\end{figure}

To understand the scalings we observe, we move to analysis of the individual magnetic components, which we show in Fig.~\ref{figure:rho-B-phase-diagram--components}. The $x$-component, which is parallel to the direction of shock propagation, should initially experience no changes across the shock, due to the $\bm{\nabla} \bm{\cdot} \bm{\rmn{B}} = 0$ constraint. On the other hand, following flux conservation, the $y$- and $z$- components should be amplified by the shock compression ratio, $x_\rmn{s}$. This would lead to scalings of $B=\text{const.}$ and $B\propto n$, respectively. Indeed, by looking at individual tracer trajectories, we find that the magnetic field is indeed initially amplified in this way\footnote{Not shown here due to space constraints.}. It can be seen on the left-hand side of Fig.~\ref{figure:rho-B-phase-diagram--components}, however, that for the {Flat-TurbB} simulation we ultimately recover scalings of $B_x\propto n^{-0.1}$ and $B_{y,z}\propto  n^{0.6}$, respectively. This is a result of the aforementioned magnetic decay: the red contours, taken at $t=250$~Myr, average over regions of freshly amplified and decayed magnetic fields.

Inspecting the right-hand side of  Fig.~\ref{figure:rho-B-phase-diagram--components}, we find that there are several differences compared to the {Flat-TurbB} simulations. Firstly, the top and bottom of the contours no longer scale the same way. In fact, the $y$- and $z$-components have a more extended tail towards low $B$-values, compared to the {Flat-TurbB}. This is likely both a result of the extended Mach number distribution, which produces a distribution of $x_\rmn{s}$ values, but also due to regions of rarefaction taking place behind the shock front, as are visible in Fig.~\ref{figure:shock-tube} and the associated animated figure.

The upper contours also show much stronger scaling. Indeed, the $y$- and $z$-components are very close to $B \propto n$. This is possible if compression happens at late times, when the magnetic field has less time to decay. Regions of compression can be seen in Figs.~\ref{figure:slice-downstream} and \ref{figure:plasma-beta-slice}; they are the filament-like shapes, which are predominantly sheets seen in cross-section. Indeed, this compression now happens across all three components, as the Rayleigh-Taylor instability also leads to the gas being compressed in the $y$- and $z$-directions.

However, whilst a significant amount of amplification arises from compression, some decay will still naturally occur, and compression cannot account for the peak scaling observed for the $x$-component. Indeed, it is notable that this component experiences the strongest amplification. By applying similar analysis to our lower resolution simulations (see Appendix~\ref{appendix:magnetic-resolution-study}), we find that any difference in magnetic field strength is explained by a corresponding change in density. As a magnetic dynamo should also produce higher strengths for the same density, this implies that a dynamo is either not at all, or only weakly, active in our simulations. This is supported by the animated version of Fig.~\ref{figure:shock-tube}, where it can be seen that eddies experience few if any turnovers, and a full turbulent cascade does not have time to form.

The varying speeds of the downstream flow (see panel {v} of Fig.~\ref{figure:shock-tube}) does, however, lead naturally to the stretching and shearing of the magnetic field. This process is easily able to produce $B \propto n^{\alpha_B}$ scalings with $\alpha_B>1$. Moreover, magnetic filaments pointing in the $x$-direction are clearly seen in panel {iii)} of Fig.~\ref{figure:slice-downstream} and in Fig.~\ref{figure:plasma-beta-slice}, as would be formed by shearing motions. We hence conclude that shearing is likely critical for the production of $\upmu$G magnetic fields in radio relics\footnote{Volumetric averages are still well below this value but we shall expand on this in the following section.}.

\begin{figure*}
    \centering
    \includegraphics[width=2.0\columnwidth]{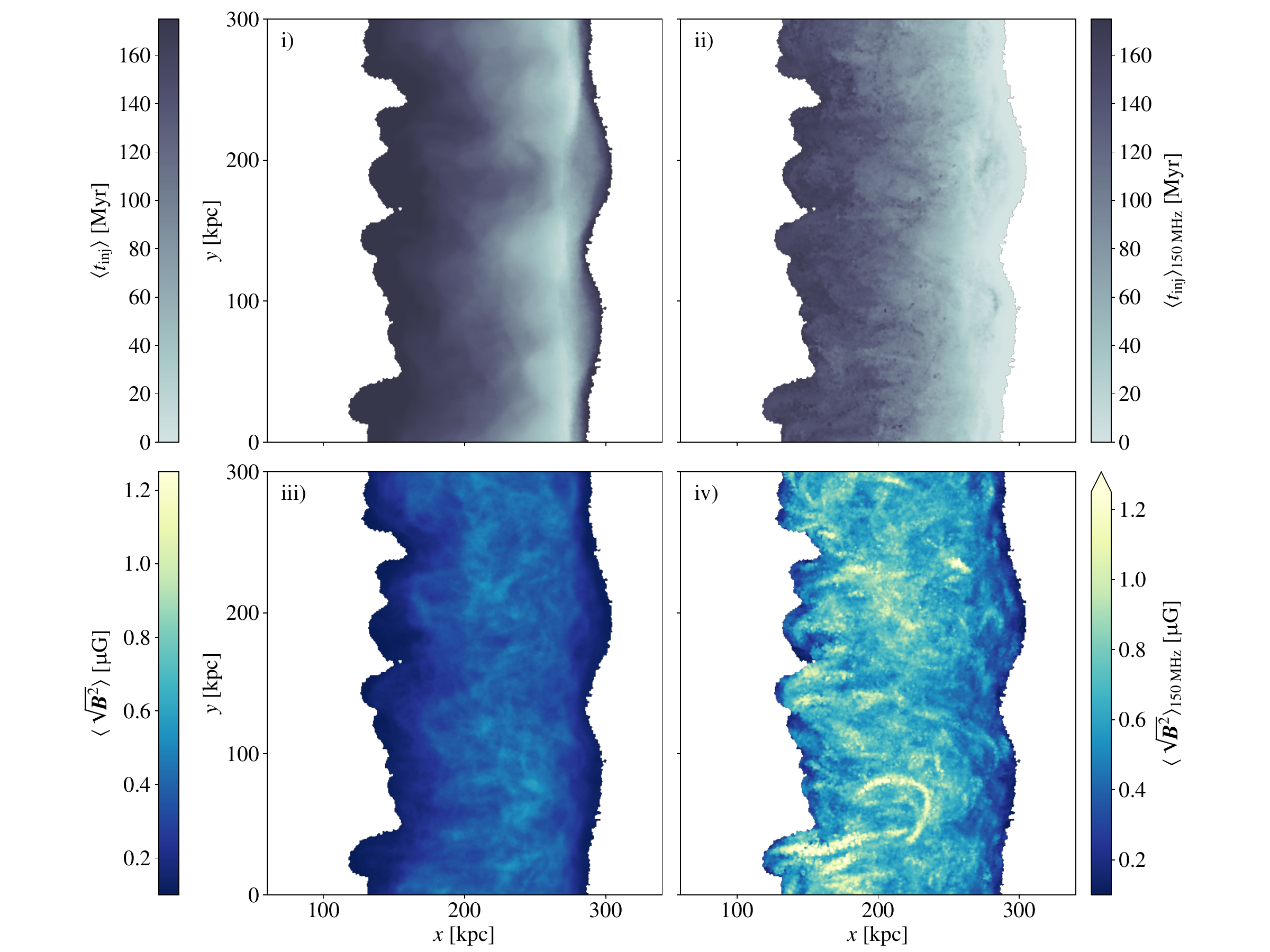}
    \caption{Projections with a depth of 300 kpc through our fiducial Mach 3 simulation at $t=180$ Myr. These show: i) volume-weighted time since injection, ii) synchrotron-weighted time since injection, iii) volume-weighted magnetic field strength, and iv) synchrotron-weighted magnetic field strength. Regions with no injected tracers have been masked (see text for further details). Synchrotron emission is calculated at $\nu = 150$~MHz and is biased towards fresher injection and stronger magnetic fields. Observations made through this channel consequently appear to show a relatively laminar flow and $\upmu$G-strength magnetic fields, even though this is not typical of the downstream.}
    \label{figure:t_inj-and-B-field-projections}
\end{figure*}

\subsubsection{Impact on observational inferences}
\label{sec:impact-on-obs}

\begin{figure*}
    \centering
    \includegraphics[width=2.0\columnwidth]{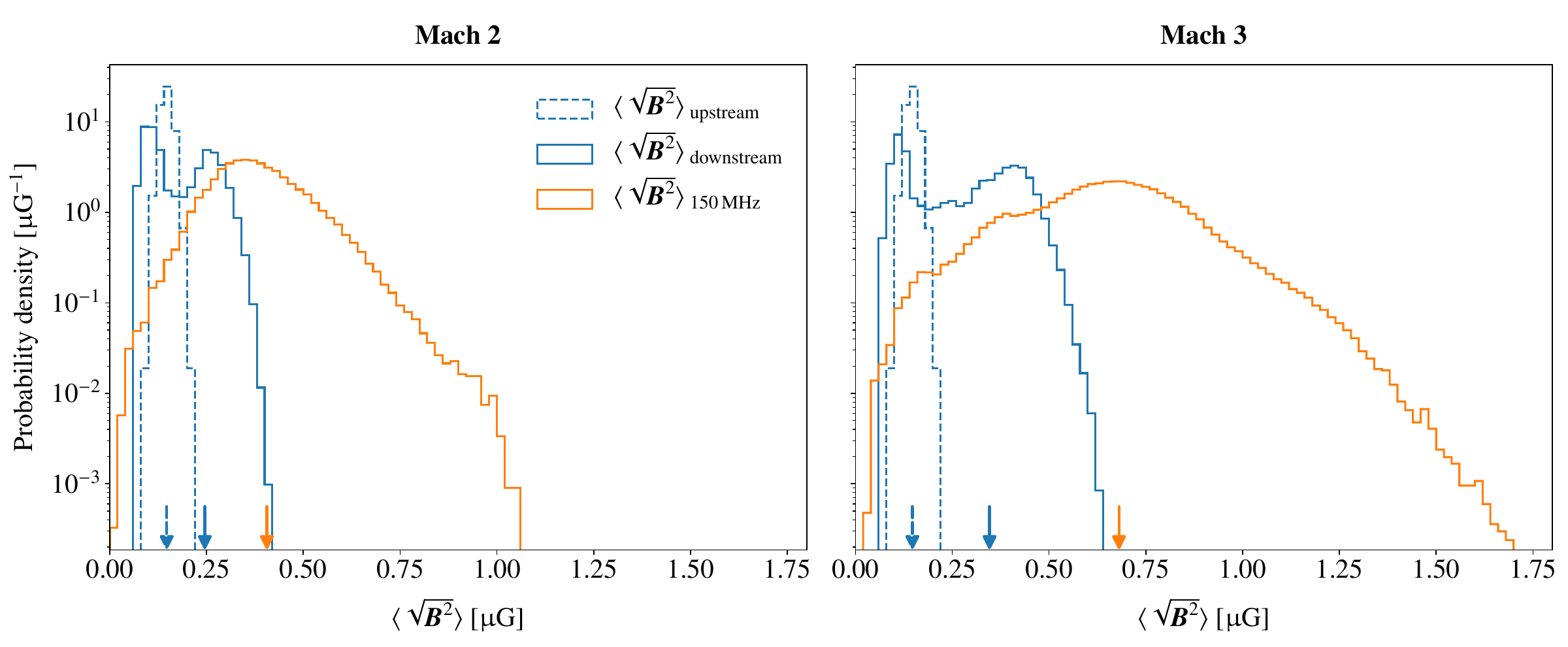}
    \caption{{Left:} Probability density distributions of the projected magnetic field strength in our fiducial Mach 2 simulation. Solid (dashed) lines represent projections through the shock-compressed region at $t=250$ Myr (high-resolution region at $t=0$ Myr). Blues lines represent volume-weighting, whilst orange lines represent synchrotron-weighting, where the synchrotron emission is calculated at $\nu = 150$~MHz. Arrows are placed at the RMS magnetic field strength calculated for each distribution. {Right:} As previous, except data come from the fiducial Mach 3 simulation. Synchrotron-weighting significantly overestimates the average magnetic field strength. Magnetic field values are able to reach $\upmu$G values even in weaker shocks, although amplification is more effective at higher Mach number shocks.}
    \label{figure:projected-B-histograms}
\end{figure*}

Returning to Fig.~\ref{figure:slice-downstream}, we may now inspect how the magnetic field strength affects synchrotron emissivity, which is shown in panel {iv)}. This is calculated using \textsc{Crayon+}, as explained in Sect.~\ref{subsec:crayon}, and is shown in this figure at 150 MHz. The colourbar for this panel is given the same overall range as was given to $f(p=2\times10^4)$. It can be seen that the dynamic range, however, is substantially smaller in panel {iv)}. This is due to the significant amplification that takes place towards the back of the shock-compressed region. There are two ways in which this impacts the synchrotron emissivity. Firstly, as can be seen by inspecting Eq.~\eqref{eq:synchrotron-emissivity}, there is a linear scaling of the emission with the component perpendicular to the line of sight. Secondly, through Eq.~\eqref{eq:critical-frequency}, a change in the magnetic field strength shifts the momenta at which most emission takes place. Specifically, as $\nu_\rmn{c} \propto B \gamma^2$, an increased magnetic field strength means that lower momenta, which have higher values of $f(p)$, contribute more at the same frequency of emission.  Together these effects boost emission from the rear of the shock-compressed zone, thereby evening out some of the effects of cooling. Indeed, in some regions where $f(p)$ is low, such as towards the back of the shock at $y\approx210$ kpc, the synchrotron emissivity is now higher than at the front of the shock. In general, however, the emissivity still decreases from front to back. We also note that the brightest regions are still at the front of the shock, at the parts where we previously identified the highest Mach number shocks. The front of the shock undergoes milder magnetic field amplification and hence milder modulation, and so our analysis in Sect.~\ref{sec:laminar-flow} regarding features close to the shock front holds for the synchrotron emission too.

We now consider how this modulation affects how we infer variables from observations of radio relics. As stated earlier, synchrotron emission is our primary window into CR electron properties in radio relics, and so it is critical to understand how observations are biased. In Fig.~\ref{figure:t_inj-and-B-field-projections}, we investigate the impact of synchrotron-weighting on the projected time since injection and projected magnetic field strength. On the left-hand side, we show the volume-weighted values, whilst on the right-hand side we show the synchrotron-weighted projections. We use the 150 MHz channel for this, as previously. We only include values where the line of sight crosses at least one injected particle. The colourbars in each row have the same scale, so panels may be directly compared.

In panel {i)}, we project the time since injection across all tracers. We set tracers that were not injected to have values equal to the simulation run-time ($t=180$ Myr). This leads to a gradual fade-out towards the left-hand side, as tracers age and as we project through non-injected tracers as well. This effect also highlights the Rayleigh-Taylor fingers, where the majority of injected tracers are. These are not as clear as in Figs.~\ref{figure:shock-tube} and \ref{figure:slice-downstream} as  we are now projecting through significant amounts of turbulence. On the right-hand side of panel {i)}, the non-injected tracers reduce the projected average in the most advanced parts of the shock, leading to these regions being darker than expected. This is a useful feature, however, as it shows us that the back of the shock is predominantly planar; it is the front of the shock that corrugates, as it advances into regions of lower density. Indeed, we could see this effect in cross-section in panel {iii)} of Fig.~\ref{figure:shock-tube}. We should consequently expect the brightest part of the simulated radio relic to also be relatively flat, as it is here that the shock-dissipated energy is highest (see Sect.~\ref{sec:mach-no-dist-origin} and Appendix~\ref{appendix:mach-in-projection}). We revisit this analysis in Sect.~\ref{sec:SI-and-intensity-maps}.

When we weight the projection by the 150 MHz channel, as in panel {ii)}, we start to recover the relationship between distance from the shock and time since injection; on large enough scales, distance once again becomes a reasonable proxy for age (see Appendix~\ref{appendix:spectra_by_dist}). On kiloparsec-scales, however, there is still substantial variance. This typically takes the form of horizontal striations. An exception to this is towards the very front of the shock, where two darker, curved regions are visible behind the shock corrugations. If we compare this panel with Fig.~\ref{figure:shock-tube}, which shows the same simulation at the same time, we can see that this is due to the turbulent transport of older material towards the shock front. As we showed in Sect.~\ref{sec:RT}, turbulence is most effective behind the regions where the shock corrugates. 

In panel {iii)}, we show the volume-weighted projected magnetic field strength. This reaches typical values of around 0.4~$\upmu$G, with peak values reaching closer to 0.65~$\upmu$G, as expected from Fig.~\ref{figure:rho-B-phase-diagram}. When we weight the projection by synchrotron emission at 150 MHz, as shown in panel {iv)}, we find a significant shift towards higher field strengths. Now, a substantial amount of the projection is above 1~$\upmu$G, more closely in line with observational inferences from radio relics. Once again, the emission-weighted projection predominantly shows horizontal striations, especially towards the rear of the shock-compressed zone. Indeed, this explains the patterns seen in panel {ii)}: the synchrotron emission is proportional to $B_\perp^{1-\alpha_\nu}$ (see Eq.~\ref{eq:synchrotron-emissivity}), hence CR electrons moving along these features dominate the projected quantities.

Magnetic field features stretched parallel to the $x$-axis could also be seen in Figs.~\ref{figure:slice-downstream} and \ref{figure:plasma-beta-slice}. By comparing these plots with the velocity structures evident in Fig.~\ref{figure:shock-tube}, we reinforce our previous conclusion that these features originate from the shearing of gas. Curved filaments of high magnetic strength are also evident, however. Through similar comparisons, it can be seen that these form in the troughs between Rayleigh-Taylor fingers, which may point to adiabatic compression (see  Fig.~\ref{figure:slice-downstream} again). All of these mechanisms require turbulence, and this is mainly generated at the contact discontinuity. Consequently, the strongest magnetic field values in Fig.~\ref{figure:t_inj-and-B-field-projections} are seen towards the back of the shock-compressed region, with minimal extra amplification at the very front of it.

\begin{figure*}
    \centering    \includegraphics[width=2.0\columnwidth]{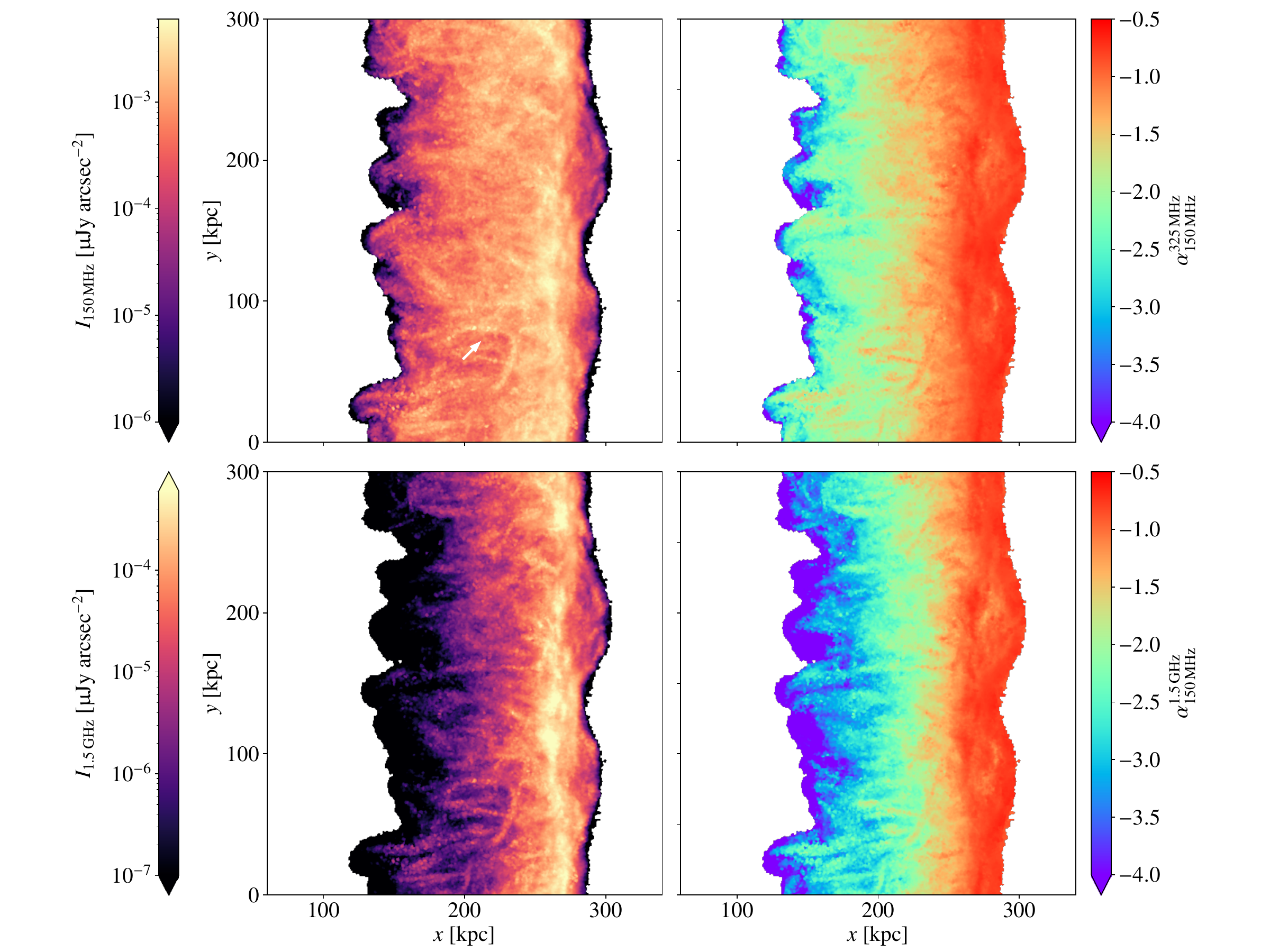}
    \caption{{Left column:} Synchrotron intensity at 150 MHz (top) and 1.5 GHz (bottom), respectively, for our fiducial Mach~3 simulation at $t=180$ Myr. The projection depth is 300 kpc. {Right column:} Spectral index maps between 325 MHz and 150 MHz (top) and 1.5 GHz and 150 MHz (bottom), respectively. Intensity fluctuations in this region are a consequence of Mach number variations, whilst spectral index variations towards the shock front are predominantly a result of the corrugated shock front. Both intensity and spectral index maps show long striations predominantly orientated along the $x$ axis, resulting from the magnetic filaments observed in Fig.~\ref{figure:t_inj-and-B-field-projections}. The white arrow indicates a particularly clear example caused by adiabatic compression of the post-shock magnetic field. An animated version of this figure can be found \href{https://www.youtube.com/watch?v=wyc2RryP3yg}{here}.}
    \label{figure:intensity-and-spectral-index}
\end{figure*}

In Fig.~\ref{figure:projected-B-histograms}, we quantify just how much the synchrotron-weighted projections are biased relative to their volume-weighted counterparts. We show the initial state in the high-resolution region (region III) in dotted lines, and the final state in the shock-compressed region at $t=250$ Myr in solid lines, with blue indicating volume-weighting and orange indicating synchrotron-weighting. The distribution is initially narrow and peaks at approximately 0.16~$\upmu$G, matching the chosen RMS value (see Sect.~\ref{subsec:generating_turb}). By construction, the initial distribution is the same in both Mach 2 and Mach 3 simulations. At the end of each simulation run, the volume-weighted distribution has broadened and now shows a double peak. This is a line-of-sight effect; with the left-hand peak representing lines of sight that predominantly pass through regions outside the shock-compressed region, and the right-hand peak representing the opposite scenario\footnote{Weakening of the field below the initial distribution is likely due to regions of rarefaction, as previously noted.}. This effect biases these projections slightly low relative to the true volume-averages in the injected region (see values in Fig.~\ref{figure:rho-B-phase-diagram}). The projected volume-weighted RMS magnetic field strengths in the Mach 2 and Mach 3 simulations are 0.23~$\upmu$G and 0.33~$\upmu$G, respectively.

The synchrotron-weighted distribution, however, shows significantly higher RMS values of 0.48~$\upmu$G and 0.78~$\upmu$G, respectively, which is more than double the volume-weighted values. This is due to the strong tail towards higher magnetic field strengths, which reach approximately 1.07~$\upmu$G and  1.73~$\upmu$G in the Mach 2 and Mach 3 simulations. Figure~\ref{figure:projected-B-histograms} shows clearly that strong caution should be exercised when using CR electron observations to infer the average magnetic field strength in radio relics.

\subsection{Impact on emission}
\label{sec:synchrotron-emission}

\subsubsection{Spectral index and intensity maps}
\label{sec:SI-and-intensity-maps}

We now turn our attention to the projected synchrotron emission itself, and how features in intensity and spectral index maps can be explained with the knowledge we have gained over the last sections. We first focus on the intensity maps, which we show for our Mach 3 simulation at $t=180$ Myr in the upper panels of Fig.~\ref{figure:intensity-and-spectral-index}. The intensity of the emission is calculated for each pixel using Eq.~\eqref{eq:intensity}. In the left-hand panel, we focus on emission at 150 MHz. The maximum values here reach approximately $6 \times 10^{-3}$ $\upmu$Jy arcsec$^{-2}$. Typical radio relic surface brightnesses, meanwhile, are $0.1 - 1$ $\upmu$Jy arcsec$^{-2}$ \citep{vanweeren2019}. This is not problematic, however, as we are not modelling first-order Fermi re-acceleration of a fossil relativistic electron population in our simulations (see Sect.~\ref{subsec:crest}). Modelling this is expected to increase the emission at Mach 3 by a factor of 10 to 100 \citep{pinzke2013}, helping bring our simulations into line with observations.

\begin{figure*}
    \centering
    \includegraphics[width=2.0\columnwidth]{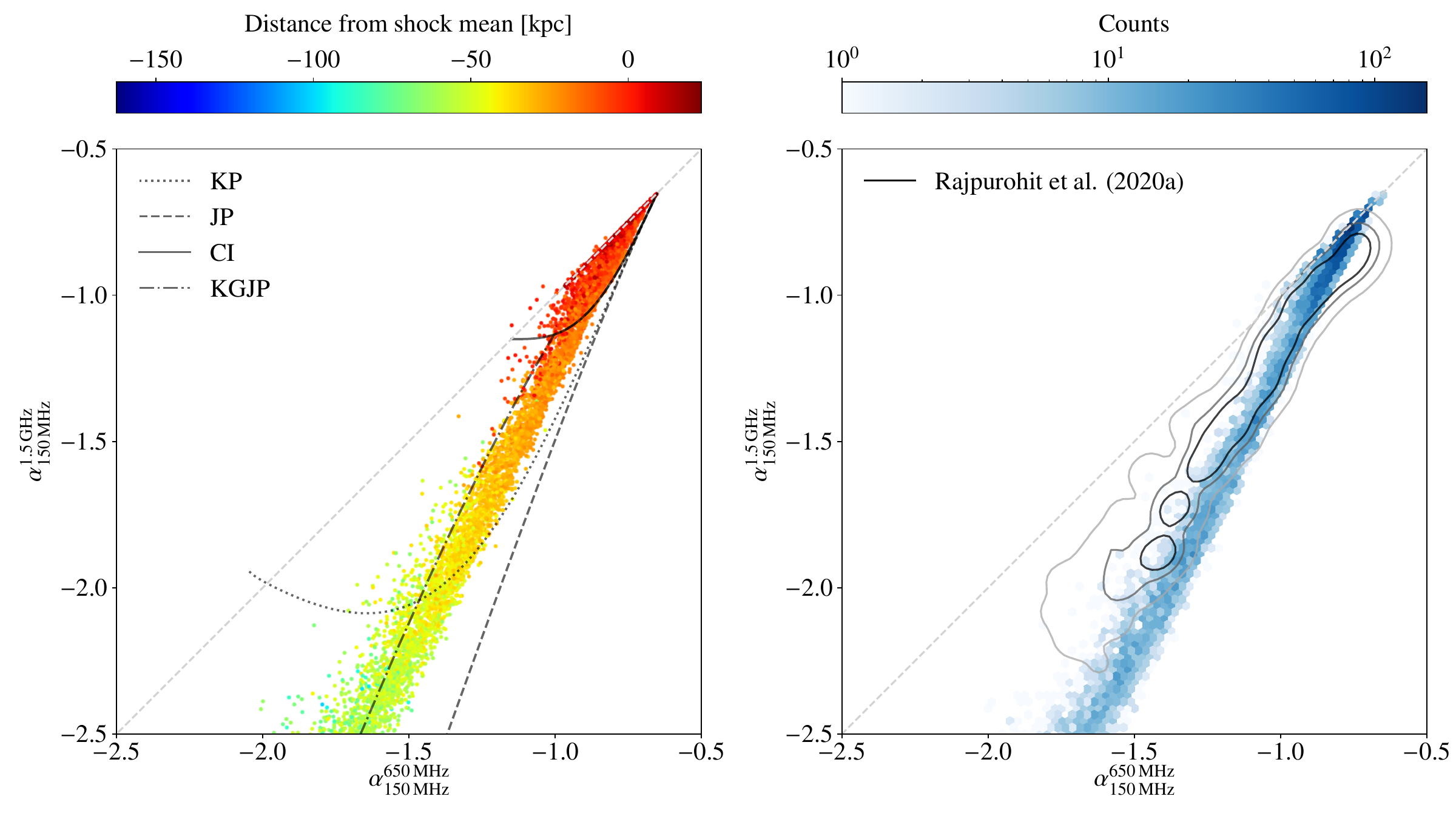}
    \caption{{Left:} Colour-colour diagram created from our fiducial Mach 3 simulation at $t=180$ Myr using spectral indices between 1.5 GHz and 150 MHz, and 650 MHz and 150 MHz, respectively. Points are coloured based on their distance to the shock median. The diagonal grey line indicates zero spectral curvature. The remaining curves indicate spectral cooling models fitted to the B1 region of the Toothbrush radio relic by \citet[][see text for details]{rajpurohit2020}. All models begin at $\alpha^{1.5\,\rmn{GHz}}_{150\,\rmn{MHz}} = \alpha^{650\,\rmn{MHz}}_{150\,\rmn{MHz}}=-0.65$. {Right:} As previous, but points have been binned over such that colour now represents the phase space density. Contours show the analogous data observed for the B1 region by \citet{rajpurohit2020} and contain 50\%, 75\%, and 95\% of the scatter points from their figure 10  (represented by darker to lighter shades, respectively). Data from the simulation contradict traditional cooling models but are in rough agreement with the observed colour-colour diagram for a real radio relic.}
    \label{figure:colour-colour}
\end{figure*}

It can be seen that the peak of the emission traces an approximately straight line, with the shock corrugation being visible as promontories ahead of it. This follows from our observation in the previous section that the back of the shock is relatively flat, especially when seen in projection (see Figs.~\ref{figure:t_inj-and-B-field-projections} and~\ref{figure:mach-number-projection}). The most advanced regions of the shock in Fig.~\ref{figure:intensity-and-spectral-index} show emission that is a factor of $10-100$ weaker than the main body of the relic, indicating that they could be trickier to locate in observations; if the shock is not observed edge-on, this feature could be overpowered by the emission behind it. We will investigate the full impact of orientation on such features in future work.

Regions of higher intensity can be seen at the shock front at $y \approx130$ kpc and $y \approx270$ kpc. These regions are especially evident in the 1.5 GHz observations, where cooling acts faster. Such regions are produced by the more intense emission created when the shock is strongest, as can be seen by comparing these features with the position of the highest projected Mach numbers in Fig.~\ref{figure:mach-number-projection}. We note that the spatial separation of these `hotspots' is approximately equivalent to the turbulent injection scale ($l_\rmn{inj} = 150$ kpc). We postpone a more detailed investigation of this correlation to the next paper in this series.

Finally, we observe that the striations towards the back of the injected region align well with the patterns seen in the synchrotron-weighted magnetic field in panel {iv)} of Fig.~\ref{figure:t_inj-and-B-field-projections}. Indeed, individual features in both plots can be matched up with one another. This implies that such synchrotron `striations' are the result of particularly strong magnetic field lines. Evidence for the origin of these striations can be seen in the spectral index maps, which we show in the right-hand panels of Fig.~\ref{figure:intensity-and-spectral-index}. We calculate the spectral index in each pixel for these maps as
\begin{equation}
    \alpha^{\nu_2}_{\nu_1} = \frac{\log_{10}(I_{\nu_2} / I_{\nu_1})}{\log_{10}({\nu_2 / \nu_1})},
\end{equation}
where $\nu_2>\nu_1$ so that $\alpha^{\nu_2}_{\nu_1}<0$ for a spectrum that decreases with increasing frequency.

The brightest filaments in the left-hand panels are evident as red-orange colours in the spectral index maps, indicating relatively fresh injection. Filaments with weaker emission are also evident, however, in greenish colours. It is notable that each filament shows a relatively uniform colour; the strong magnetic  fields in them should lead to rapid cooling, creating a colour gradient if tracers travelled along them. Indeed, the lack of colour gradient indicates that this is not the case. We find instead that the magnetic field is amplified simultaneously along its length, temporarily brightening CR electrons of the same age. This happens in one of two main ways: i) the filament is formed close to the shock front and is then advected downstream, typically being rotated vertically due to the gas flow, or ii) compression and shearing at the contact discontinuity amplifies a region of the magnetic field simultaneously, as is the case for the horseshoe-like filament evident where the white arrow is located in the top left panel. Both formation mechanisms can be witnessed in the animated version of Fig.~\ref{figure:intensity-and-spectral-index}, linked \href{https://youtu.be/A-jQvWVClyg}{here}.

Towards the front of the injected region, it can be seen that the spectral index undergoes fluctuations. Specifically, behind the most advanced part of the shock, the spectral index temporarily reduces in value before increasing again. This feature is best seen in the $\alpha^{1.5\,\rmn{GHz}}_{150\,\rmn{MHz}}$ data, in which the effects of cooling are particularly strong. This indicates that more aged material temporarily dominates the emission, as could also be seen in the synchrotron-weighted projection of the time since injection in Fig.~\ref{figure:t_inj-and-B-field-projections}. As explained in Sect.~\ref{sec:impact-on-obs}, this feature is produced by the transport of older material towards the shock front due to the Rayleigh-Taylor instability (see Fig.~\ref{figure:shock-tube}). As previously, the observation of this feature will be easiest when the relic is seen edge-on.

The spectral index values in Fig.~\ref{figure:intensity-and-spectral-index} approximately match the values and trends observed for the Toothbrush relic \citep[see e.g.][]{vanweeren2012, rajpurohit2020}. To quantify this statement, however, we turn to the use of colour-colour diagrams.

\subsubsection{Colour-colour diagrams}

\begin{figure*}
    \centering
    \includegraphics[width=1.4\columnwidth]{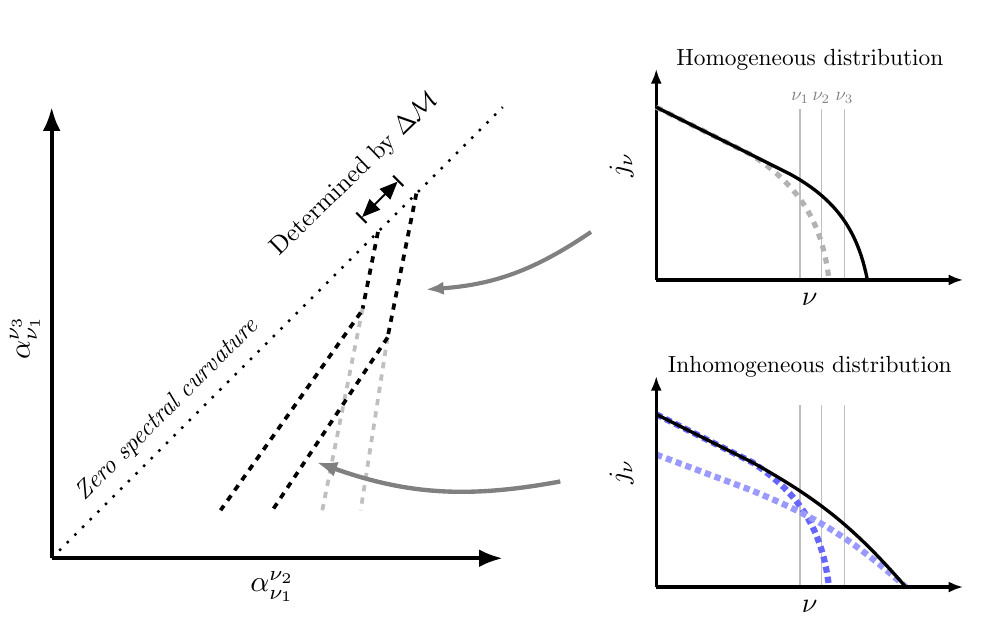}
    \caption{Schematic showing why the trajectory in the colour-colour plane flattens, as shown by the break in the dashed black lines in the left-hand panel, rather than following the standard JP model (dashed grey lines). Lines of sight initially intersect homogeneous CR electron populations (top right panel), with the same steep spectral curvature. Due to turbulence, later lines of sight intersect inhomogeneous populations (bottom right panel), made up of superpositions of cooled and fresher spectra (dashed purple and lilac lines, respectively). The resultant weaker spectral curvature correspondingly flattens the trajectory in the colour-colour plane. The effect is exaggerated here for explanatory purposes.}
    \label{figure:colour-colour-schematic}
\end{figure*}

Colour-colour diagrams are often used in radio observations to understand the underlying acceleration and cooling processes. They are created by plotting two spectral index maps against each other. Points on this plane then measure the spectral curvature, i.e. the functional behaviour of $\rmn{d}\log_{10} j_\nu / \rmn{d}\log_{10}\nu$ in the given frequency range, which, in turn, is a measure of the spectral shape. As the CR electrons age, they trace a trajectory in the plane, indicating how the spectral shape evolves with time. This can then be compared against the trajectories produced by various models.

As discussed in Sect.~\ref{sec:intro}, four models are typically used. These are abbreviated: KP, JP, CI, and KGJP\footnote{Examples of the characteristic spectral shapes formed by each model can be seen in figure~11 of \citet{vanweeren2012}.}. The first two of these model a one-time injection, assuming that the pitch angle distribution in each CR electron population is either fixed or continuously re-isotropised, respectively. The difference, here, is that the KP model leads to an increasingly inhomogeneously cooled population along the line of sight, leading to flatter spectral curvature at later times. This generates a curved trajectory in the colour-colour plane. In contrast, the JP model produces homogeneous populations along the line of sight, generating relatively straight lines in the colour-colour plane, indicative of approximately constant spectral curvature. The CI and KGJP models are extensions of the JP model, and assume continuous injection and injection for a limited time only, respectively. These models also produce (initially) curved trajectories in the colour-colour plane, for the same reasons as above.

In theory, such tracks are insensitive to the rate of cooling, as this only affects the rate at which the spectra develops, not its overall shape. Varying the density and magnetic field strength\footnote{The magnetic field strength is able to affect the `break' frequency, above which cooling dominates. As explained in Sect.~\ref{subsec:crest}, however, we expect this frequency to be predominantly set by inverse Compton cooling.} should therefore only change the distance of the points from the shock, not their locus on the plane. The key assumption underpinning each model, however, is that lines of sight intersect CR electron populations that have aged for the same amount of time. This implicitly assumes an edge-on perspective and that distance from the shock front, $d$, is a reliable measure of time cooled, $t$; specifically, that $d = \bupsilon_\rmn{post} t$, where $\bupsilon_\rmn{post}$ is the post-shock velocity given by the jump conditions.

In Fig.~\ref{figure:colour-colour}, we show a colour-colour diagram for our fiducial Mach 3 simulation at $t=180$ Myr, created from the data presented in Fig.~\ref{figure:intensity-and-spectral-index}. In the left-hand panel, we show a scatter plot, where each point matches a pixel in the projected emission maps. We colour each point by its distance from the shock median. For reference, we include the cooling models fitted to the B1 region of the Toothbrush relic by \citet{rajpurohit2020}. We do not specifically aim to replicate this radio relic in the current work, but by coincidence our simulation approximately matches the same maximum spectral index values, with both observed and simulated values reaching $\alpha \approx -0.65$. We should therefore expect such models to also fit the simulation. It is evident, however, that none of them do.

In particular, the KP and CI models produce shapes fundamentally incompatible with the simulations. The KGJP model appears to fit within the scatter, but this, too, shows an initially curving trajectory, not evident in the simulated data, and is too steep once continuous injection is turned off. Moreover, as the models start at the data point with the highest spectral index, their trajectories should be compared with the bottom of the scatter, not its mid-point. With this in mind, it can be seen that the JP model is the closest to describing the data. This is perhaps to be expected, as our setup is a one-time injection with cooling independent of pitch angle (see Sect.~\ref{subsec:crest}), which matches the theoretical basis of the model. However, even the JP model has flaws; firstly the trajectory produced is too steep, and secondly, this model is unable to replicate the subtle flattening of the simulated trajectory, which begins at $\alpha^{1.5\,\rmn{GHz}}_{150\,\rmn{MHz}} \approx -1.3$. 

This `flattening' feature is perhaps best seen in the right-hand panel of Fig.~\ref{figure:colour-colour}, where we show the phase space density. This feature appears in the Sausage relic \citep[see figure 19 of][]{stroe2013} and in the Toothbrush relic. To show this, we overlay contours indicating the phase space density observed for the B1 region of the Toothbrush relic, as given in figure 10 of \citet{rajpurohit2020}. To produce the contours, we have binned the scatter points provided and smoothed the resulting 2D histogram with a Gaussian kernel. It can be seen that the trajectory of the observed data also flattens at approximately the same point. To help explain the origin of the feature, we present a schematic in Fig.~\ref{figure:colour-colour-schematic}.

Here, the distribution of points in the colour-colour plane is represented by the two dashed, black lines in the left-hand panel. These lines represent the approximate boundaries of the distribution of spectral indices resulting from the Mach number distribution. The trajectory is initially steep, but constant, as lines of sight intersect homogeneous populations of CR electrons, which have all cooled for the same amount of time. This scenario is shown in the top right panel. As time increases, the `break' frequency moves leftwards, as shown by the dashed grey line in the same panel. The spectrum thereby becomes increasingly steep when measured at the same frequencies. However, the spectral curvature, which is the relation of the logarithmic slope of the radio surface brightness measured between $\nu_3$ and $\nu_1$, and $\nu_2$ and $\nu_1$, stays the same. If this evolution continued, it would result in the continuation of the straight trajectory in the left-hand panel, shown by the light grey, dashed lines.

Instead, however, the trajectory flattens. This is due to the turbulence in our simulations, which results in lines of sight intersecting inhomogeneous populations that have cooled for different amounts of time. The result of this scenario is shown in the bottom right panel. Lines of sight are now contributed to by both cooled spectra (purple line) and by more freshly injected spectra (lilac line). This weakens the curvature of the integrated spectra, reducing the difference between $\alpha^{\nu_3}_{\nu_1}$ and $\alpha^{\nu_2}_{\nu_1}$ and hence flattening the trajectory in the colour-colour plane as well. This effect can be seen by comparing the binned spectra in Fig.~\ref{figure:spectra-binned-by-tinj} and Fig.~\ref{figure:spectra-binned-by-distance}, which bin by age and by distance respectively. It is clear that binning by distance produces weaker spectral curvature, as we no longer bin over homogeneously cooled populations. 

Under this interpretation, the trajectory will flatten when the intensity of the more freshly injected spectra contributes significantly to the overall intensity of the pixel. Whilst this effect begins in Fig.~\ref{figure:colour-colour} at approximately the same point for both the simulated and the observed data, we note that the overall trajectory of the observed data at $\alpha^{1.5\,\rmn{GHz}}_{150\,\rmn{MHz}} < -1.3$ is even flatter than that seen in our simulation. There are multiple possibilities to explain this, including smoothing and projection effects \citep{rajpurohit2020}, or potentially still missing physics, such as Mach number dependent electron acceleration efficiencies, insufficient mixing due to post-shock turbulence not being strong enough, pitch-angle dependent cooling in combination with an incomplete pitch-angle isotropisation process, or turbulent re-acceleration \citep{fujita2015, kang2024}. We defer a more detailed analysis of the differences to future work, where we may compare with a more like-for-like simulation.

Finally, in the right-hand panel of Fig.~\ref{figure:colour-colour}, it can be seen that both the observed and simulated data produce trajectories with similar widths. This remains fairly constant over the length of the trajectory, and will initially be set by the range of Mach numbers at the shock front, as these produce different spectral indices (see Eq.~\ref{eq:mach-no-slope}). Points lying close to the dashed, grey line, which marks zero spectral curvature, have experienced minimal cooling, and should therefore be able to constrain the overall distribution. We will investigate the use of this as a diagnostic in future work as well.

\section{Discussion}
\label{sec:discussion}

\subsection{What affects the formation of the Rayleigh-Taylor instability?}
\label{subsection:impact-on-RTI}

The Rayleigh-Taylor instability has played a large role in the results shown in this paper. Specifically, in Sect.~\ref{sec:magnetic-field} we showed that it plays a fundamental role in amplifying the magnetic field through compression and shearing, whilst in Sect.~\ref{sec:synchrotron-emission} we showed that it breaks the commonly used assumption that $d = \bupsilon_\rmn{post} t$, thereby helping to explain observed trends in spectral features. Accordingly, it is important to understand under what circumstances the instability forms. 

As explained in Sect.~\ref{sec:RT}, in order to generate a Rayleigh-Taylor instability, we require the gradients of density and pressure to be misaligned with one another. Specifically, they must be misaligned such that the induced vorticity amplifies the perturbation. In our case, this means that the density gradient must point across the contact discontinuity towards the shock-compressed region. Furthermore, the amount of vorticity generated is proportional to the size of the gradients. Indeed, the growth rate of the instability follows
\begin{equation}
    \sigma = \sqrt{|\mathbf{a}| \left( \frac{\rho_2 - \rho_3}{\rho_2 + \rho_3} \right) k},
    \label{eq:RTI-growth-rate}
\end{equation}
where $\mathbf{a} \propto -\bnabla P$ is the acceleration of the lighter material into the denser material, the density terms in brackets constitute the Atwood number, the subscripts `2' and `3' indicate post-shock gas and gas to the left of the contact discontinuity, respectively, and $k$ is the wave number of the perturbation \citep{chandrasekhar1961}. In the scenarios presented in this paper, the most influential variable in this equation is the density.

The difference between $\rho_2$ and $\rho_3$ is amplified in our simulations due to the rarefaction wave produced (see Fig.~\ref{figure:1D-shock tube-profiles}). This wave exists in cosmological simulations as well, as can be seen in the \href{https://youtu.be/Ka-Odrwwamo}{movie} of Fig.~\ref{figure:cosmological}, but is somewhat masked by the overall radial density gradient found in the cluster. We would therefore expect a slightly slower growth of the instability in a real galaxy cluster; for example, if all other variables remained the same, but $\rho_3$ increased to $1\times 10^{-28}$~g~cm$^{-3}$, we would expect the growth rate to be halved.

Slower shocks will also reduce the growth rate, as this affects $a$ in Eq.~\eqref{eq:RTI-growth-rate}. These can be generated by increasing the upstream average density or weakening the Mach number. Indeed, we have predominantly shown results from our Mach 3 simulations throughout this paper, as the growth rate here is 1.5 times faster than in the corresponding Mach 2 simulation. The turbulence is subsequently better developed in these simulations at the end of the 250 Myr runtime. Nonetheless, as we show in Fig.~\ref{figure:projected-B-histograms}, even the Mach 2 simulations are able to produce sufficient turbulence to generate $\upmu$G strength magnetic fields. 

The instability will be totally suppressed when $\rho_3 > \rho_2$; i.e. when the density achieved due to shock compression is lower than the initial density downstream. As can be seen in the \href{https://youtu.be/Ka-Odrwwamo}{movie} of Fig.~\ref{figure:cosmological}, this is the case for the propagating merger shock wave during the initial stages of the cluster merger, before it reaches an accretion shock. Along with geometric and shock-dissipated energy arguments \citep{vazza2012}, as well as the density argument given above, this may help explain why radio relics are so rarely observed close to the cluster centre: the mechanism presented in this paper requires conditions only available at the cluster outskirts.

\subsection{How does our method for generating turbulence impact the results?}

Many authors have used shock tubes to probe resolution-dependent physics. In particular, they have frequently been used to investigate how upstream turbulence affects downstream magnetic field amplification \citep[see e.g.][]{ji2016, hu2022, hew2023}. Few papers, however, have focused on the gas and shock properties relevant to radio relics, with the notable exception of \citet{dominguez-fernandez2021}. In contrast to our own work, the authors of this paper used the Ornstein–Uhlenbeck method to generate upstream fluctuations. This process creates density, velocity, and magnetic turbulence simultaneously \citep[see][and references therein]{federrath2010}. The turbulence then decays over time as the shock progresses through it.

The key advantage of the Ornstein–Uhlenbeck method is that turbulent properties are self-consistently modelled. Hence fluctuations are correlated, unlike in the simulations we present. Moreover, the magnetic field will display intermittency, which is not true of the Gaussian-random fields used in our initial conditions. Such intermittency is produced naturally by small-scale dynamos \citep{sur2021}, which are expected to generate the magnetic field in the ICM \citep{tevlin2024}.  Although density turbulence dominates the effects shown in this paper, magnetic fluctuations could be influential for the striations analysed in Sect.~\ref{sec:synchrotron-emission}. We will investigate this effect in a future study.

As already stated, in our own simulations, there is no initial velocity turbulence (see Sect.~\ref{subsec:generating_turb}). Such turbulence should exist in the ICM, however, which will also result in Mach number variations. The velocity fluctuations would need to be a substantial fraction of the shock speed, however, before they could outweigh the effect of the density fluctuations shown in this paper. We leave a precise parameter study of this effect to future work as well. 

\section{Conclusions}
\label{sec:conclusions}

A standard theory of radio relics has emerged in the past two decades. This states that radio relics are formed by the megaparsec-sized shocks generated during cluster mergers and that such shocks \mbox{(re-)accelerate} electrons, enabling them to emit in the radio band. This framework has been largely successful in explaining observational data, however, several significant challenges remain. This study has focused on three of those problems:
\begin{enumerate}[label=\roman*)]
    \item What is the origin of the $\mathcal{M}_\rmn{radio} - \mathcal{M}_\rmn{X-ray}$ discrepancy if both measurement methods trace the same physical shock?
    \item How do magnetic fields reach $\upmu$G strength in radio relics (as inferred from cooling length arguments) if the surrounding ICM has field strengths an order of magnitude smaller?
    \item Why do spectral variations fail to match standard cooling models in colour-colour diagrams?
\end{enumerate}

To answer these problems, we took a hybrid approach. First we identified typical conditions in cosmological zoom-in simulations of cluster mergers before applying our findings to significantly higher-resolution idealised shock tubes. In this last step, we used the CR electron spectral code \textsc{Crest} and the emission code \textsc{Crayon+}, thereby producing synchrotron data {ab-initio}.

From our cosmological simulations, we have identified that the merger shock typically meets an accretion shock at distances commonly recorded for radio relics. This leads to the production of a narrow, shock-compressed density sheet (Fig.~\ref{figure:cosmological}). This scenario can be modelled as a shock tube problem (see Sect.~\ref{subsec:shock-tube-setup}). In addition, we included upstream density turbulence in our shock tubes, which is not well-resolved at the outskirts of our zoom-in simulations but is evident in many previous studies and observations. We chose a log-normal distribution and set the variance of the distribution according to results given in \citet{zhuravleva2013}. 

We find that density fluctuations are directly responsible for the following effects:
\begin{enumerate}[label=\roman*)]
    \item A Mach number distribution forming at the shock front (Figs.~\ref{figure:shock-tube} and~\ref{figure:mach-no-pdf}),
    \item A Rayleigh-Taylor instability forming at the contact discontinuity (Figs.~\ref{figure:shock-tube} and~\ref{figure:RTI-schematic}), and
    \item Shock corrugation (Fig.~\ref{figure:shock-tube}).
\end{enumerate}
We show how these features form in Sect.~\ref{sec:RT}. These effects provide answers to our three main questions:

\begin{itemize}

\item {X-ray versus\ radio discrepancy:} In Sect.~\ref{sec:spectra}, we show that the tail of the Mach number distribution dominates the higher frequencies. In particular, it flattens the integrated injected spectra and results in the slope at radio emitting frequencies being shallower than the theoretical value of $\alpha_e - 1$ (Fig.~\ref{figure:spectra-binned-by-tinj}). Both of these effects lead to an over-estimation of the Mach number with respect to the actual peak of the distribution (Table~\ref{tab:inferred_mach_no}). The effect is particularly strong in weak shocks ($\mathcal{M} \lesssim 2$).
\vspace{0.2cm}

\item {Magnetic field amplification:} The Rayleigh-Taylor instability causes higher levels of amplification than would be expected due to the standard theory of shock-compression alone. This boosts the magnetic field strength from ICM-like conditions up to $\upmu$G levels, although the volume-average remains significantly below this (Figs.~\ref{figure:slice-downstream} and~\ref{figure:plasma-beta-slice}). We find that the amplification is predominantly driven by the time-varying adiabatic compression and shearing that results from the instability. (Figs.~\ref{figure:rho-B-phase-diagram} and~\ref{figure:rho-B-phase-diagram--components}).
\vspace{0.2cm}

Additionally, we find that the projected synchrotron emission is dominated by the strongest magnetic field values encountered along the line of sight. This means that estimations of the magnetic field made from emission are heavily biased by these regions; values inferred from cooling length arguments are only representative of the tail of the distribution, not the volume-average (Figs.~\ref{figure:t_inj-and-B-field-projections} and~\ref{figure:projected-B-histograms}).
\vspace{0.2cm}

\item {Cooling models:} The turbulence generated by the Rayleigh-Taylor instability breaks the laminar flow assumption. This is especially true on an electron-by-electron basis, where distance from the shock front is a poor indicator of time since injection (Figs.~\ref{figure:spectra-binned-by-tinj} and~\ref{figure:slice-downstream}). The relationship is somewhat better recovered when projecting the synchrotron-weighted time since injection, but even here $d = \bupsilon_\rmn{post} t$ no longer holds true (Fig.~\ref{figure:t_inj-and-B-field-projections}). Turbulence leads to the mixing of spectra along the line of sight, which acts to flatten the spectral curvature and explains why standard cooling models do not work for radio relics (Figs.~\ref{figure:colour-colour} and~\ref{figure:colour-colour-schematic}).

\end{itemize}

In addition, we find that features in surface brightness and spectral index maps can be attributed to specific mechanisms (Fig.\ref{figure:intensity-and-spectral-index}). Specifically, the corrugation of the shock front leads to filamentary emission in projection. Mach number fluctuations, on the other hand, cause fluctuations in intensity along the shock front that spatially correlate with the injection scale. The Rayleigh-Taylor instability can produce spectral index fluctuations towards the front of the shock. Finally, magnetic fluctuations, either advected downstream, sheared, or compressed at the contact discontinuity, produce striated emission aligned with the shock normal. In the next papers in the series, we will show how these effects can be used to determine ICM conditions and will address the critical Mach number question.

\begin{acknowledgements}
    The authors thank Kamlesh Rajpurohit for providing data used in Fig. 13 of this paper. They thank Thomas Berlok for help with the analysis tools package \textsc{Paicos} \citep{berlok2024} and support in accessing the \textsc{Pico-Cluster} simulations. They would also like to thank R\"udiger Pakmor, Anatoly Spitkovsky, and the organisers and participants of {Galaxy Clusters \& Radio Relics II} for stimulating discussions and presentations, which helped improve this paper. JW acknowledges support by the German Science Foundation (DFG) under grant 444932369. CP and JW acknowledge support by the European Research Council under ERC-AdG grant PICOGAL-101019746. CP and LJ acknowledge support by the DFG Research Unit FOR-5195. PG acknowledges financial support from the European Research Council via the ERC Synergy Grant `ECOGAL' (project ID 855130). The authors gratefully acknowledge the Gauss Centre for Supercomputing (GCS) for providing computing time on the GCS Supercomputer SuperMUC-NG at the Leibniz Supercomputing Centre (LRZ) in Garching, Germany, under project pn68cu.
\end{acknowledgements}

\section*{Data Availability}

The data underlying this article will be shared on reasonable request to the corresponding author.

\bibliographystyle{aa}
\bibliography{bibliography}

\begin{appendix}

\section[Appendix A: Numerical stability]{Numerical stability}
\label{appendix:numerical-stability}

\begin{figure}
    \includegraphics[width=1.0\columnwidth]{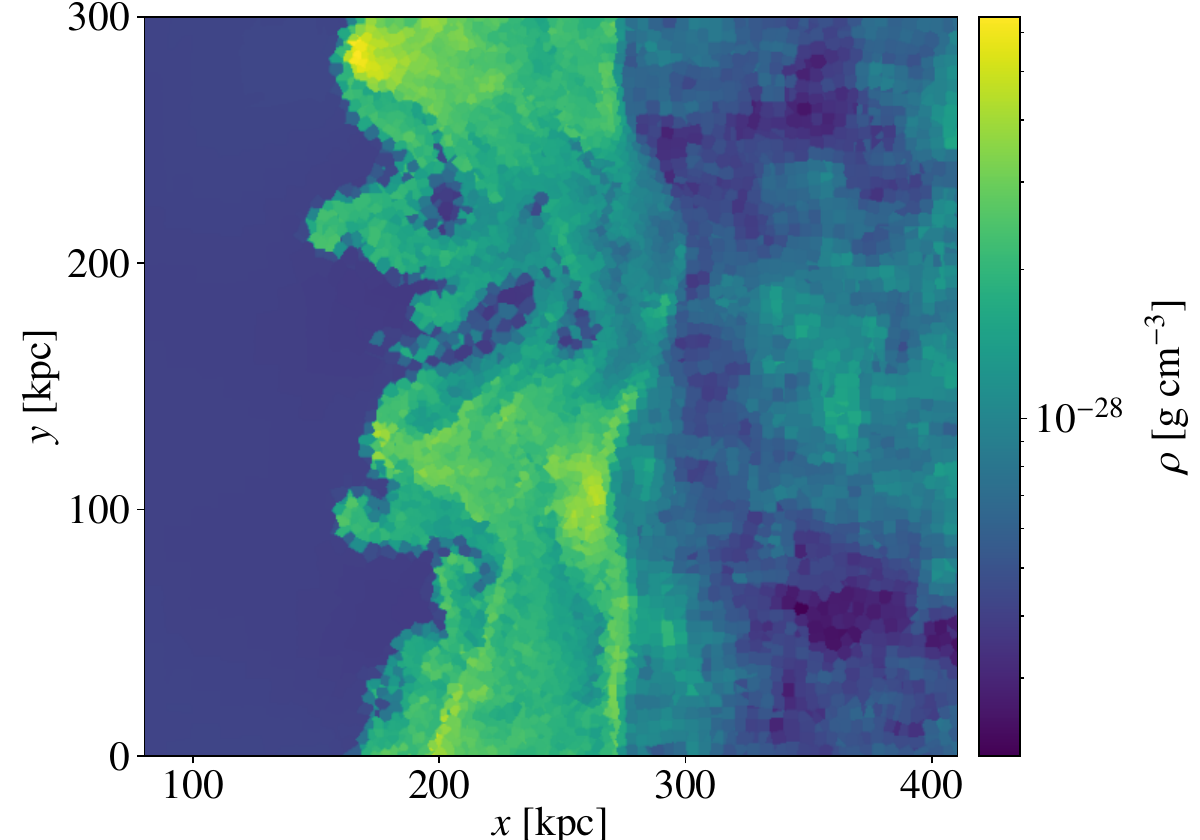}
    \caption{Version of our fiducial simulation, run with 8 times lower mass resolution. Colours indicate density, with the same limits and bounds as shown in Fig.~\ref{figure:shock-tube}. Higher resolution only leads to smaller-scale structure, indicating the numerical stability of our simulations.}
    \label{figure:low-res-density-slice}
\end{figure}

In order to trust our results, it is critical that the simulations are numerically converged. For this purpose, we re-ran our fiducial simulation, {Turb}, using a target gas mass of $m_\rmn{gas} \approx 1.2 \times 10^5 \,\rmn{M}_\odot$. This is eight times lower in mass resolution than the fiducial run or, equivalently, two times lower in spatial resolution. We find that all of our results continue to hold, although the magnetic field amplification is slightly weaker, as discussed in Sect.~\ref{subsec:mag-field-origin} and in App.~\ref{appendix:magnetic-resolution-study}.

As evidence for convergence, we present a slice through the simulation at $t=180$ Myr in Fig.~\ref{figure:low-res-density-slice}, with colours indicating density. This may be directly compared with panel {ii)} in Fig.~\ref{figure:shock-tube}. It can be seen that all structures continue to be evident, except with lower resolution. 

\section[Appendix B: The lack of a Rayleigh-Taylor instability in the Flat run]{The lack of a Rayleigh-Taylor instability in the Flat run}
\label{appendix:RTI-without-density-perturbations}
\begin{figure*}
    \includegraphics[width=2.0\columnwidth]{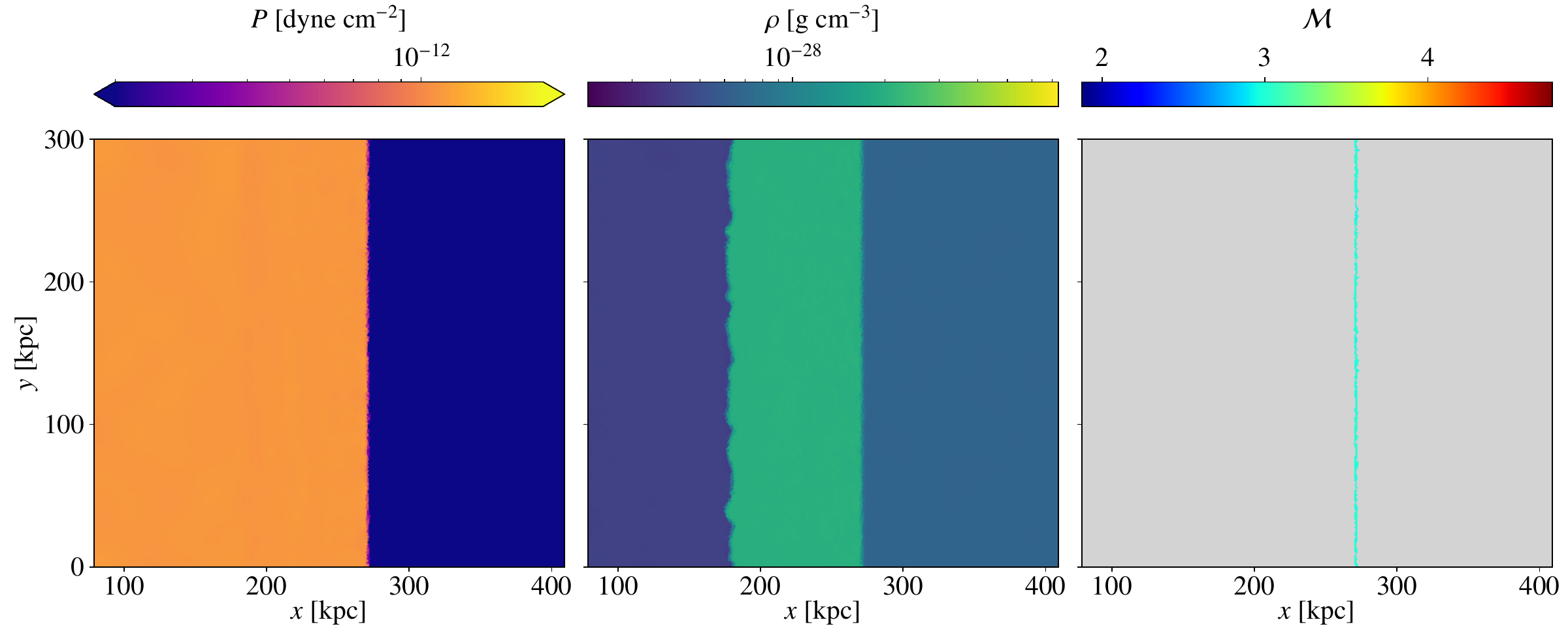}
    \caption{As Fig.~\ref{figure:shock-tube}, but only the top row is shown, and data are taken from the {Flat} simulation. Without upstream density fluctuations, the Rayleigh-Taylor instability is restricted to a very low growth rate.}
    \label{figure:shock-tube-flat}
\end{figure*}

In Sect.~\ref{sec:RT}, we claimed that density perturbations were necessary for the formation of the Rayleigh-Taylor instability in our simulations. This instability forms a critical part of our mechanism for solving the magnetic field and cooling model problems outlined in Sect.~\ref{sec:intro}. To show that perturbations are necessary, we show slices through our {Flat} shock tube simulation in Fig.~\ref{figure:shock-tube-flat}. This simulation has no density turbulence in the upstream initial conditions, but includes magnetic turbulence (see Sect.~\ref{subsec:sim-vars}). The panels shown are analogous to the panels in the top row of Fig.~\ref{figure:shock-tube} and are shown at the same time (i.e. $t = 180$ Myr).

It can be seen that without upstream density turbulence, the shock front shows no curvature. Magnetic pressure fluctuations, however, lead to mild variations in density and pressure gradients at the contact discontinuity. Consequently, the Rayleigh-Taylor instability still takes place here; indeed, this forms the bumps seen in the middle panel at the trailing edge of the shock-compressed zone. The growth-rate of this instability has been severely restricted, however. This is partially due to the significantly smaller pressure gradient, and partially due to the lack of corrugation at the contact discontinuity, which helped to seed the Rayleigh-Taylor instability in Fig.~\ref{figure:shock-tube}.
 
\section[Appendix C: Mach numbers in projection]{Mach numbers in projection}
\label{appendix:mach-in-projection}

\begin{figure}
    \vspace{-0.05cm}
    \includegraphics[width=1.0\columnwidth]{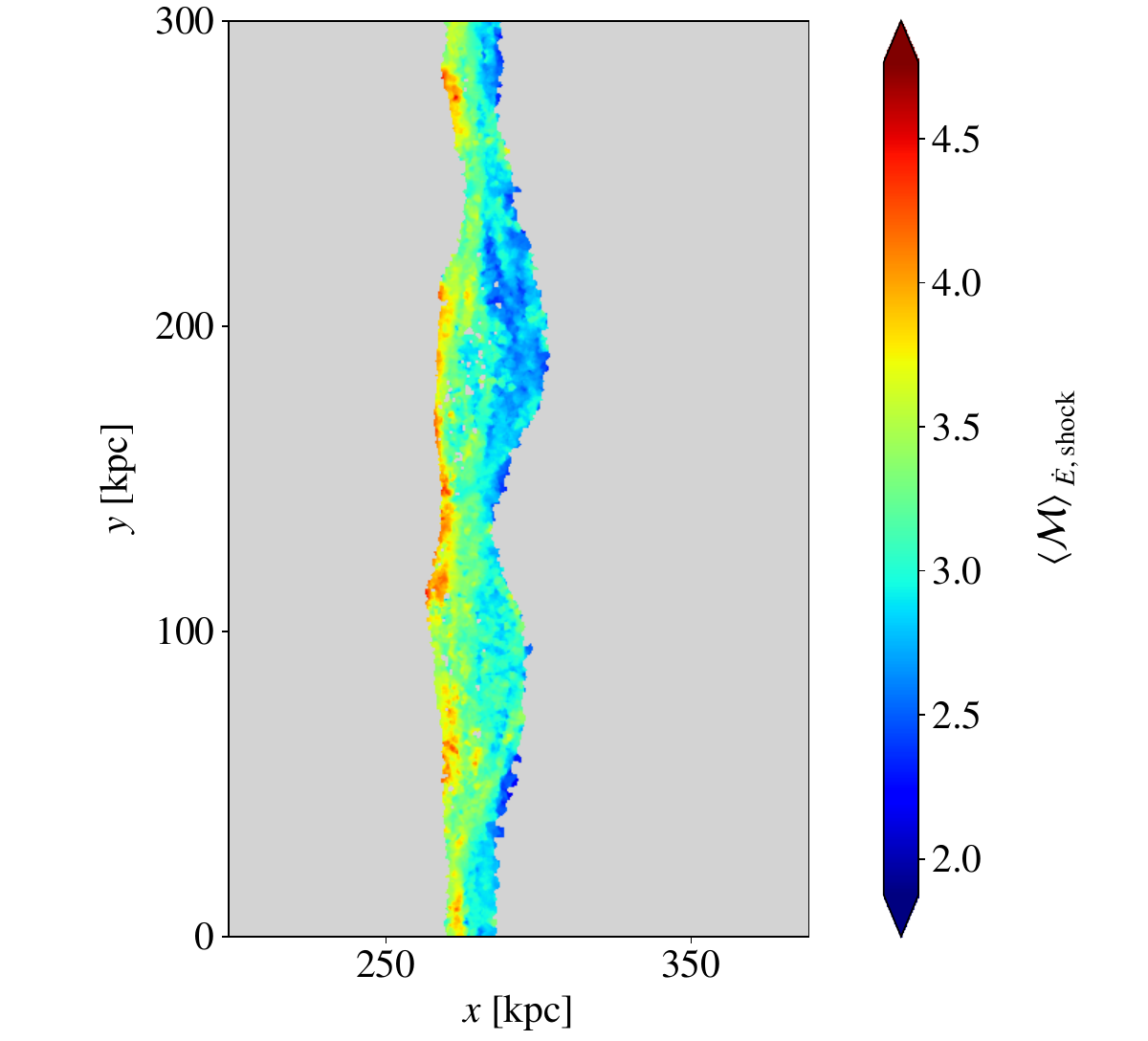}
    \caption{Projected dissipation-weighted Mach number, shown for the same simulation and time as in Fig.~\ref{figure:shock-tube}. The projection depth is 300 kpc. Weaker Mach numbers tend to form where the shock advances ahead of the median position, whilst stronger ones form where it lags behind.}
    \label{figure:mach-number-projection}
\end{figure}

In Sect.~\ref{subsec:mach-dist}, we argued that lower Mach numbers should be found typically in a more advanced position, and that stronger Mach numbers should be found towards the back of the shock. In panel {iii)} of Fig.~\ref{figure:shock-tube}, specifically, we analysed a thin projection and found the result to be consistent. In Fig.~\ref{figure:mach-number-projection}, we show the projected dissipated-energy weighted Mach number in our Mach 3 {Turb} simulation at $t=180$ Myr. This is analogous to the previous projection, except the projection depth has increased from 35 kpc to 300 kpc (i.e. the box size). It can be seen that our argument continues to hold.

We have already shown in Sect.~\ref{subsec:mach-dist} that the properties of the Mach distribution remain relatively stable over time. With this said, however, we must also emphasise that our argument only holds true in a statistical sense; Mach numbers do not decrease monotonically in value towards the most advanced part of the shock. This can also be seen in Fig.~\ref{figure:mach-number-projection}.

\section[Appendix D: Spectra binned by distance]{Spectra binned by distance}
\label{appendix:spectra_by_dist}

\begin{figure*}
    \includegraphics[width=2.0\columnwidth]{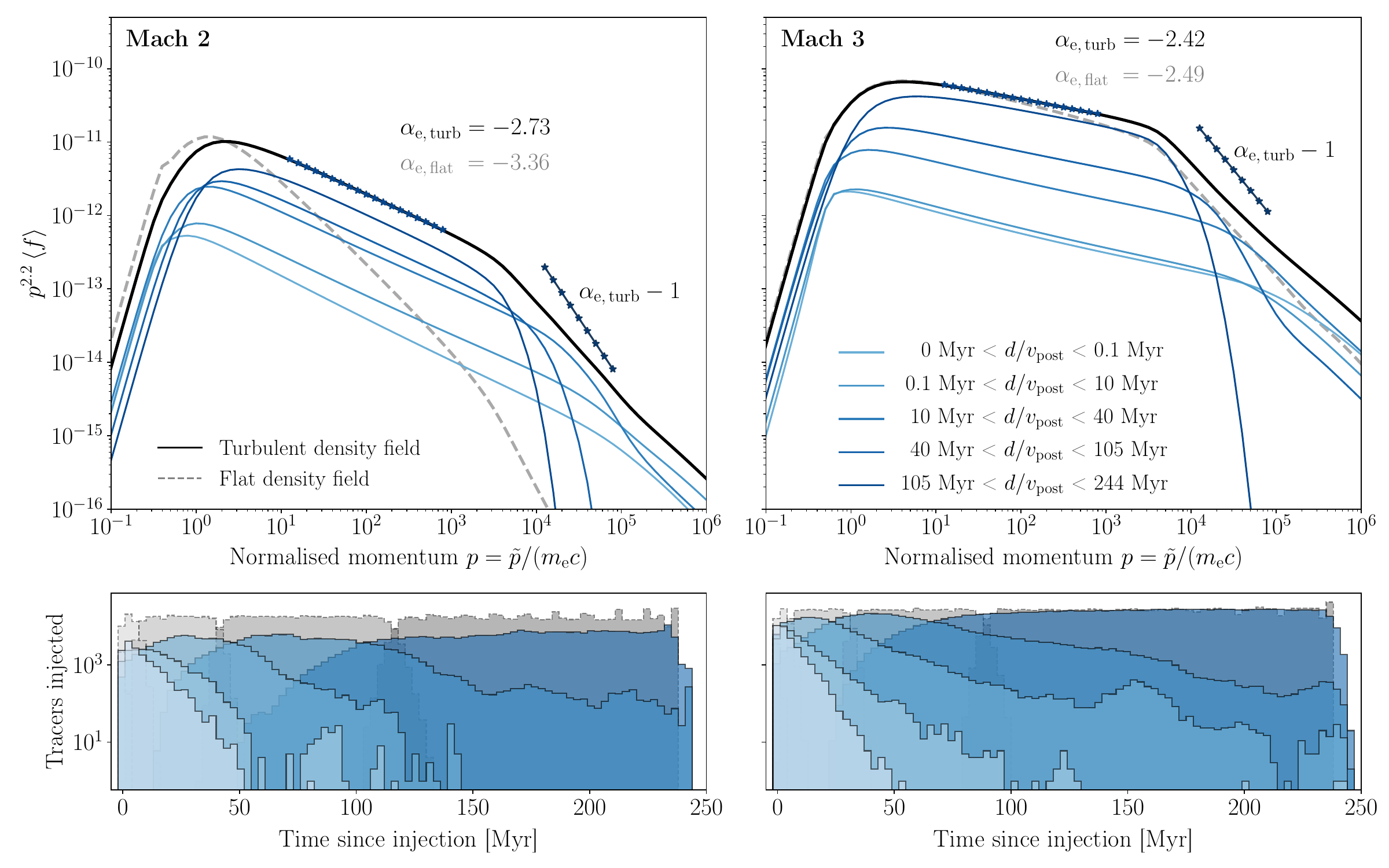}
    \caption{As Fig.~\ref{figure:spectra-binned-by-tinj}, except tracers are now binned by distance from the median shock position, rather than time since injection. Distance bins are equal to the time bins in Fig.~\ref{figure:spectra-binned-by-tinj} multiplied by the relevant post-shock velocity (approximately 440 and 500 km/s for Mach 2 and Mach 3 shocks, respectively). Contributions binned by distance do not match those binned by time since injection due to the highly non-laminar flow. This can also be seen in the strongly overlapping distributions of time since injection within each distance bin ({bottom row}).}
    \label{figure:spectra-binned-by-distance}
\end{figure*}

In Sect.~\ref{sec:laminar-flow} we presented evidence that distance from the shock is not a reliable proxy for time since injection, at least on a tracer-by-tracer basis (see caveats given in Sect.~\ref{sec:impact-on-obs}). We showed, in particular, that the volume-averaged spectrum is created through the layering of tracers binned by age, and that tracers in these bins overlap significantly when their spatial distributions are overplotted (see Fig.~\ref{figure:spectra-binned-by-tinj}). In Fig.~\ref{figure:spectra-binned-by-distance}, we plot the inverse; that is, we bin tracers by their distance and make histograms of the temporal distributions. To remove some degree of arbitrariness, we use the same time bins as previously, but now multiply out by the theoretical post-shock speed $\bupsilon_\rmn{post}$. We have calculated this by taking the actual shock speeds of approximately $1,000$ km~s$^{-1}$ and $1,500$ km~s$^{-1}$ in the Mach 2 and Mach 3 simulations, respectively, and applying the standard jump conditions, giving $\bupsilon_\rmn{post} \approx 440$ km~s$^{-1}$ and $\bupsilon_\rmn{post} \approx 500$ km~s$^{-1}$. This equates to distance intervals of approximately 0, 0.04, 4, 18, 46, and 107 kpc in the Mach 2 simulations, and intervals 1.14 times higher in the Mach 3 simulations. 

The layering in Fig.~\ref{figure:spectra-binned-by-distance} is clearly less well-defined than in Fig.~\ref{figure:spectra-binned-by-tinj}. Nonetheless, it still remains true that CR electrons further from the shock are, on average, more cooled. Consequently, it is still possible to pick distance intervals that dominate a particular momentum range. This is especially true at later ages, when the cooled part of the spectra is steeper. This is despite the fact that the tracers are strongly mixed, as can be seen by inspecting the bottom row of Fig.~\ref{figure:spectra-binned-by-distance}. 

The layering of spectra can be partially improved by removing the assumption that the distance bins in Fig.~\ref{figure:spectra-binned-by-distance} must be proportional to the time bins given in Fig.~\ref{figure:spectra-binned-by-tinj}. However, in order to see an improvement, the spacing of bins must be picked independently for Mach 2 and Mach 3 simulations. Moreover, we find that, regardless of binning, it is not possible to produce a figure entirely analogous to Fig.~\ref{figure:spectra-binned-by-tinj}. This implies that the standard formula of $d = \bupsilon_\rmn{post} t$ cannot be adapted in a linear fashion, and that laminar flow is only a good assumption for simulations without density turbulence.

\section[Appendix E: Plasma beta distribution]{Plasma beta distribution}
\label{appendix:beta}

\begin{figure}
    \centering
    \includegraphics[width=1.0\columnwidth]{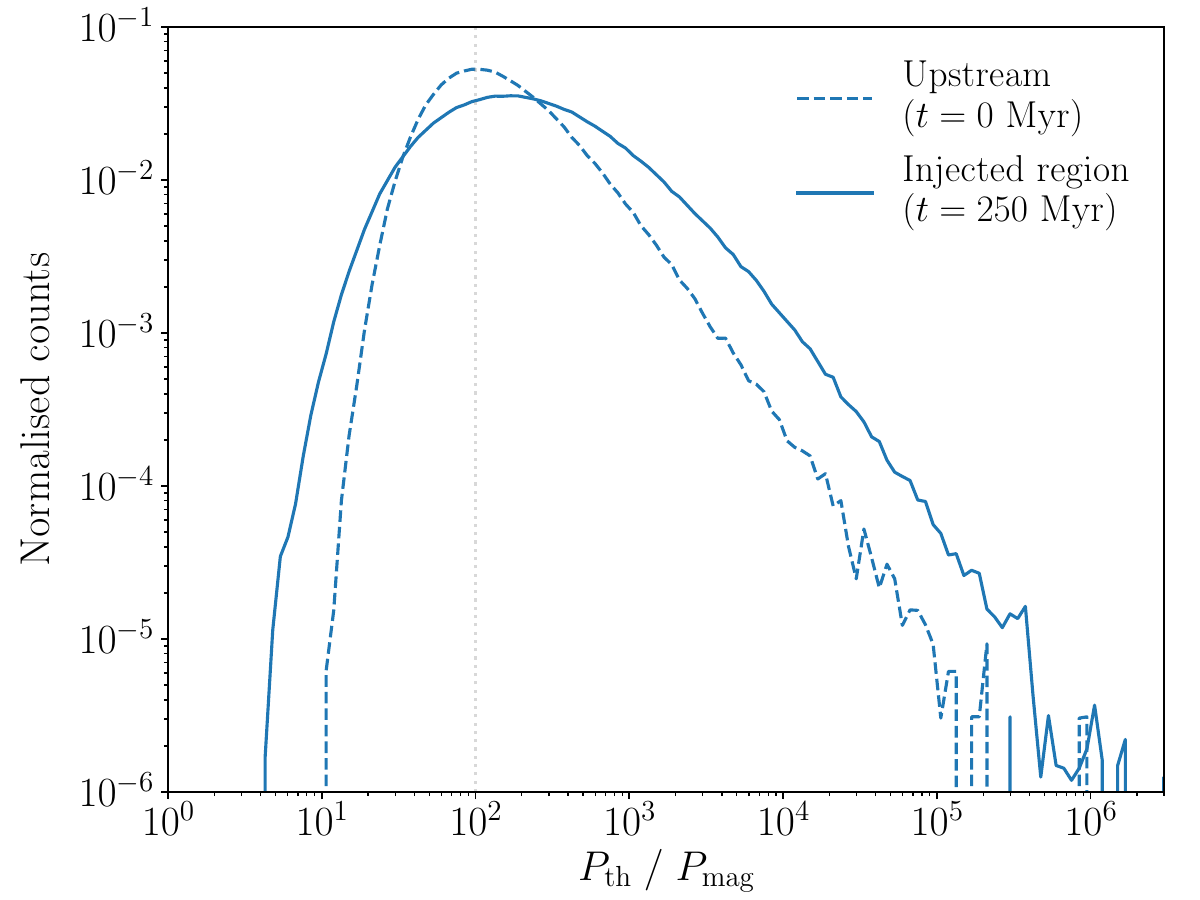}
    \caption{Histograms of the plasma beta values for all gas cells upstream at $t=0$ Myr (dashed) and in the shock-injected region at $t=250$ Myr (solid) for our Mach 3 fiducial simulation, weighted by the cell volume. The distribution initially peaks at 100 (marked by a dotted, grey line). Amplification downstream is able to increase the magnetic field pressure up to $\sim$10\% of the gas pressure in a fraction of the cells.}
    \label{figure:plasma-beta-hist}
\end{figure}

In Fig.~\ref{figure:plasma-beta-hist} we quantify our earlier statement in Sect.~\ref{sec:magnetic-field-strength} that despite significant amplification, the magnetic field does not play a major dynamical role. To do this, we show histograms of the plasma beta values in the initial upstream region (region III, in the nomenclature given in Sect.~\ref{subsec:shock-tube-setup-resolution}) at $t=0$ Myr and in the shock-compressed region at $t=250$ Myr. As discussed in Sect.~\ref{subsec:generating_turb}, we implemented density and magnetic field turbulence in our simulations independently of one another. This leads to a wide initial distribution with a strong tail towards higher plasma beta values. The RMS magnetic field strength was picked, however, to produce a peak at $P_\rmn{th} / P_B = 100$, as can be seen in the figure.

After the gas has undergone shock compression, the width of the distribution in Fig.~\ref{figure:plasma-beta-hist} increases. A minority of cells reach values $P_\rmn{th} / P_B < 10$, but no cell reaches below $P_\rmn{th} / P_B = 4$. We may therefore safely treat the downstream region as being in the kinetic regime, meaning that hydrodynamic models of the downstream remain valid.

\section[Appendix F: Magnetic field decay]{Magnetic field decay}

\label{appendix:decay}

In Sect.~\ref{sec:magnetic-field}, we claimed that the reduced scaling of the magnetic field with density in the {Flat-TurbB} simulation was due to the decay of the magnetic field. We also claimed that, in the absence of additional amplification mechanisms, this was an expected physical process in the post-shock region. We justify this claim here.

\begin{figure}
    \centering
    \includegraphics[width=1.0\columnwidth]{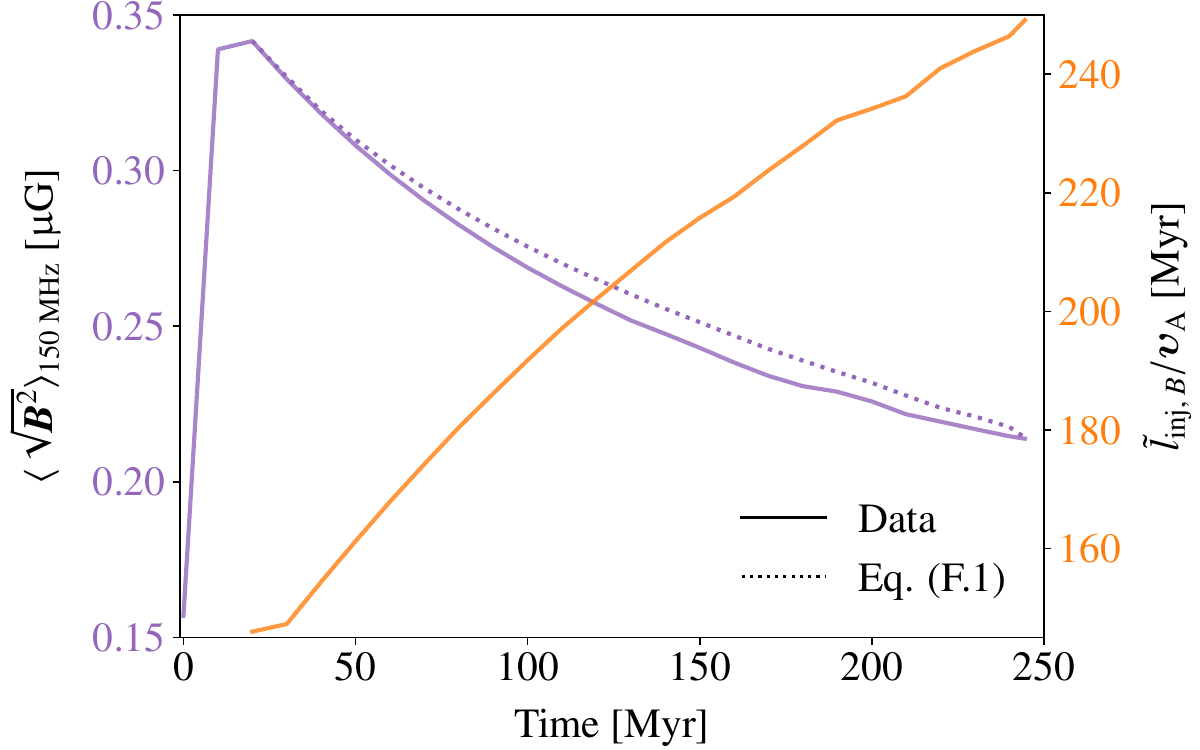}
    \caption{RMS magnetic field strength (solid lilac) and the effective Alfv\'{e}n timescale (solid orange) as a function of time for tracers in the {Flat-TurbB} simulation. Averages are calculated using all tracers that passed the shock before $t=10$~Myr. The magnetic field initially experiences the expected level of amplification, but then subsequently decays. This decay is well-described by a simple model (Eq.~\ref{eq:magnetic-decay}), which replicates the untangling of the turbulent magnetic field (dotted lilac).}
    \label{figure:magnetic-untangling}
\end{figure}

In Fig.~\ref{figure:magnetic-untangling}, we show the RMS magnetic field strength over time, calculating the average using all tracers that passed the shock before $t=10$~Myr. It can be seen that the solid lilac line starts at the upstream average value of $0.16~\upmu$G before being strongly amplified by the next snapshot\footnote{This peaks below the expected value of $0.4$ $\upmu$G, as we are averaging over a 10~Myr window. Consequently, some magnetic field values have already begun to decay at this point.}. Shortly thereafter, however, the magnetic field strength enters a prolonged decline. This decline is naturally explained by the untangling of the magnetic field: a tangled magnetic field feels the effects of magnetic tension, which is communicated at the Alfv\'en speed, $\bupsilon_\rmn{A}$. As a result, the magnetic field will untangle itself over a timescale of $\tau = l_{\rmn{inj},\,B} / \bupsilon_\rmn{A}$, where $l_{\rmn{inj},\,B}$ is the outer scale of the magnetic turbulence. Without a process to keep the field amplified, this leads to the overall magnetic energy decaying as $E_{B} \propto \exp{(- t / \tau)}$, and hence the magnetic field strength decays as $B \propto \exp{(- 0.5 \, t / \tau)}$.

In our initial conditions, we set $l_{\rmn{inj},\,B} = 40$~kpc. However, post-shock, the field is compressed along the $x$-direction by the shock compression ratio, $x_\rmn{s}$. For a Mach 3 shock, the effective injection scale hence becomes $\tilde{l}_{\rmn{inj},\,B} = 40 / 3$~kpc. This leads to a model for the observed decay of the field strength:
\begin{equation}
    \label{eq:magnetic-decay}
    B = B_2 \exp\left(- 0.5 \,\frac{\bupsilon_\rmn{A}(t)}{\tilde{l}_{\rmn{inj},\,B}} \, t\right),
\end{equation}
where $B_2$ is the peak magnetic field strength reached post-shock.

We plot this as the dotted lilac line in Fig.~\ref{figure:magnetic-untangling}. To calculate it, we keep $\tilde{l}_{\rmn{inj},\,B}$ fixed, but set $\bupsilon_\rmn{A}(t)$ to the magnetic-energy weighted mean Alfv\'en speed calculated for the same group of tracers as used for the solid line. We also renormalise $B_2$ to the time-averaged post-shock value. It can be seen that this simple model does a remarkably good job of predicting the decay of the field strength.

As the magnetic field becomes weaker, the decay slows down. This is because $\bupsilon_\rmn{A} = |B| / \sqrt{4 \pi \rho}$, where $\rho$ is the gas density. Since the post-shock gas density is effectively a constant in the {Flat} simulations, a weaker magnetic field results in a slower Alfv\'en speed\footnote{Post-shock, this quantity ranges from 90~km~s$^{-1}$ to 50~km~s$^{-1}$.} and hence a longer Alfv\'en timescale on the outer scale of magnetic turbulence. We plot this quantity in orange in Fig.~\ref{figure:magnetic-untangling}.

We note that the timescale ranges from $\sim$150$-$250~Myr, whilst the system evolves for a comparable amount of time. This explains why we see magnetic decay in the first place; were the timescale much greater, there would be no time for the decay to set in. In contrast, in the upstream, the Alfv\'en timescale is closer to 450~Myr, owing predominantly to the larger value of $l_{\rmn{inj},\,B}$.

Finally, we compute the average post-amplification RMS magnetic field strength in Fig.~\ref{figure:magnetic-untangling} to be 0.27~$\upmu$G. This means that the effective magnetic field increase is by only a factor of 1.7, and hence we recover a scaling of $B\propto n^{\alpha_B}$, with $\alpha_B = \log (1.7) / \log(3) \approx 0.5$, as observed in Fig.~\ref{figure:rho-B-phase-diagram}.

\section[Appendix G: Magnetic amplification at lower resolution]{Magnetic amplification at lower resolution}
\label{appendix:magnetic-resolution-study}

\begin{figure}
    \centering
    \vspace{-0.05cm}
    \includegraphics[width=1.0\columnwidth]{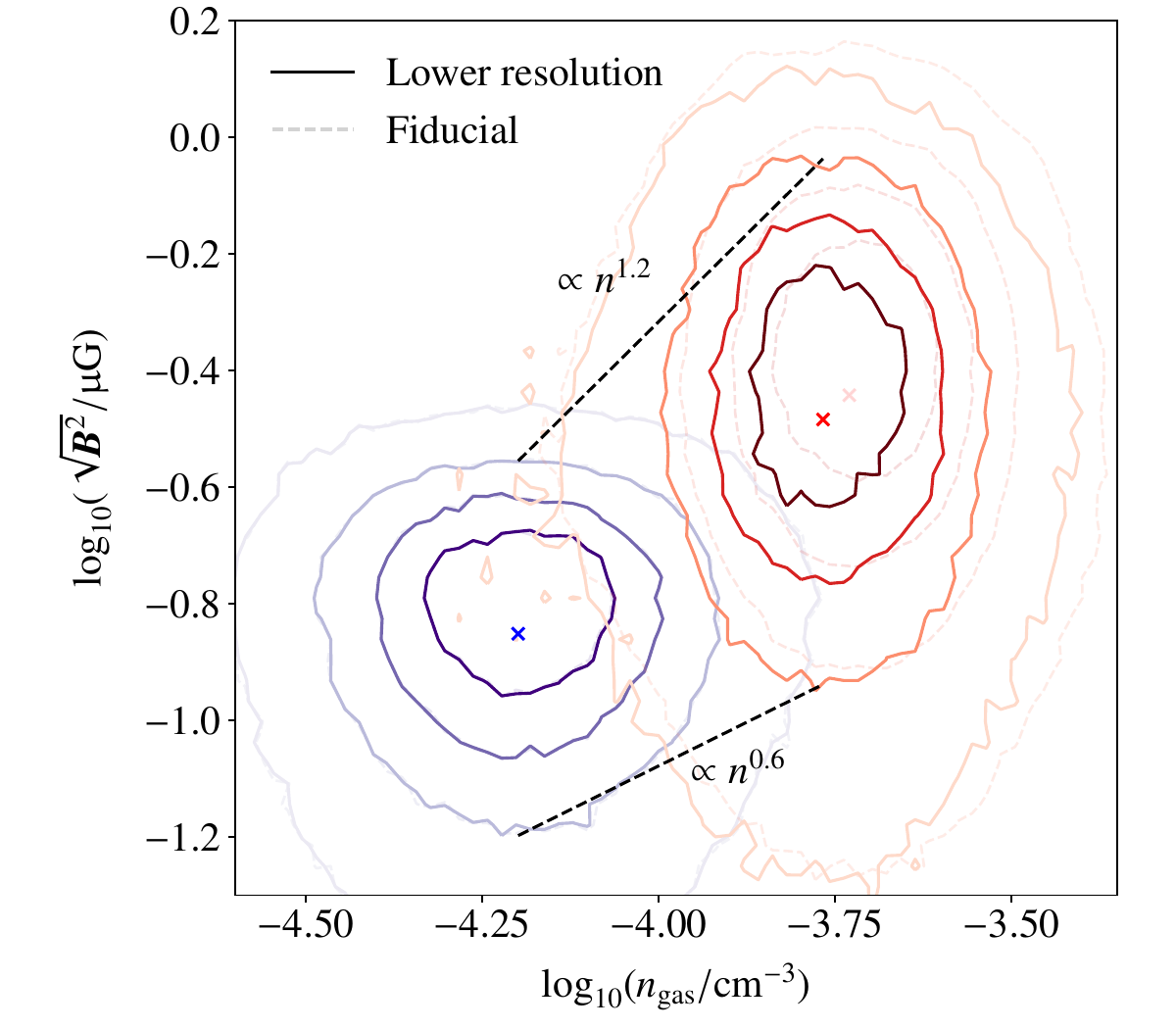}
    \caption{As Fig.~\ref{figure:rho-B-phase-diagram}, except solid lines now show data from our lower resolution Mach 3 fiducial simulation. The faint, dashed lines behind indicate data from the original, higher resolution version. Dashed, black lines show how the 75\% percentile contour scales. It can be seen that the scalings remain the same as in Fig.~\ref{figure:rho-B-phase-diagram}. This is evidence against the existence of a dynamo in our simulations.}
    \label{figure:magnetic-amplification--resolution-study}
\end{figure}

In Sect.~\ref{subsec:mag-field-origin}, we claimed that a magnetic dynamo was unlikely to be active in our simulations. We show evidence for this in Fig.~\ref{figure:magnetic-amplification--resolution-study}. Here, we show how density and magnetic field strength scales for the lower resolution version of our Mach 3 fiducial simulation, as initially introduced in App.~\ref{appendix:numerical-stability}. For comparison, we show data from the original, higher resolution simulation behind in fainter, dashed lines.

It is evident that, whilst the lower resolution simulation does not quite reach the same peak magnetic field strength values, the empirical scalings are identical. In particular, any increase in magnetic field strength is compensated for by an additional increase in density. For subsonic turbulence, however, a magnetic dynamo should also result in additional amplification at the same density, thereby increasing $\alpha_B$, where $B \propto n^{\alpha_B}$. We have repeated this analysis on the individual magnetic field components and find the same behaviour here as well. We therefore conclude that a dynamo is not relevant for magnetic field amplification in our simulations.

\end{appendix}

\end{document}